\documentclass[aps,prx,reprint,superscriptaddress,longbibliography,nofootinbib]{revtex4-2}
\usepackage{graphicx}

\usepackage[dvipsnames]{xcolor}
\usepackage{amsfonts}
\usepackage{amsmath}
\usepackage[figuresright]{rotating}  
\usepackage{amssymb}
\usepackage{bm}
\usepackage{mathtools}
\usepackage{multirow}
\usepackage{tabularx}
\usepackage{adjustbox}
\usepackage{units}
\usepackage{lipsum}
\usepackage{physics}
\usepackage{setspace}

\allowdisplaybreaks

\usepackage[bookmarksnumbered,pdfpagelabels=true,plainpages=false,colorlinks=true,linkcolor=magenta,citecolor=magenta, urlcolor=blue]{hyperref}
\begin{document}

% affiliation
\newcommand{\TUM}{\affiliation{Technical University of Munich, TUM School of Natural Sciences, Physics Department, 85748 Garching, Germany}}
\newcommand{\MCQST}{\affiliation{Munich Center for Quantum Science and Technology (MCQST), Schellingstr. 4, 80799 M{\"u}nchen, Germany}}

\title{Non-invertible Lattice 1-Form Symmetries for Non-Abelian Topological Order}
\author{Rafael Flores-Calder\'on}
\TUM \MCQST

\author{Frank Pollmann}
\TUM \MCQST

\author{Michael Knap} 
\TUM \MCQST

\date{\today}

\begin{abstract}

Higher-form symmetries generalize conventional global symmetries and act on lower-dimensional submanifolds of a quantum system.
While Abelian topological phases can be organized by 1-form symmetries that form a group, non-Abelian topological phases based on finite groups require 1-form symmetry operators governed by non-invertible fusion algebras. 
In this work, we make this statement precise in quantum double lattice models \(\mathcal D(G)\) for finite non-Abelian groups $G$. 
We construct the electric, magnetic, and dyonic 1-form operators directly at the lattice fixed point and show that together they form a complete nonlocal diagnostic algebra for the topological Hilbert space. 
Using these operators, we explicitly determine the cylinder and torus ground-state subspaces for arbitrary finite \(G\). %, and show that the anomaly of mixed electric and magnetic loops provides a diagnostic of non-contractible flux sectors. 
Furthermore, we calculate the microscopic fusion and gluing of the 1-form symmetries and show that their topological deformation properties emerge after projection to the defect-free topological subspace. 
Our results establish ground states of non-Abelian quantum double models as a concrete microscopic realization of spontaneous non-invertible 1-form symmetry breaking and provide an operator language that may be useful for characterizing such states in quantum processors.

\end{abstract}

\maketitle

\tableofcontents

\section{Introduction}
\label{sec:introduction}

Symmetries provide a fundamental organizing principle for quantum phases of matter. Generalizations of conventional global onsite symmetries have revealed surprising new structures. Among them are higher-form symmetries: a $p$-form symmetry in $d$ spatial dimensions has symmetry operators in $d-p$ dimensions and charged operators in $p$ dimensions~\cite{GaiottoKapustinSeibergWillett2015,Wen2019EmergentHigherSymmetries,McGreevy2023GeneralizedSymmetries}. In this framework, conventional global onsite symmetries are 0-form symmetries, whose symmetry operators act on the entire system and whose charged operators are local. Higher-form symmetries with $p>0$ extend this notion to symmetries acting on lower-dimensional manifolds.

Once the microscopic representation of the symmetry is specified, intrinsic Abelian topological order in two spatial dimensions can be understood by the behavior of 1-form symmetries ~\cite{Batista, Nussinov,GaiottoKapustinSeibergWillett2015,Wen2019EmergentHigherSymmetries,McGreevy2023GeneralizedSymmetries,PaceWen2023,XuRakovszkyKnapPollmann2025,xuFM, liu_1form_2025,ChermanJacobson2024,
IqbalMcGreevy2022,
SomozaSernaNahum2021,
SernaSomozaNahum2024,
BhardwajLectures2024}. These symmetries can be spontaneously broken. Moreover, distinct 1-form symmetries can also possess a mixed anomaly, which is an obstruction to a unique, trivially gapped ground state that simultaneously preserves all of them. 
A familiar example is the $\mathbb Z_2$ toric code, where electric Wilson loops and magnetic ’t Hooft loops generate an anomalous pair of $\mathbb Z_2$ 1-form symmetries. Contractible symmetry loops act trivially within the ground-state subspace, whereas non-contractible loops act nontrivially and transform between distinct topological ground states. The resulting ground-state degeneracy can thus be interpreted as a consequence of spontaneous 1-form symmetry breaking. This closely parallels conventional spontaneous symmetry breaking of a 0-form symmetry: just as the global $\mathbb Z_2$ symmetry exchanges the two symmetry-breaking ground states of the ordered Ising model, non-contractible 1-form symmetry operators exchange distinct topological ground states.

In recent years, there has also been an intense effort to realize and probe topologically ordered phases and lattice gauge theories using quantum simulators and quantum processors. Much of this progress has focused on realizing Abelian topological orders in two dimensions, including $\mathbb Z_2$~\cite{Satzinger2021TopologicalOrderProcessor,Cochran2025ChargesStrings,Semeghini} and $U(1)$ topological order~\cite{Haghshenas2026, BrowaeysGroup2026, LukinGroup2026, karch2026}. 

For non-Abelian topological order, however, both the underlying structure and the potential applications are considerably richer, and experimental realizations have only recently begun to emerge~\cite{Iqbal2024NonAbelianTO,XuSun,Evered, Will2025NonEquilibriumTopologicalOrder,Lo2026UniversalS3}.
The extension from Abelian to non-Abelian topological order requires a corresponding extension of the symmetry framework. Non-Abelian anyons do not fuse according to a group: the fusion of two anyon types can produce multiple superselection sectors. Consequently, the associated extended symmetry operators cannot, in general, form an ordinary group of invertible symmetries. Instead, their multiplication is governed by non-invertible fusion algebras~\cite{Kaidi2022HigherCentralCharges,Yu2021GaugingCategoricalSymmetries,Benini2023Factorization,McGreevy2023GeneralizedSymmetries,BhardwajBottiniSchaferNamekiTiwari2023,Shao2023}. Non-Abelian topological order has been studied for instance with
tensor-network methods
\cite{Buerschaper2009, Gu2009, Schuch2010,Bultinck2017, Sahinoglu2021, Cirac2021}, 
and by employing gauging and duality mappings
\cite{delcamp_higher_2024,InamuraOhmori2024,
Eck2025FusionSurfaceDualities,MoradiAksoyBardarsonTiwari2025,
GarreRubio2024,ChoiSanghaviShaoZheng2025,
BhardwajBottiniSchaferNamekiTiwari2025,
KongLanWenZhangZheng2020,BarkeshliBondersonChengWang2019,
LootensFuchsHaegemanSchweigertVerstraete2021,KitaevKong2012}. These advances suggest that non-Abelian topological order can be understood as spontaneous breaking of non-invertible higher-form symmetry; recently this question has been explored in the context of strong-to-weak symmetry breaking ~\cite{SongZhang2025} and for certain Honeycomb models based on fusion categories \cite{KitaevCategory}. Relatedly, in the context of decoherence, recent work has looked at the symmetries of non-Abelian anyon proliferation \cite{Vadali2026StatisticalMechanics}. Despite this conceptual expectation, a microscopic realization of this perspective in concrete non-Abelian lattice models has remained open. In particular, it is pertinent to identify the full set of lattice 1-form symmetry operators and establish explicitly how their algebra and action organize the topological ground-state space.

\begin{figure}
\includegraphics[width=0.48\textwidth]{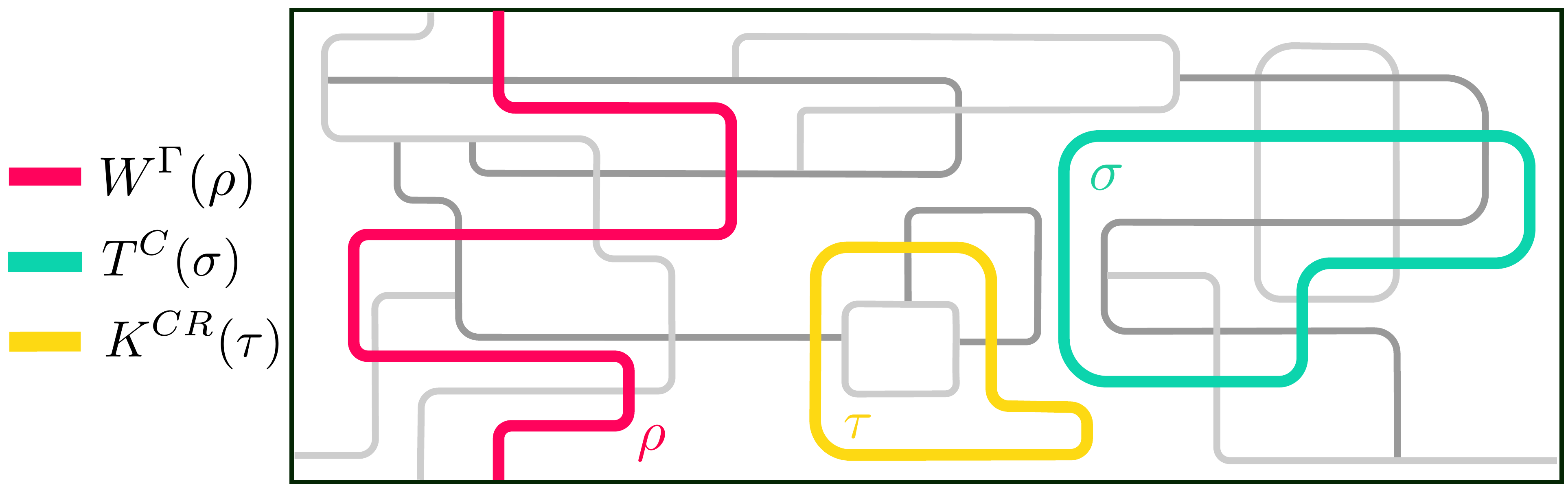}
\caption{\textbf{Illustration of non-invertible higher-form symmetries.} 
String-net picture of a torus ground state of a quantum double model with non-Abelian group  $G$ visualized as a condensate of closed magnetic strings (thin gray lines). The thick colored ribbons represent the non-invertible 1-form symmetry operators studied in this work. Electric Wilson loops $W^\Gamma(\rho)$ are labeled by irreducible representations $\Gamma$ of $G$, magnetic ’t Hooft loops $T^C(\sigma)$ by conjugacy classes $C$, and dyonic operators $K^{CR}(\tau)$ by a conjugacy class $C$ and an irreducible representation $R$ of the corresponding centralizer. Non-contractible loops connect distinct ground states (see Wilson loop as an example).
}
\label{fig:ClassyStringNet}
\end{figure}

In this work, we provide such a construction for the Kitaev quantum double models $\mathcal D(G)$ of arbitrary finite groups $G$~\cite{DijkgraafWitten1990,kitaev_fault-tolerant_2003}. We show that the same group-theoretic data that classify the topological sectors of $\mathcal D(G)$ also organize the microscopic non-invertible 1-form symmetry operators of the lattice model; see Fig.~\ref{fig:ClassyStringNet}. The electric operators are Wilson loops $W^\Gamma$ labeled by irreducible representations $\Gamma$ of $G$, the magnetic operators are ’t Hooft loops $T^C$ labeled by conjugacy classes $C$, and 
dyonic ribbon operators $K^{CR}$, labeled by a conjugacy class $C$ and an irreducible representation $R$ of the corresponding centralizer.
Together, these operators form a complete nonlocal diagnostic algebra for the topological Hilbert space. Crucially, for any given loop, the 1-form operators commute and are therefore Abelian as symmetry operators, while their fusion is non-invertible and can contain multiple channels. More subtle properties are found for operators supported on distinct loops. In that case, electric 1-form operators and magnetic 1-form operators fail to commute even in the flux-free sector, serving as a diagnostic of a mixed 't Hooft anomaly. Moreover, magnetic 1-form operators commute only in the flux-free sector while electric 1-form operators generally commute.

We use these operators to reconstruct the complete cylinder and torus ground-state subspaces for arbitrary finite $G$, and show that the expected topological fusion and gluing relations of the 1-form symmetries emerge from the microscopic lattice operators upon projection to the defect-free topological subspace. Throughout the paper, we illustrate the general construction using $\mathcal D(S_3)$, the quantum double of the smallest non-Abelian finite group. Our results thereby provide a microscopic symmetry-based framework for non-Abelian topological order and an operator language that can help characterize such states in quantum processors.

The remainder of the paper is organized as follows. In Sec.~\ref{SecII:Abelian}, we review the toric code as the Abelian prototype, where topological ground-state degeneracy can be understood in terms of spontaneous breaking of an anomalous pair of group-like 1-form symmetries. In Sec.~\ref{sec:main-results}, we summarize our main results and contrast the Abelian and non-Abelian cases in Table~\ref{tab:Abelian-nonAbelian-summary}. In Sec.~\ref{sec:non-ab-KQD}, we introduce the non-Abelian quantum double model $\mathcal D(G)$ and its string-net ground-state structure. We also introduce $\mathcal D(S_3)$ as an explicit example used throughout the paper. In Sec.~\ref{sec:electric-1form}, we construct the electric Wilson-loop operators and derive their $\mathrm{Rep}(G)$ fusion algebra. In Sec.~\ref{sec:ribbons}, we review open and closed ribbon operators and the conjugacy-class and centralizer data required to organize non-Abelian closed ribbons. In Sec.~\ref{sec:magnetic-1form}, we construct the magnetic conjugacy-class ’t Hooft loops and derive their fusion algebra, emerging in the flux-free sector. In Sec.~\ref{sec:anomaly}, we determine the mixed Wilson–’t Hooft algebra and introduce the dyonic closed-ribbon operators that complete the pure electric and magnetic sectors. In Sec.~\ref{sec:ground-states}, we use these non-contractible operators to reconstruct the cylinder and torus ground-state subspaces, including an explicit analysis of $\mathcal D(S_3)$. We conclude in Sec.~\ref{sec:discussion} with a discussion of the implications of our results and future directions.

\section{Abelian Topological Order: $\mathbb{Z}_2$ toric code} \label{SecII:Abelian}
As a warm-up for the non-Abelian case, let us summarize the results for the simplest Abelian topological order: the $\mathbb{Z}_2$ toric code \cite{kitaev_fault-tolerant_2003}. In the following we work on the square lattice; the total Hilbert space is composed of qubits living on the edges, denoted by $j$ %or $\ell$, 
of the lattice $\mathcal{H}=\otimes_j \mathbb{C}^2\equiv\otimes_j \mathbb{C}[\mathbb{Z}_2]$. Here, the group $\mathbb{Z}_2=\{1,-1\}$ labels the two linearly independent states for a single edge qubit.  The Hamiltonian is given by a sum of star and plaquette projectors
\begin{align}
    H_\text{TC}=-\sum_s A(s)-\sum_p B(p),\label{H_TC}
\end{align}
where  $A(s)=(1+A_s)/2,\ B(p)=(1+B_p)/2$ and we have the usual star and plaquette operators $A_s= \prod_{j\in \text{star}(s)}X_j$ given by the product of Pauli $X$ operators surrounding a vertex $s$, and $ B_p=\prod_{j\in \partial p}Z_j$ is the product of Pauli $Z$ operators around a plaquette $p$. The model is exactly solvable since the Hamiltonian is a sum of commuting projectors. Therefore, the ground state subspace is given by $\mathcal{L}=\{ \ket{\psi}\in \mathcal{H} \ | \ A(s)\ket{\psi}=\ket{\psi}, B(p)\ket{\psi}=\ket{\psi}, \forall \ s,p  \}$. The excitations on top of the ground state are composed of four Abelian anyons labeled $1,e,m,\epsilon$, where the trivial anyon/vacuum is denoted $1$ and the $e$ ($m$) anyon violates a star (plaquette) constraint~\cite{kitaev_fault-tolerant_2003}. The bound state of the $e$ and $m$ anyons is $\epsilon$, and it has fermionic statistics. The anyonic excitations of the model do not only determine the spectrum of the Hamiltonian but they also determine together with the topology of the space the ground state degeneracy. 

\begin{table}[t]
\centering
\renewcommand{\arraystretch}{1.2}
\setlength{\tabcolsep}{14pt}
\begin{tabular}{cc|cc}
\hline\hline
& & \multicolumn{2}{c}{\({W_x^e}\)} \\
& & \(1\) & \(-1\) \\
\hline
\multirow{2}{*}{\({W_y^e}\)}
& \(1\)  & \(\ket{[++]} \) & \(\ket{[+-]}\) \\
& \(-1\) & \(\ket{[-+]} \) & \(\ket{[--]}\) \\
\hline\hline
\end{tabular}

\caption{Electric 1-form symmetry eigenvalues for the fourfold degenerate ground states of the $\mathbb{Z}_2$ toric code on a torus.} \label{table:TC_wilson}
\end{table}
If the system is placed on a sphere or the plane with appropriate boundary conditions, there is a unique ground state. However, for the torus, there is a fourfold degeneracy, a hallmark of topological order. We can construct such ground states by observing that the toric code Hamiltonian~\eqref{H_TC} has a 1-form symmetry. For our (2+1)D lattice system, a 1-form operator is supported on a closed dimension one manifold, in other words a loop $\mathcal{C}$. There are two distinct sets of operators that commute with the Hamiltonian, they are the so-called electric and magnetic 1-form operators also denoted Wilson and 't Hooft loops. We define them as
\begin{align}
W^e(\mathcal{C}) = \prod_{j \in \mathcal{C}} Z_j,\quad  T^m(\tilde{\mathcal{C}}) = \prod_{\tilde{\ell} \in \tilde{\mathcal{C}}} X_\ell,
\end{align}
where $\mathcal{C}$ is a closed loop of the lattice and we defined the loop $\tilde{\mathcal{C}}$ as living on the dual lattice. The dual lattice is constructed by placing a dual site at the center of each plaquette of the direct lattice, and by connecting two dual sites whenever the corresponding plaquettes share an edge. For all direct and dual lattice loops the 1-form operators are an exact symmetry of the toric code Hamiltonian:
\begin{align}
    [H_\text{TC}, W^e(\mathcal{C})]=0, \quad [H_\text{TC}, T^m(\tilde{\mathcal{C}}) ]=0.
\end{align}
The reason for calling these operators electric and magnetic comes from considering an open string $\mathcal{C}\ (\tilde{\mathcal{C}})$. If we label the direct (dual) lattice endpoints by $s,s'$ ($p,p'$), the excitations created at the ends of the string are electric $e$ (magnetic $m$) anyons that violate the star (plaquette) constraints at the endpoints. For the Abelian case, the algebra of the 1-form operators is 
\begin{align}
    [W^e(\mathcal{C}),W^e(\mathcal{C}')]&=0, \quad [T^m(\tilde{\mathcal{C}}),T^m(\tilde{\mathcal{C}}')]=0,\\
    (W^e(\mathcal{C}))^2&=1, \quad (T^m(\tilde{\mathcal{C}}))^2=1. \label{Abelian1-formalgebra}
\end{align}
The first line establishes that the electric or magnetic operators always commute with themselves. It is important to remark that deformations of the 1-form operators are generated by smaller loops which means $W^{e}(\mathcal{C}_1\oplus \mathcal{C}_2)= W^{e}(\mathcal{C}_1)W^{e}(\mathcal{C}_2)$, where $\oplus$ denotes the operation of combining loops. Similarly bigger magnetic loops can be built from smaller ones $T^{m}(\tilde{\mathcal{C}_1}\oplus \tilde{\mathcal{C}}_2)= T^{m}(\tilde{\mathcal{C}}_1)T^{m}(\tilde{\mathcal{C}}_2)$.  For the non-Abelian case this relation will turn out to apply only in a defect free region. The last line of Eq.~\eqref{Abelian1-formalgebra} establishes that the 1-form symmetry is group-like with the group $G=\mathbb{Z}_2$. We denote the combined electric and magnetic 1-form symmetry by
\begin{align}
    \mathbb{Z}_2^{(1)}\times  \mathbb{Z}_2^{(1)}.
\end{align}
Nevertheless, this generalized symmetry is anomalous since the two distinct 1-form operators need not commute. Let us define $W_x^e=W^e(\mathcal{C}_x)$ as a 1-form operator supported on a non-contractible loop $\mathcal{C}_x$ of the torus winding around the $x$ direction and  $T^m_y=T^m(\tilde{\mathcal{C}}_y)$ analogously for the dual lattice and the $y$ direction. These operators anticommute
\begin{align}
    W_x^eT^m_y=-T^m_y W_x^e.
\end{align}
This anticommutation relation is the operator manifestation of a mixed ’t Hooft anomaly between the electric and magnetic 1-form symmetries. More generally, a mixed ’t Hooft anomaly occurs when two symmetries are separately well defined but cannot be coupled simultaneously to background gauge fields in a gauge-invariant way; equivalently, a background gauge transformation for one symmetry changes the partition function by a phase that depends on the background field of the other symmetry. Such an anomaly is preserved under renormalization-group flow and must therefore be matched by the infrared theory~\cite{tHooft1980}. In the toric code, the anomaly is directly encoded in the nontrivial algebra of intersecting electric and magnetic 1-form symmetry operators and is matched by the topological ground-state degeneracy, which can be understood as spontaneous breaking of the 1-form symmetries.
% \begin{table*}[t]
% \centering
% \renewcommand{\arraystretch}{1.35}
% \begin{tabular}{l|l|l}
% \hline\hline
% 1-form properties & toric code / Abelian order & Non-Abelian quantum double \(\mathcal D(G)\) \\
% \hline
% Wilson/Electric loops
% &
% Labeled by one-dimensional irreps
% &
% \(W^\Gamma\), \(\Gamma\in\mathrm{Rep}(G)\) at least one $\text{dim}(\Gamma)>1$
% \\

% 't Hooft/ Magnetic loops
% &
% Labeled by group elements
% &
% \(T^C\), conjugacy classes  \(C \in \mathrm{Class}(G)\)
% \\

% Fusion
% &
% Group-like and invertible (single channel)
% &
% Non-invertible (multi-channel)
% \\

% Commutativity
% &
% Abelian
% &
% Abelian (emergent at low energy)
% \\

% Topological degeneracy
% &
% Electric and magnetic loops suffice
% &
% Dyonic ribbons \(K^{CR}\) needed,\\
% && \; \((C,R)\), with \(C \in \mathrm{Class}(G)\), \(R\in\mathrm{Rep}(Z_C)\)
% \\
% \hline\hline
% \end{tabular}
% \caption{
% Comparison between Abelian and non-Abelian topological order in terms of loop
% operators and higher-form symmetries.
% }
% \label{tab:Abelian-nonAbelian-summary}
% \end{table*}
\begin{table*}[t]
\centering
\renewcommand{\arraystretch}{1.35}
\setlength{\tabcolsep}{5pt}
\newcommand{\tcell}[2]{%
  \parbox[t]{#1}{\raggedright #2\par}%
}
\begin{tabular}{lll}
\hline\hline
\tcell{0.23\textwidth}{\textbf{1-form properties}}
&
\tcell{0.31\textwidth}{\textbf{toric code / Abelian order}}
&
\tcell{0.40\textwidth}{\textbf{Non-Abelian quantum double \(\mathcal D(G)\)}}
\\
\hline
\tcell{0.23\textwidth}{Wilson/Electric loops}
&
\tcell{0.31\textwidth}{Labeled by one-dimensional irreducible representations}
&
\tcell{0.40\textwidth}{%
  \(W^\Gamma\), \(\Gamma\in\mathrm{Rep}(G)\) at least one
  \(\text{dim}(\Gamma)>1\)%
}
\\[0.4em]
\tcell{0.23\textwidth}{'t Hooft/ Magnetic loops}
&
\tcell{0.31\textwidth}{Labeled by group elements}
&
\tcell{0.40\textwidth}{%
  \(T^C\), conjugacy classes \(C \in \mathrm{Class}(G)\)%
}
\\[0.4em]
\tcell{0.23\textwidth}{Fusion}
&
\tcell{0.31\textwidth}{Group-like and invertible (single channel)}
&
\tcell{0.40\textwidth}{Non-invertible (multi-channel)}
\\[0.4em]
\tcell{0.23\textwidth}{Commutativity}
&
\tcell{0.31\textwidth}{Abelian}
&
\tcell{0.40\textwidth}{Abelian (emergent at low energy)}
\\[0.4em]
\tcell{0.23\textwidth}{Topological degeneracy}
&
\tcell{0.31\textwidth}{Electric and magnetic loops suffice}
&
\tcell{0.40\textwidth}{%
  Dyonic ribbons \(K^{CR}\) needed,\\
  \(\;(C,R)\), with \(C \in \mathrm{Class}(G)\),
  \(R\in\mathrm{Rep}(Z_C)\)%
}
\\
\hline\hline
\end{tabular}
\caption{
Comparison between Abelian and non-Abelian topological order in terms of loop
operators and higher-form symmetries.
}
\label{tab:Abelian-nonAbelian-summary}
\end{table*}

In the toric code case the anomaly implies that we cannot simultaneously diagonalize the electric and magnetic 1-form operators. We must choose a basis to explicitly find the ground state of the system. Let us choose as reference state the magnetic vacuum (zero-flux state) which is the product state $\ket{\mathbf{1}}\equiv \otimes_j \ket{1}$. We now  use the projector to the ground state given by $P_\text{GS}=\prod_s A(s) \prod_p B(p)$ where we have fixed the representation of the symmetry by the definitions of $A(s)$ and $B(p)$. Since $B(p)\ket{\mathbf{1}}=\ket{\mathbf{1}}$, we need only apply the projector of the star operators to obtain the flux-free ground state
\begin{align}
   \ket{\Psi_\text{TC}}=|\left.[++]\right>\equiv\prod_s A(s)\ket{\mathbf{1}}.
\end{align}
In the plane with proper boundary terms, this will be the unique ground state. Physically it is a massive superposition of all possible magnetic loops in the lattice built on top of the magnetic vacuum. We construct now a distinct ground state by threading a flux loop around the x-direction with a non-contractible 't Hooft loop $\ket{[-+]}= T_x^m\ket{\Psi_\text{TC}}$, similarly for the other directions we get $\ket{[+-]}= T_y^m\ket{\Psi_\text{TC}}$ and $\ket{[--]}= T_y^mT_x^m\ket{\Psi_\text{TC}}$. The four ground states can be distinguished by their eigenvalue of the electric 1-form operators because of the  mixed 't Hooft anomaly, see Table~\ref{table:TC_wilson}. By contrast, these ground states are not symmetric under the magnetic 1-form operators (as the minimally entangled states of Ref.~\cite{Minimal-entangled-states}) since they switch between ground states. Thus, one says then that the system has spontaneously broken the magnetic 1-form symmetry. 

Next, we will study extensions of these concepts to non-Abelian topological order, which as we will see, has profound consequences on the structures of the 1-form symmetries.

\section{Non-Abelian Topological Order: Main results}
\label{sec:main-results}

Before introducing the microscopic quantum-double Hamiltonian, we summarize
the organizing principle and the main contributions of this work. The basic
topological data of a finite-group quantum double \(\mathcal D(G)\) are 
known \cite{kitaev_fault-tolerant_2003}. Electric charges are labeled by irreducible representations $\Gamma \in \mathrm{Rep}(G)$ of the group \(G\),
magnetic fluxes are labeled by conjugacy classes $C$ of \(G\), and general dyonic
anyons are labeled by pairs \((C,R)\), where
\(C\in\mathrm{Class}(G)\) is a conjugacy class of $G$ and \(R\in\mathrm{Rep}(Z_C)\) is an irreducible
representation of the centralizer $Z_C$ of a representative of \(C\). %Hereafter, we will use $\mathrm{Rep}(G)$ as the set of irreps, which are the simple objects in the representation category. 
What we show
is that the same group-theoretic data also organize the non-invertible
1-form symmetry operators of the lattice model.

The first main result of this work is the unification of several perspectives
on non-Abelian topological order within one microscopic lattice framework.
From the string-net point of view, the ground state is a condensate of closed
magnetic strings~\cite{LevinWen2005}. From the anyon point of view, the topological sectors are
classified by the quantum-double labels \((C,R)\). From the generalized
symmetry point of view, the relevant extended operators form non-invertible
higher-form symmetry algebras. We connect these viewpoints explicitly in the
lattice quantum double by deriving the electric Wilson loops, magnetic
't Hooft loops, and dyonic closed ribbons from the microscopic \(G\)-qudit
operators. In this way, the string-net condensate, the anyon labels, the loop
algebras, the anomaly, and the topological ground states are all tied to the
same underlying group-theoretic structure. A summary of the differences for the 1-form properties between Abelian and non-Abelian topological orders is given in Table~\ref{tab:Abelian-nonAbelian-summary}. In both cases the 1-form symmetries commute; however, while their fusion is invertible for Abelian topological order it is non-invertible for non-Abelian topological order. %Note that the commutativity does not imply invertible operators. 
The non-invertibility is reflected in the multi-channel fusion of operators for non-Abelian topological order. The relevant 1-form operators that detect the 1-form symmetry broken states are schematically shown in Fig.~\ref{fig:ClassyStringNet}.

The second main result is an explicit construction of the complete
topological ground-state subspace for arbitrary finite \(G\) directly in terms
of non-contractible 1-form operators. On the cylinder, non-contractible magnetic loops generate
the sectors labeled by conjugacy classes. On the torus, the full ground-state
subspace is obtained by threading compatible 1-form operators along the two
non-contractible cycles. Equivalently, the resulting sectors are labeled by
commuting pairs \((g,h)\in G\times G\), modulo simultaneous conjugation. This
gives a direct operator-level construction of the full finite-group quantum double
\(\mathcal D(G)\) torus degeneracy, rather than only a counting argument or an
abstract categorical classification. In the \(\mathcal D(S_3)\) example discussed below, the
construction makes explicit why pure electric and magnetic loops only almost
resolve the topological sectors, and why one additional dyonic operator is
needed to distinguish the residual centralizer-charge data.

The third main result concerns a microscopic subtlety of 1-form operator fusion. At the level of the topological field theory, line operators can be fused,
deformed, and composed according to their topological fusion rules \cite{Kaidi2022HigherCentralCharges,McGreevy2023GeneralizedSymmetries}. On the
lattice, however, these relations are not always exact identities on the full
Hilbert space. When loops are fused or deformed in the presence of local
defects, residual microscopic operators can remain. The expected topological
relations are recovered only after projection back to the defect-free
topological subspace. We make this projection explicit for both electric and
magnetic 1-form operators. This provides a concrete bridge between microscopic
lattice operators and the emergent non-invertible topological 1-form symmetries. 

\section{Kitaev quantum double} \label{sec:non-ab-KQD}
\subsection{General formalism}
\begin{figure}[t]
    \centering
    \includegraphics[width=1\linewidth]{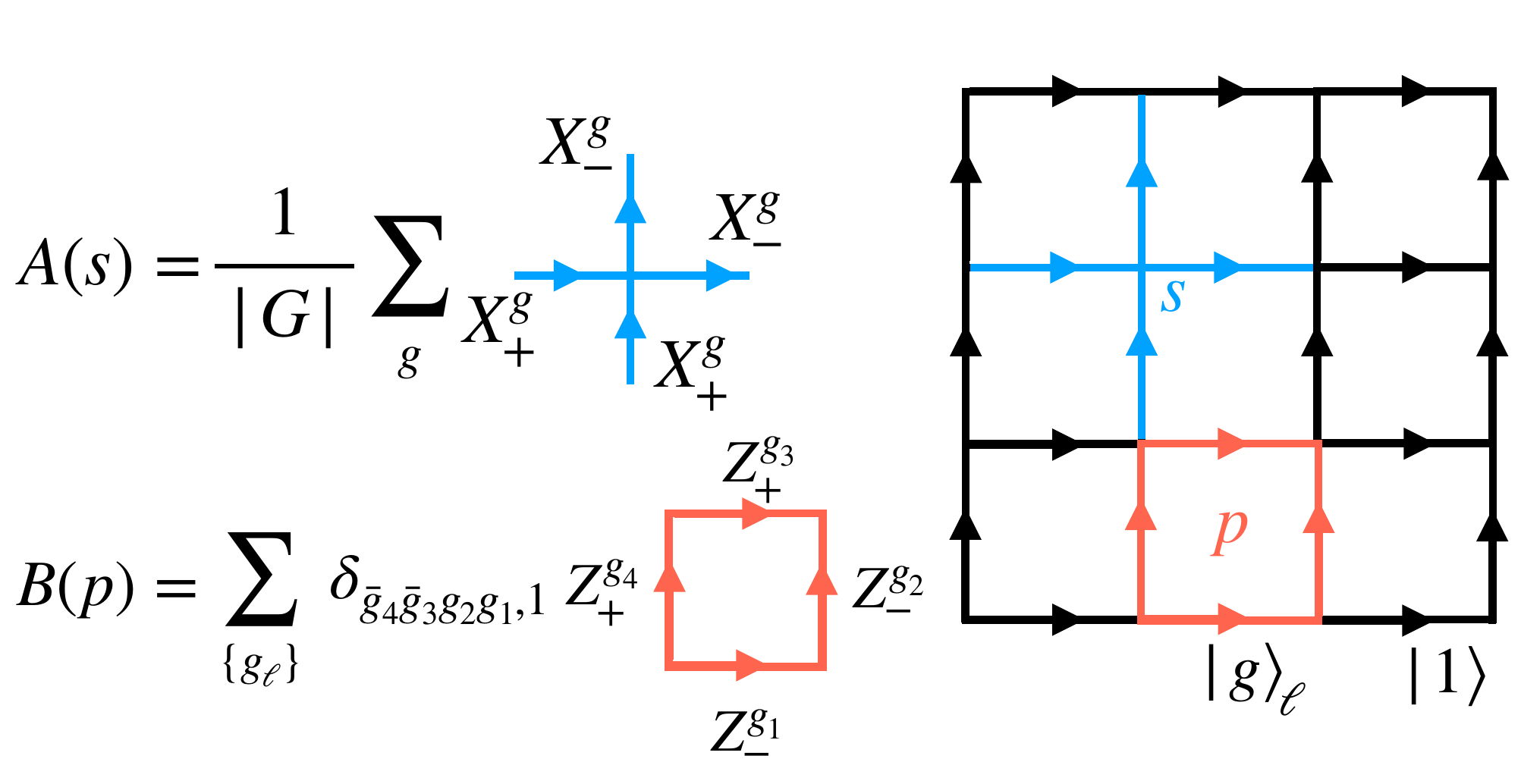}
    \caption{\textbf{Star and plaquette operators.} Star $A(s)$ and plaquette $B(p)$ operators for the $G$ quantum double model on the square lattice with all horizontal arrows pointing to the right and vertical arrows to the top. A $G$-qudit resides on the edges of the lattice, a basis is given by all group elements $\ket{g},\ g\in G$.}
    \label{fig:GQDM}
\end{figure}

Having recapitulated the 1-form symmetry properties for an Abelian case, let us now focus on the general non-Abelian case. 
Let us consider a finite group $G$ with total number of group elements $\abs{G}$. We define a $G$-qudit as a state in the local Hilbert space $\mathbb{C}[G]$ generated by all group elements $\{\ket{g}, \forall g \in G\}$. We denote the inverse of the group element by a bar $\bar{g}\equiv g^{-1}$. Consider a 2D square lattice with a fixed orientation for each edge We place a $G$-qudit at each edge see Fig.~\ref{fig:GQDM} (changing the direction of the arrow corresponds to the basis change $\ket{g}\rightarrow \ket{\bar{g}}$, $g \in G$). The total Hilbert space is a tensor product of the edges $\mathcal{H}=\otimes_{j}\mathcal{H}_j$ with $\mathcal{H}_j=\mathbb{C}[G]$. Operators in $\mathbb{C}[G]$ acting on the qudits can return the same element with some multiplicative factor or switch between group elements. The simplest operators accomplishing this are generalizations of the Pauli matrices $X,Z$. For a general non-Abelian group we define $X^g_{\pm}$ and $Z^g_{\pm}$ indexed by a group element $g\in G$. They act on the basis states as follows
\begin{align}
    &X^g_+ \ket{h} = \ket{gh} , \qquad \; \;\;\;    X^g_- \ket{h} = \ket{h\bar{g}} \\
   &Z^g_+ \ket{h} = \delta_{h,g}\ket{h} , \qquad   Z^g_- \ket{h} = \delta_{\bar{h},g}\ket{h} ,
\end{align}
where the sign indicates if one acts on the left or right of a qudit, for details see Ref. \cite{kitaev_fault-tolerant_2003} and App.~\ref{A:qudits}. For a given edge $j$ let $s$ denote one of the endpoints. Define $X^g(j,s)=X^g_\pm (j)$ with $+$ if $s$ is the endpoint of the arrow (orientation of the link), otherwise if it is the origin choose $-$. Similarly, if $p$ is the left (the right) adjacent face of the edge $j$, then $Z^h(j,p)$  is $Z_-^h(j)$ ($Z_+^h(j)$) respectively. In analogy to the Abelian case, we define the star and plaquette operators as (see Fig.~\ref{fig:GQDM}):
\begin{align}
&A(s)=\dfrac{1}{\abs{G}} \sum_{g \in G}  \prod_{j \in \operatorname{star}(s)} X^g(j, s), \\ 
&B(p)= \sum_{h_1 \cdots h_4=1} \prod_{m=1}^4 Z^{h_m}\left(j_m, p\right),\label{ABops}
\end{align}
where $j_1,\dots, j_4$ are boundary edges of the plaquette $p$ listed in counterclockwise notation starting from and ending at $s$. The sum is taken over all group element combinations satisfying $h_1\dots h_4=1$.  More generally one can define operators $A_g(s)$ and $B_h(p)$ (see App.~\ref{A:gKqdm}) which generate an algebra denoted $\mathcal{D}(G)$ called \textit{ Drinfeld's quantum double} of the group algebra $\mathbb{C}[G]$. 
For a general group $G$ the quantum double fixed point Hamiltonian is  \cite{kitaev_fault-tolerant_2003}
\begin{align}
    H_\text{QD}= \sum_s(1-A(s))+ \sum_p (1-B(p)). \label{KQDM}
\end{align}

The space of ground states with zero energy is the protected subspace 
\begin{align}
    \mathcal{L}=\{|\psi\rangle \in \mathcal{H}: \ A(s)|\psi\rangle=|\psi\rangle, \; B(p)|\psi\rangle=|\psi\rangle \ \forall \ s, p\}.\label{KQDM_GS}
\end{align}
All excitations have finite positive energy, as in the Abelian case, and the $A,B$ operators are mutually commuting orthogonal projectors.

\begin{figure}[t]
    \centering
    \includegraphics[width=1\linewidth]{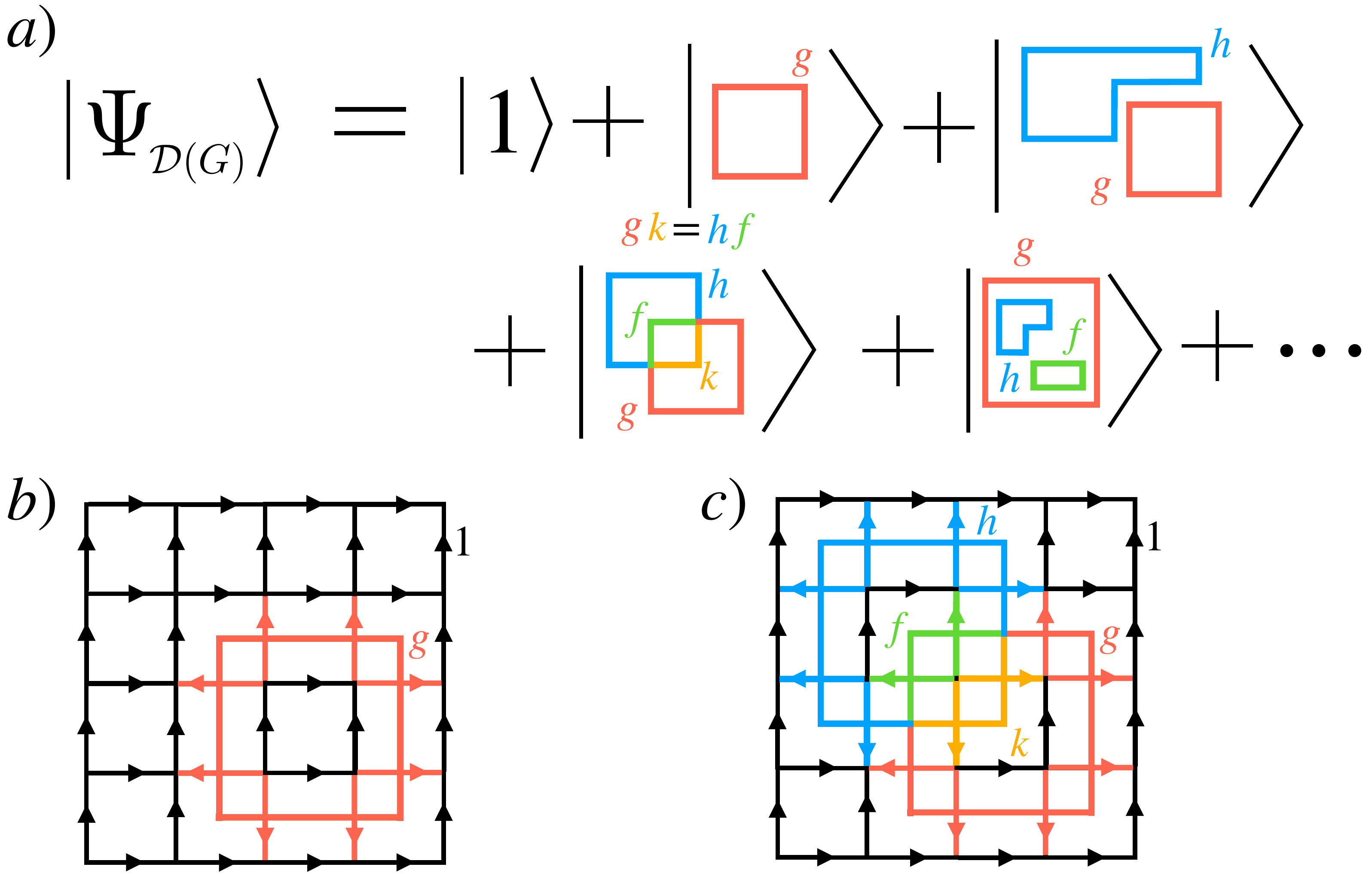}
    \caption{\textbf{Structure of the ground state wave function.}  a) Ground state wavefunction for the quantum double model in the $G$ basis; $1$ is the identity element, and all group elements are summed over. b) Detailed signs of a lattice with one $g\in G$ dual lattice loop. c) When two loops $g,h$ intersect, the Gauss law allows for transmutation of the inner paths into different group elements $k,f$ as long as $gk=hf$. }
    \label{fig:psi_G}
\end{figure}

Before studying further properties of the quantum double algebra, let us construct the ground state of Eq.~\eqref{KQDM} on the infinite plane or sphere, illustrated in Fig.~\ref{fig:psi_G} a). As in the toric code, the ground state is non-degenerate and satisfies $B(p)\ket{\Psi}=\ket{\Psi}$
for every plaquette $p$. The plaquette operator acts diagonally in the group-element basis and enforces vanishing flux through each plaquette $B(p)\ket{g_1,g_2,g_3,g_4} = \delta_{\bar{g}_4\bar{g}_3g_2g_1,1}\ket{g_1,g_2,g_3,g_4}$  (see the red square in Fig.~\ref{fig:GQDM}). To express this constraint compactly, we associate a group element $g_\ell\in G$ with each oriented edge $\ell$, with reversal of the edge orientation corresponding to $g_{\bar\ell}=\bar{g}_\ell$. The zero-flux constraint then takes the form
$\prod_{\ell\in\partial p} g_\ell = 1,$
where the product is ordered along the oriented boundary $\partial p$ of the plaquette. This is the zero-flux constraint or magnetic Gauss law for a classical $G$ lattice gauge theory, see Ref.~\cite{henley2013}. We can construct the ground state starting from the projector to the ground state subspace $P_\text{GS}= \prod_{s}A(s)\prod_{p}B(p)$ acting on the trivial magnetic vacuum $\ket{\textbf{1}}=\otimes_j \ket{1}_j$ then $\ket{\Psi_{\mathcal{D}(G)}} = P_\text{GS}\ket{\textbf{1}} = \prod_{s}A(s) \ket{\textbf{1}}$.
Different configurations can be generated by loops on top of the magnetic vacuum in the dual lattice with a label $g\in G$. Due to the magnetic Gauss law, the loops cannot change their group element, labelling them. Indeed, a corner of the loop will change the group element of one edge, but only into its inverse $\bar{g}$, which can then be turned back into $g$ by flipping the sign of the arrow convention. The resulting loop will have all arrows pointing out as shown in Fig.~\ref{fig:psi_G} b). The oriented red dual lattice links are in the state $\ket{g}$.

The ground state wavefunction will have not only a superposition of all possible loops with distinct group elements, but also loops that intersect. If the two loops that intersect have labels $g,h$, then after the intersection, the inner sections can change group element or \textit{color} into $k,f$ for each strand of the path. The allowed new elements must satisfy the Gauss law at the intersections, which implies the equation $gk = hf$.
Any such $(k,f)$ is an allowed loop transmutation and will be included in the wave function $\ket{\Psi_{\mathcal{D}(G)} }$  as shown in Fig.~\ref{fig:psi_G} c). One example is if the loop $g$ continues being $g$ in that case $f=g$ and this implies $k=\bar{g}hg$, which is just the conjugate of $h$ by $\bar{g}$. Alternatively, $h$ can continue in which case $k=h$ and $f=\bar{h}gh$, now $g$ is conjugated by $\bar{h}$. 
\begin{align}
   \ket{\Psi_{\mathcal{D}(G)}} = \prod_{s}A(s) \ket{\textbf{1}}= \sum_{\{ g_\ell\}} \prod_p \delta_{\prod_{\ell\in \partial p}g_\ell,1 }\ket{\{ g_\ell\}}. \label{QD_gs}
\end{align}

We close this section by recapitulating the non-Abelian excitations of the quantum double model. Let us consider a site composed of a vertex and one of the four neighbouring plaquettes $z=(s,p)$ where $A(s)\neq 1$ or $B(p)\neq 1$. Kitaev showed that the products and sums of local operators form a finite-dimensional $C^*$-algebra isomorphic for all sites $z$, depending on the group only, denoted $\mathcal{D}(G)$, the \textit{quantum double} of the group \cite{kitaev_fault-tolerant_2003}. The irreducible representations of this algebra determine the quasiparticles or excitations of the system, they carry a topological label or superselection sector that cannot be changed by local operations, and these will be the non-Abelian anyons.

The irreducible representations of $\mathcal{D}(G)$ are described by first fixing a group element $u\in G$, then denoting its conjugacy class as $[u]= \{gu\bar{g}\ |\ g\in G\}$ \cite{kitaev_fault-tolerant_2003}. We define its centralizer by $Z_u=\{g\in G\ |\ gu=ug\}$. Next, we define the conjugacy class $C=[u]\in \text{Class}(G)$ with $\text{Class}(G)$ the set of all conjugacy classes for the group and $R=\Gamma_{Z_u}$ denotes an irreducible representation of the centralizer $Z_u$. If we choose another element $v\in [u]$ then $Z_v=pZ_u\bar{p}$ since there exists $p\in G$ such that $v=pu\bar{p}$. Because the groups are related by conjugation, one can write a isomorphism between both groups and denote the group independent of $u$ but dependent on the conjugacy class. Hence, we call it $Z_C$. The irreducible representations of the group are denoted $\Gamma^{C}\equiv R \in \text{Rep}(Z_C)$ so that the number of quasiparticles, which for a finite group are generically non-Abelian anyons are given by all combinations of $\mathfrak  a=(C,R)$. The vaccum sector or trivial anyon is $([1],I)$ composed of the identity class and trivial irreducible representation. In general one may define the \textit{quantum dimension} of an anyon as 
\begin{align}
    d_{\mathfrak a}=|(C,R)|= \abs{C}\text{dim}(R).
\end{align}
If the quantum dimension is bigger than one $(d_{\mathfrak  a}>1)$ then the $\mathfrak  a$ anyon is a non-Abelian anyon.

\subsection{$S_3$ quantum double} \label{sec:S3-example}
\begin{figure}
\includegraphics[width=0.9\linewidth]{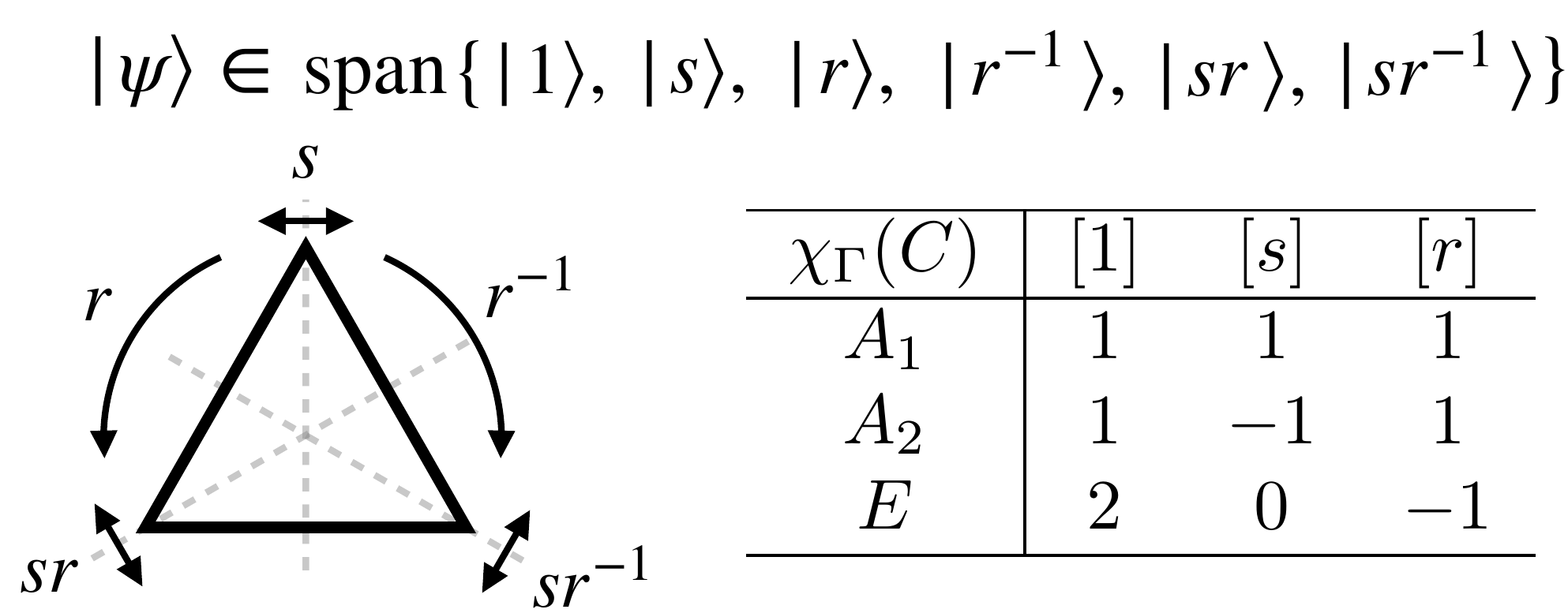}
\caption{\textbf{$S_3$-qudit and character table.} Elements of the group $S_3$ are the symmetry operations of an equilateral triangle (bottom left) and the properties of the group are summarized in its character table (bottom right). A general $S_3$-qudit $\ket{\psi}$ is given by a linear combination of all group elements (top).} \label{fig:s3_triangle}
\end{figure}

To illustrate the concepts in a clear way we will consider as a guiding example the smallest non-Abelian group $S_3$ throughout this paper. Here, we give some details on its group structure and the resulting quantum double $\mathcal{D}(S_3)$.
The group $S_3$ is the permutation group of three objects and
has order $|S_3|=6$. Equivalently, it is the dihedral group $D_3$ of symmetries of an equilateral triangle. We use the presentation

\begin{align}
S_3=\langle r,s \mid r^3=e,\ s^2=e,\ srs=r^{-1}\rangle ,
\end{align}
where $r$ is a rotation by $2\pi/3$ and $s$ is a reflection as seen in Fig.~\ref{fig:s3_triangle}. Its
elements are $ S_3=\{e,r,r^2,s,sr,sr^2\}$ which label the basis of the local Hilbert space; see Fig.~\ref{fig:s3_triangle} b).
The conjugacy classes are given by
\begin{align}
[1]&=\{e\},&
[s]&=\{s,sr,sr^2\},&
[r]&=\{r,r^2\}.
\end{align}
Thus $S_3$ has three conjugacy classes and hence three irreducible
representations $\Gamma$, conventionally denoted by $A_1$, $A_2$, and $E$. Their dimensions $d_\Gamma\equiv \text{dim}(\Gamma)$ are
$d_{A_1}=1,\  d_{A_2}=1, \ d_E=2$,
satisfying
$\sum_{\Gamma} d_\Gamma^2
=
1^2+1^2+2^2
=
|S_3|$.
Here $A_1$ is the trivial representation, $A_2$ is the sign
representation, and $E$ is the two-dimensional representation. The characters of $S_3$ are calculated from the definition $\chi_\Gamma(g)=\tr [\Gamma(g)]$ for $g\in S_3$ and $\Gamma \in \text{Rep}(S_3)$. Since character is a function of class we may write instead $\chi_\Gamma(C)$, $C\in \text{Class}(S_3)$, the resulting character table is presented in Fig.~\ref{fig:s3_triangle}. 
This will be important when talking about the mixed 't Hooft anomaly.

We also summarize the centralizer subgroups $Z_h$ of $S_3$. Recall they are defined to be the group elements that commute with a given $h\in G$ therefore
\begin{align}
Z_{1}=S_3,\ Z_{s}=\{1,s\}\simeq\mathbb{Z}_2,\ Z_{r}=\{1,r,r^2\}\simeq\mathbb{Z}_3. \label{eq:S3_central}
\end{align}
All other group elements have a centralizer subgroup that is isomorphic to one from Eq.~\eqref{eq:S3_central}, which is why we may denote them as $Z_C$ for $C$ a conjugacy class. Following the logic of the previous section we have eight anyons with labels
\begin{align}
&\mathrm{A}= ([1],A_1), \; \mathrm{B}=([1],A_2),\; \mathrm{C}= ([1],E),\\
&\mathrm{D}= ([s],A_1),\; \mathrm{E}=([s],A_2)\\
&\mathrm{F}=([r],A_1), \;  \mathrm{G}=([r],\omega),\; \mathrm{H}=([r],\bar{\omega}),
\end{align}
where we used $A_1=I$ as the trivial irreducible representation (irrep) and $\omega=e^{i2\pi/3}$ serves to label the irrep of $\mathbb{Z}_3$. We note that for $S_3$ the centralizer of proper subgroups are Abelian, which implies the irreps are given by phases. 
Here, the vacuum is $A$, the electric anyons are $\mathrm{B},\mathrm{C}$ ($C=[1]$) and the magnetic ones are $\mathrm{D},\mathrm{F}$ ($R=A_1$), while the dyons satisfy $R\neq A_1$ and $C\neq[1]$, which means there are three dyons for $S_3$ with labels $\mathrm{E},\mathrm{G},\mathrm{H}$.

\section{Non-invertible 1-form symmetry operators}
\subsection{Electric 1-form operators}\label{sec:electric-1form}
\begin{figure}
    \centering
    \includegraphics[width=\linewidth]{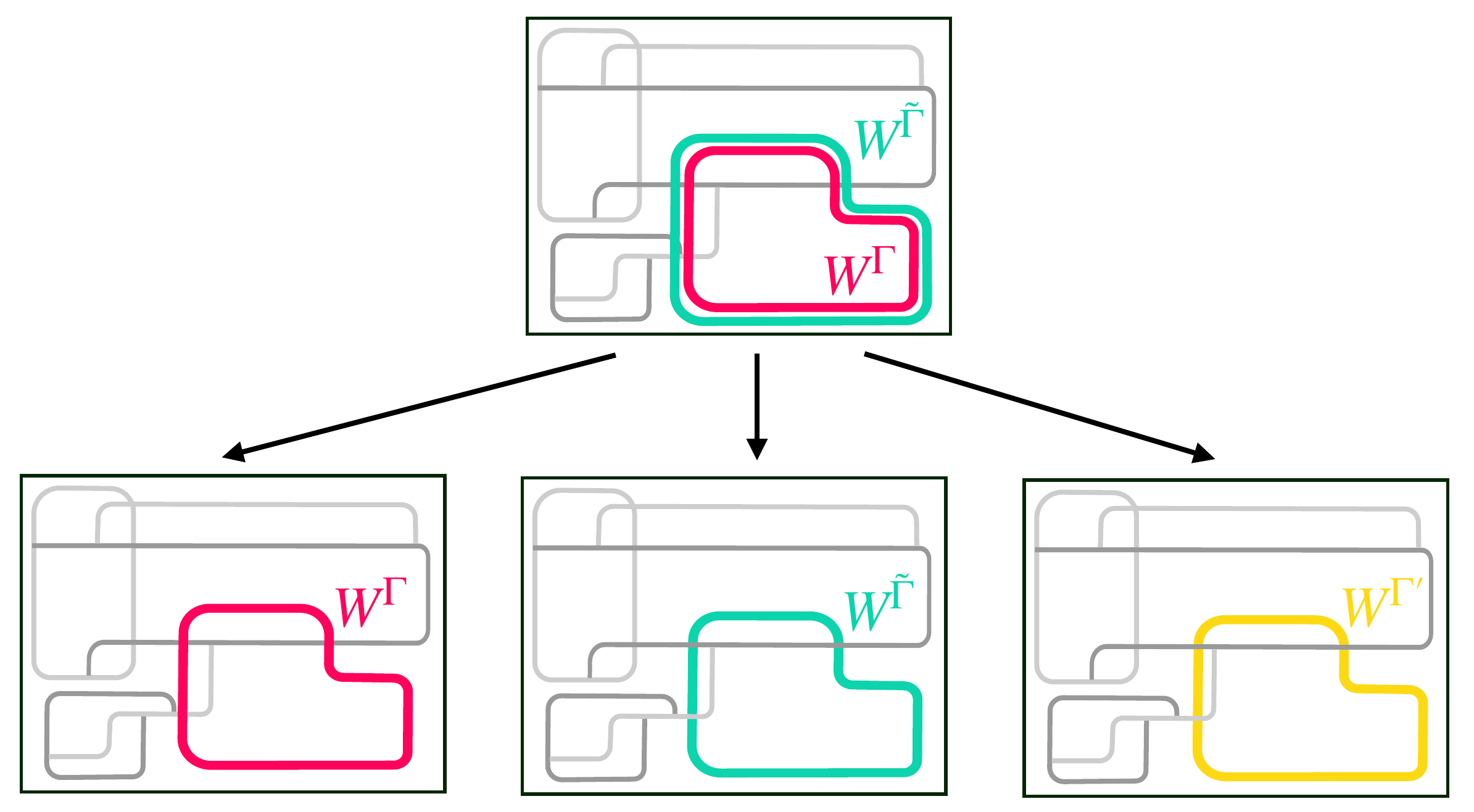}
    \caption{\textbf{Illustration of non-Abelian fusion of electric 1-form symmetries} For a non-Abelian group $G$ two Wilson loops labelled by irreps of the group $\Gamma,\tilde{\Gamma}$ fuse into multiple channels, see Eq.~\eqref{RepG_Wilson}. The multiplicity of a given channel $\Gamma'$ is $N^{\Gamma'}_{\Gamma\tilde{\Gamma}}$. }
    \label{fig:Wilsonfusion}
\end{figure}

The generalization of the Wilson loop or electric 1-form operator for non-Abelian topological order requires introducing a dual basis of the local Hilbert space $\mathbb{C}[G]$ where the electric charges are defined. This dual basis is an analog of the Fourier transform and involves the irreducible representations of the group, denoted $\Gamma$, which labels electric anyons. We define an orthonormal basis by $\ket{\Gamma_{\alpha\beta}}= \sqrt{\tfrac{d_\Gamma}{\abs{G}}} \sum_g \Gamma(g)_{\alpha\beta} \ket{g}$ with $\alpha,\beta$ indices for the irrep $\Gamma$. Dual $Z$ operators can be defined by $Z^\Gamma_{\alpha\beta}\ket{g}=\Gamma(g)_{\alpha\beta} \ket{g}$, details are given in App.~\ref{A:irrepbasis}. The Wilson loop, given by a product of $Z$ operators for the Abelian case, generalizes to 
\begin{align}
    W^\Gamma(\mathcal{C}) = \text{tr} \Big[\mathcal{P}\prod_{\ell\in \mathcal{C}}(\bm{Z}^{\Gamma}(\ell))^{O_\ell}\Big ]\equiv  \text{tr}[\bm{W}^\Gamma(\mathcal{C})] , \label{Wilson-Loops}
\end{align}
where we defined $O_\ell$ as the orientation of the edge $\ell$ with respect to the loop $\mathcal{C}$. If the orientations match $O_\ell=1$, otherwise, it acts as the adjoint $O_\ell=\dagger$ realizing the conjugate representation $\bm{Z}^{\bar{\Gamma}}$. We introduced a matrix Wilson loop $\bm{W}^\Gamma(\mathcal{C})$ with components $\alpha,\beta=1,\dots,d_\Gamma$ and $\mathcal{P}$ denotes path-ordering. In contrast to the Abelian case the electric 1-form operators do not form a group. Instead for two arbitrary irreps $\Gamma_1,\Gamma_2$ and a fixed closed loop $\mathcal{C}$ the product of Wilson loops reduces to computing the tensor product of irreps $\Gamma_1\otimes \Gamma_2$, as shown in App.~\ref{A:Wilsonfusion}, which can be decomposed into a sum of irreps:
\begin{align}
   W^{\Gamma_1}(\mathcal{C})W^{\Gamma_2}(\mathcal{C}) = \sum_\Gamma N^{\Gamma}_{\Gamma_1\Gamma_2} W^\Gamma(\mathcal{C}).\label{RepG_Wilson}
\end{align}
The fusion coefficients $N^{\Gamma}_{\Gamma_1\Gamma_2}$ are given by the multiplicity of the irrep $\Gamma$ in the tensor-product representation $\Gamma_1\otimes \Gamma_2$. A schematic illustration is given in Fig.~\ref{fig:Wilsonfusion}.
Equivalently, using character orthogonality,
they can be written as \cite{Frobenius1899Composition}
\begin{align}
N^{\Gamma}_{\Gamma_1\Gamma_2}
=
\frac{1}{|G|}
\sum_{g\in G}
\chi_{\Gamma_1}(g)\chi_{\Gamma_2}(g)
\chi_{\Gamma}^{*}(g),\label{fuse_electro}
\end{align}
where we defined the character of a representation as $\chi_\Gamma(g)=\tr(\Gamma(g))$ as introduced above. It will be useful to remember that character is a function of class. Equation~\eqref{RepG_Wilson} shows that the traced Wilson loops realize
the fusion algebra of $\mathrm{Rep}(G)$. Moreover, these operators are symmetries of  the quantum double Hamiltonian of Eq.~\eqref{KQDM}, meaning
\begin{align}
    [H_\text{QD},W^\Gamma(\mathcal{C})]=0.
\end{align}
The operators $W^\Gamma(\mathcal{C})$ are the electric
non-invertible 1-form symmetry algebra of the quantum double model. The trivial representation acts as the identity element,
$W^{\mathbf{1}}(\mathcal{C})W^{\Gamma}(\mathcal{C})
=
W^{\Gamma}(\mathcal{C})$ but a generic Wilson loop need not have an inverse. Instead, its product with another Wilson loop can split into a direct sum of several representations. This is an essential distinction from the Abelian case. If $G$ is
Abelian, all irreducible representations are one-dimensional and the
tensor product of two irreps is again a single irrep. The fusion
coefficients therefore reduce to $
N^{\Gamma}_{\Gamma_1\Gamma_2}=\delta_{\Gamma,\Gamma_1\Gamma_2}$ and the Wilson loops form a group-like 1-form symmetry. For non-Abelian
$G$, a higher-dimensional irrep $E$ always exists ($d_E>1$) and tensor products are generally reducible. 

The Wilson loops satisfy additionally some important relations. First, they give rise to an \textit{Abelian 1-form symmetry} since for two closed loops $\mathcal{C}_1,\mathcal{C}_2$ we have
\begin{align}
     [W^{\Gamma_1}(\mathcal{C}_1), W^{\Gamma_2}(\mathcal{C}_2)]=0,
\end{align}
by virtue of being diagonal in the group basis. Indeed, if the closed loops are the same loop the relation holds because the fusion coefficients have the symmetry
\begin{align}
    N^\Gamma_{\Gamma_1\Gamma_2}= N^\Gamma_{\Gamma_2\Gamma_1}.
\end{align}
Second, Wilson loops have an \textit{emergent} topological gluing property inside the defect-free subspace. This gluing property is not an operator identity on the full Hilbert space; it holds only after projecting out electric and magnetic defects in the region swept out by the deformation.
Let $\mathcal R$ be the region containing two adjacent plaquettes and their shared edge, and define the local defect-free projector
\begin{align}
P_{\mathcal R}^0
=
\prod_{s\in \mathcal R}A(s)
\prod_{p\in \mathcal R}B(p).
\end{align}
Let $\mathcal C_1$ and $\mathcal C_2$ be two elementary plaquette loops sharing
one edge with opposite orientations, and let
$\mathcal C=\mathcal C_1\oplus\mathcal C_2$ denote the boundary of the union of
the two plaquettes, with the common edge removed. Then, for normalized Wilson
loops
\begin{align}
\widetilde W^\Gamma(C)=\frac{1}{d_\Gamma}W^\Gamma(C),
\end{align}
one has the projected gluing relation
\begin{align}
P_{\mathcal R}^0\,
\widetilde W^\Gamma(\mathcal C_1)
\widetilde W^\Gamma(\mathcal C_2)
\,P_{\mathcal R}^0
=
P_{\mathcal R}^0\,
\widetilde W^\Gamma(\mathcal C_1\oplus\mathcal C_2)
\,P_{\mathcal R}^0 .
\end{align}
By gluing multiple plaquettes, we can reproduce this identity for any contractible loops $\mathcal{C}_1$ and $\mathcal{C}_2$. The obstruction to making this an exact operator identity is precisely the possible presence of magnetic or dyonic defects\footnote{The electric anyons of the quantum double braid trivially within themselves, therefore the $A(s)\equiv 1$ is not strictly required, see App.~\ref{A:Wilsonfusion}.} in the region $\mathcal R$. In the Abelian case, all anyons braid trivially with themselves and therefore the identity becomes exact in the full Hilbert space. Alternatively, we can interpret the projector as implementing the ground-state constraint locally; the low-energy or infrared (IR) theory must be topological, and therefore, contractible loops should not distinguish the state. Indeed, in the ground state
\begin{align}
\widetilde{W}^\Gamma(\mathcal{C})\ket{\Psi_{\mathcal{D}(G)}} = \ket{\Psi_{\mathcal{D}(G)}}, \quad \mathcal{C} \; \text{contractible}.
\end{align}
This property will hold for the whole ground state subspace, since any given manifold $\mathcal{M}$ where the full lattice is embedded, has $P_{\mathcal{M}}^0=P_\text{GS}$ by definition.

We conclude this subsection by presenting the example of $S_3$ explicitly. As we have seen there are three irreps $\Gamma\in \{A_1,A_2,E\}$ with $A_1=I$ the trivial irrep. The corresponding Wilson loop operators fuse according to Eq.~\eqref{RepG_Wilson}; from the fusion coefficients Eq.~\eqref{fuse_electro} we have for all loops $\mathcal{C}$ according to the Rep$(S_3)$ algebra
\begin{align}
&W^{A_1}(\mathcal{C})=1\nonumber\\
&W^{A_2}(\mathcal{C})W^{A_2}(\mathcal{C})= W^{A_1}(\mathcal{C})\nonumber\\
&W^{E}(\mathcal{C})W^{A_1}(\mathcal{C})= W^{E}(\mathcal{C})\nonumber\\
&W^{E}(\mathcal{C})W^{A_2}(\mathcal{C}) = W^{E}(\mathcal{C})\nonumber\\
&W^{E}(\mathcal{C})W^{E}(\mathcal{C}) =W^{A_1}(\mathcal{C})+W^{A_2}(\mathcal{C})+W^{E}(\mathcal{C}).
\end{align}
These relations reflect the decomposition in terms of irreps of the tensor product of representations. We observe that the electric charges fuse in an Abelian way except for the $E$ anyon which fuses to all channels.

\begin{figure}
    \centering
    \includegraphics[width=0.9\linewidth]{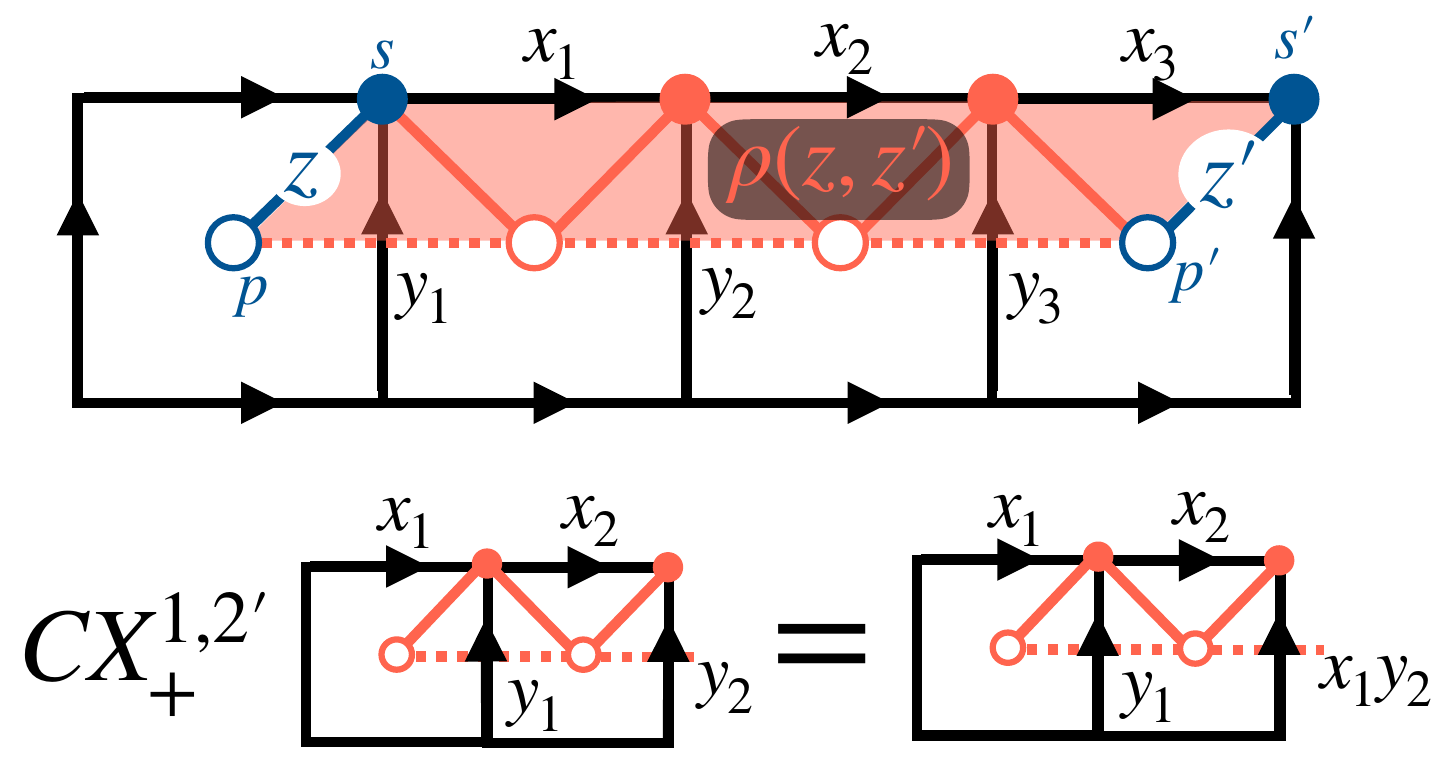}
    \caption{\textbf{Geometry of the ribbon and control operation.} Top: A ribbon $\rho(z,z')$ is shown in orange shade. The blue sites $z=(s,p),\; z'=(s',p')$ are defined as the endpoint sites of the ribbon. Empty circles denote plaquette positions that the ribbon touches, colored circles denote the vertices. Orange lines denote the sites of the ribbon composed of an interior vertex and a plaquette. The ribbon has associated direct lattice edges $x_1,x_2,x_3$ and dual lattice edges $y_1,y_2,y_3$. Bottom: Action of the control operation $CX_+^{1,2'}$ defined in the main text. The $G$-qudit of the first position has group element $x_1$ and stays fixed while the dual lattice link get multiplied $y_2 \rightarrow x_1y_2$}
    \label{fig:ribbon}
\end{figure}

\subsection{Open and closed ribbon operators} \label{sec:ribbons}

To investigate the magnetic 1-form operators, we first need to generalize the product of Pauli $X$ in the $\mathbb{Z}_2$ case to the non-Abelian setting. We require the string operators to commute with all star and plaquette projectors $A(s),B(p)$ except at the endpoints. Since the simple product of $\prod_{\tilde{\ell}}X^g_+(\ell)$ along dual lattice edges does not commute with $B(p)$, the direct lattice links surrounding the string must be considered in defining a 1-form symmetry operator. 

The correct objects to consider are ribbons and sites. A site is defined as a joined vertex and plaquette, $z=(s,p)$. A ribbon $\rho(z,z')$ is composed of sites denoted by orange lines in Fig.~\ref{fig:ribbon} with endpoints $z,z'$ and has a finite width.  The inner sites lead to triangles pointing up or down. Each down triangle has a direct lattice edge as the top side and every up triangle has a dual-lattice link as the bottom side. We labeled the direct lattice edge group elements by $x_\ell$ and the dual lattice edge group elements by $y_\ell$. Next, we define control X gates acting on two $G$-qudits labeled 1,2 by the group multiplication $ CX^{1,2}_+\ket{g_1,g_2} = \ket{g_1,g_1g_2}$ and correspondingly $CX^{1,2}_-\ket{g_1,g_2} = \ket{g_1,g_2\bar{g}_1}$ (see Fig.~\ref{fig:ribbon}). 

Let us now label the edges of the direct lattice by $1,\dots , j ,\dots, L_\rho$, with $L_\rho$ the number of ribbon direct lattice links and the dual edges of the oriented lattice by $1',\dots , j' ,\dots, L_\rho'$. Let us define a string operator (assuming positively oriented links) for a dual lattice edge $j'$ of the ribbon as
\begin{align}
    S_j = CX^{j-1,j'}_+CX^{j-2,j'}_+\cdots CX^{1,j'}_+.
\end{align}
The adjoint operator acts by multiplying instead by the inverse group element $(CX^{i,j}_+)^\dagger=\bar{C}X^{i,j}_+$. We introduce now a modified $X^g_+$ operator carrying a ghost string of control X gates
\begin{align}
    \tilde{X}_+^h(j) = S_jX_+^h(j)S_j^\dagger.
\end{align}
Since this is a unitary transformation, all the properties of the original $X^g_{\pm}$ are preserved. Nevertheless, the unitary transformation does not commute with translation. Let us define a path-ordered product of modified $\tilde X$'s, along a ribbon $\rho$ 
\begin{align}
T^g(\rho)=\mathcal{P}\prod_{\tilde{\ell} \in \rho}\tilde{X}_{O_{\tilde{\ell}}}^g(\tilde{\ell}),
   \label{eq:Tloop}
\end{align}
where we labeled by $\tilde{\ell}$ the dual lattice edges crossed by the ribbon $\rho$ and used $O_{\tilde{\ell}}=\pm$ with a positive (negative) sign if the links crossed a point to the left (right) of the ribbon direction. 

We are now in a position to introduce ribbon operators $F_\rho$. They are a product of Wilson loops and path-ordered products of $\tilde X$'s.
\begin{align}
    F_\rho^{(h,\Gamma)}
    =
    W^\Gamma(\rho)T^h(\rho).
\end{align}
The label $\Gamma$ denotes an irrep of the group. The operator $W^\Gamma(\rho)$ is an open Wilson string obtained from the same path-ordered product as the closed Wilson loop in Eq.~\eqref{Wilson-Loops}, with the $\textbf{Z}^\Gamma$ operators acting on the direct lattice sites of the ribbon. These operators are related to the ribbon operators $F^{(g,h)}_\rho$ from Kitaev in Ref.~\cite{kitaev_fault-tolerant_2003} via the relation
\begin{align}
    F_\rho^{(h,\Gamma)}
    =
    \sum_{g\in G}\chi_\Gamma(\bar{g})F_\rho^{(h,g)}.
\end{align}
Here $h,g\in G$ are group elements and $\rho$ is an arbitrary ribbon. The ribbon operators $F_\rho^{(h,g)}$ are the desired generalization of the string operators in the $\mathbb{Z}_2$ case since they commute with all $A(s),B(p)$ except at the end sites. Physically, we can interpret the first group element as
threading a $h$ flux through the dual lattice links consistent with the plaquette and star projectors. The second group element ensures that the holonomy of the direct lattice links is fixed $U_\sigma\equiv\prod_{\ell\in\sigma} x_\ell=g$. Open ribbons create endpoint excitations and, therefore, are not symmetry operators. Proper higher-form symmetry operators are obtained only from closed ribbons $\sigma$. In the non-Abelian case, however, a closed ribbon labeled by a single group element $h$ is not generally well defined: changing the base point conjugates $h$. To ensure the open ribbon operators $F_\sigma$  are symmetries, we must additionally ensure that the resulting closed ribbon operators satisfy an inner translational invariance. 

The proper closed ribbon will then be a linear combination of the $F_\sigma$ operators such that the sum acts the same way on any state, independent of the initial base point of $\sigma$. In App.~\ref{A:closedribbons} we show how this physical condition automatically leads to a combination of $F_\sigma$ operators that is now indexed by a conjugacy class of the group $C\in \text{Class}(G)$ and a conjugacy class of the centralizer of the class $D\in \text{Class}(Z_C)$. 
Since the centralizer group is isomorphic for all $r_C\in C$ we define a closed ribbon operator that depends only on the conjugacy class of the group and the class of the centralizer $Z_C$ denoted $D$. We label the elements of the conjugacy class $C$ by $c_1=r_C$ and $c_i=\bar{p}_ir_C p_i$ with $i=1,\dots,\abs{C}$ for some $p_i\in G$ then the closed ribbons operators are
\begin{align}
  F^{CD}_\sigma= \dfrac{1}{\abs{D}}\sum_{i=1}^{\abs{C}} \sum_{k\in D} F^{(\bar{p}_ir_Cp_i,\bar{p}_i k p_i)}_{\sigma}.
\end{align}
An alternative basis is obtained by considering the irreps of $Z_C$, which we denote by $R\in\text{Rep}(Z_C)$ of dimension $d_R$. In this basis, we define the most general closed ribbon operator
\begin{align}
    K^{CR}_\sigma = \sum_{i=1}^{\abs{C}} \sum_{k\in Z_C} \chi_{R}^*(k)F^{(\bar{p}_ir_Cp_i,\bar{p}_i k p_i)}_{\sigma}. \label{Dyon-1-form}
\end{align}
The $K^{CR}_\sigma$ or $F^{CD}_\sigma$ operators commute with all star and plaquette terms of the quantum double Hamiltonian, $[K^{CR}_\sigma,H_\text{QD}]=0$. For $C=[1]$ we recover the Wilson loop operators of the previous section, since $Z_C=G$ leads to $ K^{[1]R}_\sigma = K^{[1]\Gamma}_\sigma = W^\Gamma(\mathcal{C}) $ with $\mathcal{C}$ the direct lattice loop of the ribbon $\sigma$.

\begin{figure*}[t]
\centering
\includegraphics[width=0.75\textwidth]{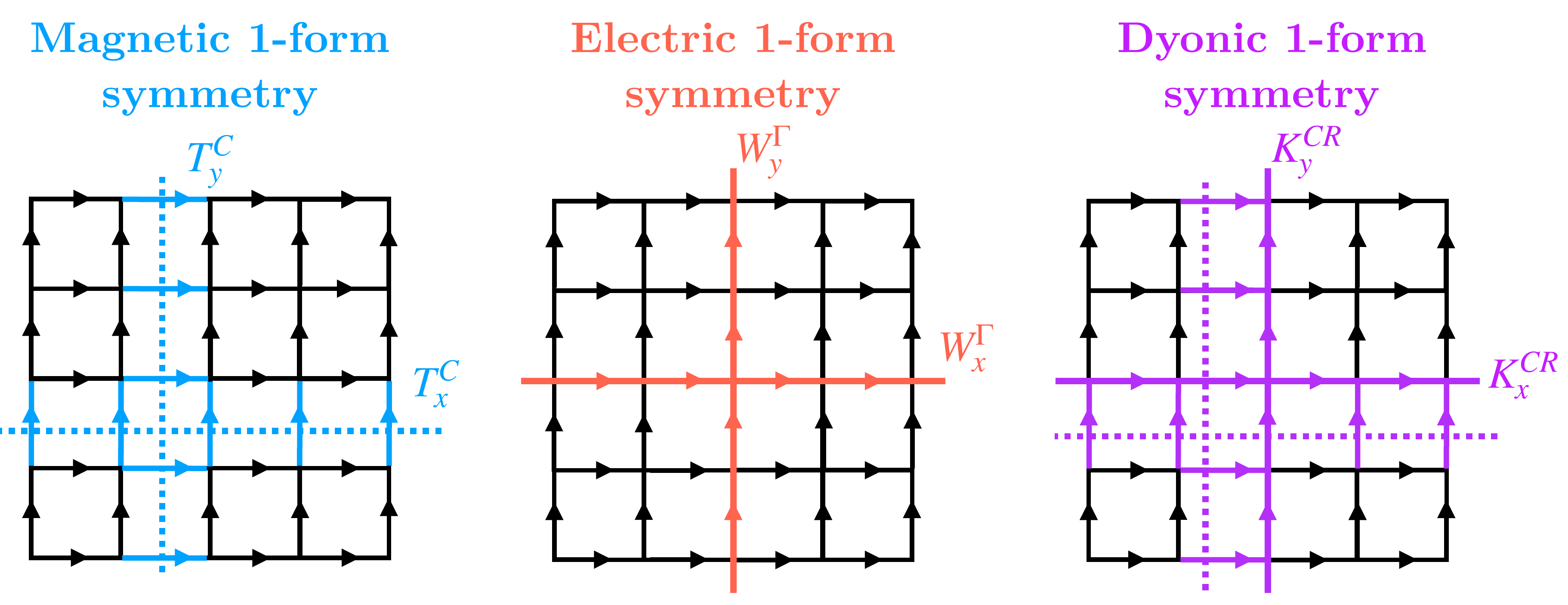}
\caption{\textbf{Magnetic, electric and dyonic $1$-form symmetries.} Schematic illustration of the three types of non-contractible 1-form operators on the torus for the quantum double. Left: The magnetic symmetry operator $T^C_{x,y}$ carries a magnetic flux $C\in\text{Class}(G)$ and acts along a loop whose dual lattice path is indicated by the dashed blue line. Middle: The electric 1-form symmetry operator $W^\Gamma_{x,y}$ acts only in the direct lattice links, indicated by the orange line. Its charges are irreducible representations $\Gamma=\text{Rep}(G)$ of the group. Right: The general dyonic 1-form symmetry operator $K^{CR}_{x,y}$ carries both a magnetic flux $C$ and an electric charge $R\in \text{Rep}(Z_C)$ is defined on a non-trivial ribbon. Both magnetic and electric 1-form operators can be interpreted as ribbon operators as well: For the magnetic 1-form symmetry, the loop is augmented by a trivial identity loop acting on the primal lattice, while for the electric 1-form symmetry, the identity acts on the dual lattice.
}
\label{fig:non-contract-loops}
\end{figure*}
\subsection{Magnetic 1-form operators}\label{sec:magnetic-1form}
We define the magnetic 1-form operators, or 't Hooft loops, by choosing
the trivial centralizer representation,
$T^C(\sigma)\equiv K^{CI}_\sigma$. In a non-Abelian theory, closing a
fixed-label magnetic ribbon requires more than a product of $X$
operators along the dual lattice. The resulting operator must be
independent of the arbitrary choice of base point. If
$\displaystyle U_\sigma=\prod_{\ell\in\sigma}^{\leftarrow}x_\ell$ is the
direct-lattice holonomy, where we use the leftarrow to denote path ordering from right to left,  transporting an inserted label $g$ once around
the ribbon maps it to $U_\sigma gU_\sigma^{-1}$. Closure without a
residual endpoint or seam defect therefore requires
$[U_\sigma,g]=1$. Moreover, a gauge transformation at the base point
simultaneously conjugates $U_\sigma$ and $g$, so a gauge-invariant
closed operator is obtained by summing $g$ over its conjugacy class.
These two requirements lead to
\begin{equation}
T^C(\sigma)=\sum_{g\in C}P^g_{\mathrm{com}}(\sigma)T^g(\sigma),\label{thooft_loop}
\end{equation}
Here $P_{\mathrm{com}}^g(\sigma)$ restricts the operator to configurations
with $U_\sigma\in Z_g$, or equivalently $[U_\sigma,g]=1$. The projector
therefore enforces closed-ribbon base-point consistency, while the sum
over $g\in C$ makes the result invariant under conjugation at the base
point. This construction should not be interpreted as assuming that
magnetic excitations braid trivially. To understand the properties of these magnetic 1-form operators, we introduce in App.~\ref{A:magentic1form} the conjugacy class algebra $\text{Class}(G)\equiv Z(\mathbb{C}[G])$, constructed as the center of the group algebra $\mathbb{C}[G]$. The elements of $e_{C}\in \text{Class}(G)$ are labeled by conjugacy classes and satisfy a non-invertible fusion $e_{C_1}e_{C_2}=\sum_{C}N^{C}_{C_1C_2}e_C$. The coefficients $N^{C}_{C_1C_2}\in \mathbb{Z}_{\ge0}$ are the class-algebra structure constants. In the flux-free sector, they govern the fusion of the magnetic anyon-line operators $T^C$, and as we will show below label the composition of the ground-state sectors by conjugacy classes. This projected fusion action should be distinguished from unrestricted \(\mathcal D(G)\) anyon fusion, which can contain dyonic centralizer-charge channels, and as we will show in Sec. \ref{sec:anomaly} is recovered from the complete \(K^{CR}\) algebra.
%are the magnetic anyon fusion coefficients. 
By virtue of the unitary string operator defining the modified $X$ operators, we note that the 't Hooft loops are representations of the class algebra, implying that for the same ribbon, we get a non-invertible fusion relation on the flux-free sector
\begin{align}
    T^{C_1}(\sigma) T^{C_2}(\sigma) \ket{\Psi_{\mathcal{D}(G)}}= \sum_{C}N^{C}_{C_1C_2} T^{C}(\sigma)\ket{\Psi_{\mathcal{D}(G)}}.
\end{align}
The fusion coefficients of the algebra can be expressed in terms of the characters of the group and the size of the classes
\begin{align}
N^{C}_{C_1 C_2}
=
\frac{|C_1|\,|C_2|}{\abs{G}}
\sum_{\Gamma}
\frac{
\chi_\Gamma(C_1)\,
\chi_\Gamma(C_2)\,
\chi^*_\Gamma(C)
}{d_\Gamma}. \label{eq:mag_fusion}
\end{align}
The class algebra relation is valid on the flux-free state, which is the ground state for the plane or sphere, since as we will see in the torus case different ground states will be created by application of non-contractible magnetic or dyon operators. Then the fusion relation above will be modified by the presence of these magnetic/dyon anyon loops and their fusion with the magnetic loop operators above. The magnetic 1-form operators are symmetries of the quantum double model and as such satisfy
\begin{equation}
[T^C(\sigma),H_\text{QD}]=0,
\end{equation}
We remark here that in the Abelian topological order case, the classes are in one-to-one correspondence with the elements of the group. Therefore, the magnetic fluxes are equally labeled by group elements. Moreover, by writing the product of group elements as $gh=f$, the multiplication of 't Hooft loops results in a single loop, meaning $N^{C}_{C_1C_2}=\delta_{C,C_1C_2}$. In contrast, for non-Abelian topological order the fluxes are given by conjugacy classes and fuse non-trivially. Nevertheless, the 't Hooft loops realize an \textit{Abelian 1-form symmetry} since \(Z(\mathbb C[G])\) is a commutative so that
\begin{align}
N^{C}_{C_1 C_2}
=N^{C}_{C_2C_1}.
\end{align}
For magnetic 1-form operators that act on different contractible loops that do not intersect, we have
\begin{align}
&[T^{C_1}(\sigma),T^{C_2}(\tau)]=0,\;  \sigma \cap \tau = \emptyset.
\end{align}
Nevertheless, for intersecting loops, the commutation of the magnetic 1-form operators becomes valid only in a defect-free region (App.~\ref{A:magentic1form}):
\begin{align}
    &  P_{\mathcal{R}}^0[T^{C_1}(\sigma),T^{C_2}(\tau)]P_{\mathcal{R}}^0=0.
\end{align}
Similarly to the electric 1-form operators, they satisfy an emergent gluing property within the defect-free subspace.
We define the normalized magnetic loop
\begin{align}
\widetilde T^C(\sigma)=\frac{1}{|C|}T^C(\sigma).
\end{align}
Let $\sigma_1$ and $\sigma_2$ be two adjacent elementary dual-lattice loops sharing one direct edge with opposite induced orientations. Let $\sigma=\sigma_1\oplus\sigma_2$ denote the joined loop obtained by removing the common segment. Let $\mathcal R$ be the region containing the two elementary loops and their shared edge, then 
\begin{align}
P_{\mathcal R}^0\,
\widetilde T^C(\sigma_1)
\widetilde T^C(\sigma_2)\,
P_{\mathcal R}^0
=
P_{\mathcal R}^0\,
\widetilde T^C(\sigma_1\oplus\sigma_2)\,
P_{\mathcal R}^0 .
\label{eq:normalized-magnetic-gluing}
\end{align}
From the perspective of the flux-free ground state, we know the $T^C(\sigma)$ operators acting on the trivial state $\otimes_\ell \ket{1}$ will create a dual lattice loop labeled by $[g]=C$ since the $T^C(\sigma)$ operators commute with the $A(s)$ and therefore act on the $\ket{\textbf{1}}$ state as pure $X$ operators. Now we know the ground state is a superposition of all possible such loops, so the resulting state is contained in $\ket{\Psi_{\mathcal{D}(G)}}$, which implies that the ground state is a condensate of closed contractible magnetic loops. In this way, applying the operator to other product states with loops will form another state in the ground state superposition. For a loop with group element label $h$ that intersects $\sigma$ one can see that $T^{[g]}_\sigma$ conjugates it by $gh\bar{g}$, resulting again in a valid state of the superposition, as this satisfies the fusion rules explained in Sec.~\ref{sec:non-ab-KQD}. For the contractible magnetic 1-form symmetry of the flux-free ground state we thus find
\begin{align}
    \tilde{T}^C(\sigma) \ket{\Psi_{\mathcal{D}(G)}} = \ket{\Psi_{\mathcal{D}(G)}}, \quad \sigma \; \text{contractible}.
\end{align}
Using the Wilson and 't Hooft loop operator we can rewrite the quantum double Hamiltonian to make explicit that the electric and magnetic 1-form operators are symmetries. We use the notation $\tilde{p}$ to denote the dual lattice plaquettes and $\partial \tilde{p}$ its boundary as the links cross in the direct lattice. Up to a constant term, we have
\begin{align}
H_\text{QD}= -\sum_{\tilde{p}} \dfrac{1}{\abs{G}} \sum_C T^C(\partial \tilde{p})- \sum_p \dfrac{1}{\abs{G}} \sum_\Gamma d_\Gamma W^\Gamma (\partial p). \label{loopHQD}
\end{align}
We again explicitly show the construction of the magnetic fusion for $S_3$. Using Eq.~\eqref{eq:mag_fusion} we must compute the fusion for $C\in \{[1],[s],[r]\}$. We note that $T^{[1]}=1$ the identity operator. Moreover, due to the fact that the inverse permutations in $S_n$ are in the same conjugacy class as their original permutation we have the hermiticity condition $(T^C(\sigma))^\dagger=T^C(\sigma)$. We denote by $\ket{\Psi_0} =\ket{\Psi_{\mathcal{D}(S_3)}} $ the flux-free sector. The non-trivial fusion products for any closed ribbon $\sigma$ in the flux-free ground state are
\begin{align}
(T^{[s]}(\sigma)T^{[s]}(\sigma))\ket{\Psi_0}
&=
(3T^{[1]}(\sigma)+3T^{[r]}(\sigma)) \ket{\Psi_0},\nonumber\\
T^{[s]}(\sigma)T^{[r]}(\sigma)\ket{\Psi_0}
&=
2T^{[s]}(\sigma)\ket{\Psi_{\mathcal{D}(G)}},\nonumber\\
T^{[r]}(\sigma)T^{[r]}(\sigma)\ket{\Psi_0}
&=
(2T^{[1]}(\sigma)+T^{[r]}(\sigma))\ket{\Psi_0}.
\end{align}
In contrast to the electric 1-form operators the magnetic 1-form operators fuse into multiple channels or one channel with higher multiplicity.

\subsection{Generalized  mixed 't Hooft anomaly and dyonic 1-form operators} \label{sec:anomaly}
Having analyzed the generalizations of the electric and magnetic 1-form operators to the non-Abelian setting, we are led to the conclusion that at the fixed-point of the quantum double, there is a non-invertible 1-form symmetry 
\begin{align}
    \text{Rep}(G)^{(1)}\times \text{Class}(G)^{(1)}.
\end{align}
This symmetry is anomalous. As we will see in this section, it has a  mixed 't Hooft anomaly, which can be detected by the non-contractible loops in the torus. Let us denote by  $\mathcal{C}_x$ and $\mathcal{C}_y$ the $x$ and $y$ direct lattice loops, respectively, see orange lines in Fig.~\ref{fig:non-contract-loops}. We define the Wilson loop electric 1-form operators
\begin{align}
    W_x^\Gamma=W_x^\Gamma(\mathcal{C}_x) = \text{tr} \Big[\mathcal{P}\prod_{\ell\in \mathcal{C}_x}(\bm{Z}^{\Gamma}(\ell))\Big ], \\
    W_y^\Gamma=W_y^\Gamma(\mathcal{C}_y) = \text{tr} \Big[\mathcal{P}\prod_{\ell\in \mathcal{C}_y}(\bm{Z}^{\Gamma}(\ell))\Big ].\label{non-contr_Wilson}
\end{align}
Analogously we define along the dual lattice handles $\tilde{\mathcal{C}}_x,\tilde{\mathcal{C}}_y$ the 't Hooft magnetic 1-form operators 
\begin{align}
    T^C_x = T^C(\tilde{\mathcal{C}}_x)=\sum_{g\in C}P^{g}_{\text{com}}(\tilde{\mathcal{C}}_x)  \mathcal{P}\prod_{\tilde{\ell} \in \tilde{\mathcal{C}}_x}\tilde{X}_{O_{\tilde{\ell}}}^g(\tilde{\ell}), \\  T^C_y = T^C(\tilde{\mathcal{C}}_y)=\sum_{g\in C}P^{g}_{\text{com}}(\tilde{\mathcal{C}}_y)  \mathcal{P}\prod_{\tilde{\ell} \in \tilde{\mathcal{C}}_y}\tilde{X}_{O_{\tilde{\ell}}}^g(\tilde{\ell})  .\label{non-contr_tHooft}
\end{align}
We can now calculate the commutator between the electric and magnetic 1-form operators threading two distinct handles of the torus. We focus on the flux-free ground state which leads after some algebra (App.~\ref{A:anomaly}) to a generalized mixed 't Hooft anomaly
\begin{align}
     W_x^\Gamma T_y^C  \ket{\Psi_{\mathcal{D}(G)}}= \dfrac{\chi_\Gamma(C)}{d_\Gamma} T_y^C W_x^\Gamma \ket{\Psi_{\mathcal{D}(G)}}.\label{Anomaly}
\end{align}
We define the generalized mixed anomaly operationally by the nontrivial crossing algebra of the non-invertible loop operators on the topological subspace; in the invertible Abelian limit, this reduces to the conventional phase-valued mixed ’t Hooft anomaly.
We interpret the normalized character of the group then as the generalization of the anticommutator for $\mathbb{Z}_2$. In general, the character can be larger than one or even zero, but $\abs{\chi_\Gamma(g)}/d_\Gamma\leq 1$. In fact, for non-Abelian groups, there must always be one zero entry in the character table of the group. The character thus encodes the braiding coefficient between the electric and the magnetic anyons. More precisely,
denote the $S$-matrix of the anyon theory by $S_{\mathfrak a \mathfrak b}$ with $\mathfrak a=(C,R)$ and $\mathfrak b=(C',R')$ anyons. The $S$-matrix element between the $\mathfrak a=([1],\Gamma)$ electric anyon and $\mathfrak b=(C,I)$ is given by $S_{\Gamma C}\equiv S_{([1],\Gamma),(C,I)}= \tfrac{\abs{C}}{\abs{G}}\chi_\Gamma(C)$ for unitary irreps. Let us define a reduced $S$-matrix or unitary character transform $\tilde{S}_{\Gamma C} = \sqrt{\frac{|C|}{\abs{G}}}\chi_\Gamma(C)$. The $\tilde{S}$ matrix is not the full modular $S$-matrix of \(\mathcal D(G)\). Nevertheless, it exactly diagonalizes the pure-electric fusion algebra, from the Verlinde formula for the full $S$ matrix \cite{Verlinde}. Moreover, the fusion coefficients for the pure-magnetic anyons in the flux-free sector are also expressed in terms of $\tilde{S}$ as follows
\begin{align}
N^{\Gamma}_{\Gamma_1 \Gamma_2}
&=
\sum_{C}
\frac{
\tilde{S}_{\Gamma_1 C}\,
\tilde{S}_{\Gamma_2 C}\,
\tilde{S}_{\Gamma C}^{*}
}{
\tilde{S}_{IC}
}.\\
N^{C}_{C_1 C_2}
&=
\sqrt{\dfrac{\abs{C_1}\abs{C_2}}{\abs{C}}}\sum_{\Gamma}
\frac{
\tilde{S}_{\Gamma C_1}\,
\tilde{S}_{\Gamma C_2}\,
\tilde{S}_{\Gamma C}^{*}
}{
\tilde{S}_{\Gamma [1]}
}.
\end{align}
The square-root factor in the magnetic fusion coefficients arises because \(N^{C}_{C_1C_2}\) is defined in the unnormalized basis of conjugacy-class sums. The full algebra of non-contractible Wilson and 't Hooft loops  is
\begin{align}
[W_x^\Gamma,T^C_x]=0, \quad [W_y^\Gamma,T^C_y]=0, \quad[W_x^\Gamma,W_y^\Gamma]=0.
\end{align}
The following relation expresses the ground-space commutativity of the
't Hooft loops in the flux-free sector 
\begin{align}
[T_x^C,T_y^{C'}]\ket{\Psi_{\mathcal{D}(G)}}=0.
\end{align}
Remarkably, the commutativity arises even though the magnetic anyons of the $\mathcal{D}(G)$ quantum double do not braid trivially. 

We finalize this section by describing the dyonic 1-form operators. They are given by the closed ribbon operators $K^{CR}_\sigma$, which commute with the quantum double Hamiltonian. They carry magnetic flux $C\neq [1]$ and non-trivial electric charge $R\neq I$ labeled by the irreps of the centralizer of the flux (for $C=[1]$ the dyonic operators reduce to the electric 1-form operators and for $R=I$ to the magnetic 1-form operators; Fig.~\ref{fig:non-contract-loops}). Since they carry both magnetic and electric charges we do not expect them to commute. In fact, we already indicated that we can write the $K^{CR}_\sigma$ operator as a mixed Wilson and 't Hooft object, see App.~\ref{A:anomaly}, with some tensor coefficients transforming to the correct labels
\begin{align}
K^{CR}_\sigma = \sum_{i=1}^{|C|} \sum_\Gamma \xi_{\Gamma\alpha\beta,i}^{CR} W_{\alpha\beta}^\Gamma(\sigma)T^{c_i}(\sigma), \label{dyon-mixWT}
\end{align} 
where $W_{\alpha\beta}^\Gamma$  are the matrix values of $\textbf{W}^\Gamma$ and we define the coefficients
\begin{align}
\xi_{\Gamma\alpha\beta,i}^{CR} =\dfrac{d_\Gamma}{\abs{G}}  \sum_{k \in Z_C} \chi_{R}(\bar{k}) \Gamma(\bar{p}_i k p_i)_{\beta\alpha}.
\end{align}
It is worth mentioning here that when considering all the dyon 1-form symmetry operators as anyon lines with Fraktur letters denoting anyons  like $\mathfrak a =(C,R)$ then for the same ribbon the fusion is exact (full-Hilbert space) at the operator level with the result, see App. \ref{A:closed-ribbon-projectors} for calculation,
\begin{align}
K_\sigma^{\mathfrak a}K_\sigma^{\mathfrak b}
=\sum_{\mathfrak c}
N_{\mathfrak a\mathfrak b}^{\mathfrak c}
K_\sigma^{\mathfrak c},
\label{eq:dyon-fusion}
\end{align}
where the fusion coefficients are now given from the full $S$-matrix \cite{coste_finite_2000}, $S_{\mathfrak a \mathfrak b}$,  containing the modular data of the $\mathcal{D}(G)$ quantum double and by virtue of the Verlinde formula \cite{Verlinde}
\begin{align}
N_{\mathfrak a\mathfrak b}^{\mathfrak c}
=
\sum_{\mathfrak d}
\frac{S_{\mathfrak a\mathfrak d}
S_{\mathfrak b\mathfrak d}
S_{\mathfrak c\mathfrak d}^*}
{S_{\mathfrak0\mathfrak d}}.
\label{eq:dyon-Verlinde}
\end{align}
The pure electric and pure magnetic non-invertible 1-form symmetries organize the Wilson and 't Hooft sectors of the quantum double. The dyonic closed ribbons $K_\sigma^{CR}$ provide the minimal centralizer-charge completion required to resolve the full $\mathcal{D}(G)$ topological degeneracy, as we will see in the next section.  

We now apply these concepts for $S_3$.
The 't Hooft anomaly in this case can be calculated from Fig.~\ref{fig:s3_triangle}, which with Eq.~\eqref{Anomaly} leads to the explicit form ($\ket{\Psi_0}\equiv \ket{\Psi_{\mathcal{D}(S_3)}}$)
\begin{align}
 W_x^{A_1} T_y^C \ket{\Psi_0}&= T_y^C W_x^{A_1} \ket{\Psi_0},\ C=\{[1],[s],[r]\} \nonumber\\
 W_x^{\Gamma} T_y^{[1]}  \ket{\Psi_0}&= T_y^{[1]}  W_x^\Gamma \ket{\Psi_0},\ 
 \Gamma=\{A_1,A_2,E\} \nonumber\\
  W_x^{A_2} T_y^{[s]}  \ket{\Psi_0}&= -T_y^{[s]} W_x^{A_2} \ket{\Psi_0} \nonumber\\
  W_x^{A_2} T_y^{[r]} \ket{\Psi_0}&=T_y^{[r]} W_x^{A_2} \ket{\Psi_0} \nonumber\\
  W_x^{E} T_y^{[r]} \ket{\Psi_0}&= -T_y^{[r]} W_x^E \ket{\Psi_0}\nonumber\\
  W_x^{E} T_y^{[s]} \ket{\Psi_0}&=0. 
\end{align}
The last equation can only happen for non-invertible 1-form symmetries since a zero eigenvalue is prohibited for unitary symmetries. Moreover, for any non-Abelian group $G$ there is at least one entry in the character table that is zero. Therefore, applying the operator $ W_x^E T_y^{[s]}$ to the quantum double ground state and  detecting zero serves as a signal of the  $\mathcal{D}(G)$ non-Abelian topological order.

\section{Ground states from 1-form symmetry operators}\label{sec:ground-states}
In this section, we construct the full ground state subspace explicitly of the $\mathcal{D}(G)$ quantum double via 1-form symmetry operators and give the $S_3$ quantum double example.
\subsection{Ground states on a cylinder}
For the cylinder, with links sticking out of the $x$-direction it is sufficient to consider the periodic direction, which we choose to be the $y$ direction. We can construct the different ground states by starting with the flux-free ground state on an infinite plane or sphere, $\ket{\Psi_{\mathcal{D}(G)}}$ given by Eq.~\eqref{QD_gs}.
The Wilson loop operators act trivially $\tr\Gamma(1)=d_\Gamma$, so to generate the ground states we must act with the magnetic 1-form operators to insert a magnetic flux. This operation can be done for all the conjugacy classes $C$
\begin{align}
\ket{\Psi_{\text{cyl}}(C)} =  T_y^C \ket{\Psi_{\mathcal{D}(G)}}=\prod_s A(s) T_y^C \ket{\textbf{1}}.
\end{align}
We note here that these states are $\text{Rep}(G)^{(1)}$ symmetric states since they are eigenstates of the Wilson loops, with the eigenvalue given by the  mixed 't Hooft anomaly coefficient
\begin{align}
W^\Gamma_x
\ket{\Psi_{\text{cyl}}(C)} = \chi_\Gamma(C) 
\ket{\Psi_{\text{cyl}}(C)} \label{Wilson-CylGS}.
\end{align}
Nevertheless, they are not $\text{Class}(G)^{(1)}$ symmetric since the magnetic 1-form symmetry operators toggle between different ground states
\begin{align}
&T_y^{C'}\ket{\Psi_{\text{cyl}}(C)}=T_y^{C'}T_y^C \ket{\Psi_{\mathcal{D}(G)}}\nonumber \\ &=\sum_{\tilde{C}}N_{C'C}^{\tilde C}T_y^{\tilde C} \ket{\Psi_{\mathcal{D}(G)}}= \sum_{\tilde{C}}N_{C'C}^{\tilde C}\ket{\Psi_{\text{cyl}}(\tilde 
C)} \label{CylGS}.
\end{align}

Therefore, we can identify the magnetic ground states in the cylinder as spontaneously breaking the magnetic part of the $\text{Rep}(G)^{(1)}\times\text{Class}(G)^{(1)}$ symmetry with degeneracy 
\begin{align}
\mathrm{GSD}_{S^1\times[0,1] }\bigl(\mathcal D(G)\bigr) = |\text{Class}(G)|.
\end{align}
For the case of $G=S_3$ we have three ground states for the cylinder, which are 
\begin{align}
&\ket{\Psi_{\text{cyl}}([1])} =  T_y^{[1]} \ket{\Psi_0} = \ket{\Psi_0}\nonumber\\
&\ket{\Psi_{\text{cyl}}([s])} =  T_y^{[s]} \ket{\Psi_0} \nonumber\\
&\ket{\Psi_{\text{cyl}}([r])} =  T_y^{[r]} \ket{\Psi_0}.
\end{align}

\subsection{Ground states on a Torus}

Next, we construct all ground states on the torus starting from the flux-free ground state $\ket{\Psi_{\mathcal{D}(G)}}$. To do this let us follow the intuition of the previous section, but now applying general closed ribbon operators along both handles of the torus, in the magnetic basis
$F^{CD}_x,F^{\tilde{C}\tilde{D}}_y$. In that case, a general eigenstate to consider is given by 
$F^{\tilde{C}\tilde{D}}_y F^{CD}_x\ket{\Psi_{\mathcal{D}(G)}}$.
We know the 1-form symmetry operators map the ground state to a subspace with the same ground state energy since all the closed ribbon operators commute with the Hamiltonian and with $A(s),B(p)$. Nevertheless, not all combinations of $C,D,\tilde{C},\tilde{D}$ are allowed and result in non-zero norm states. To resolve the degeneracy of the GS subspace, we must write their action explicitly and count the non-zero states correctly; the detailed derivation is given in App.~\ref{A:anomaly}. Here, we go through the main points.
The first ribbon operator acting on the flux-free state gives a nonzero state only when the centralizer-class label is trivial: 
\begin{align} 
F_x^{CD}\ket{\Psi_{\mathcal D(G)}} = \delta_{D,[1]}\, T_x^C\ket{\Psi_{\mathcal D(G)}}. \label{eq:first-ribbon-main} 
\end{align}
Thus, a single non-contractible ribbon simply creates a pure magnetic flux sector labeled by the conjugacy class \(C\). This is why the cylinder geometry discussed above reduces to the application of magnetic loops along the non-contractible direction. For the torus we have to consider the second non-contractible ribbon, which is more restrictive. If a flux \(g\in C\) has already been threaded along the \(x\)-cycle, then a \(y\)-cycle ribbon carrying flux \(h\) gives a nonzero flat state only when 
\begin{align} gh=hg . \label{eq:commuting-condition-main} \end{align} 
Equivalently, the centralizer label of the second ribbon must be compatible with the first flux 
\begin{align} D\subseteq [g]\cap Z_{[h]}. \label{eq:D-condition-main} \end{align} 
Here \(Z_{[h]}\) denotes the centralizer of a representative of the conjugacy class \([h]\). The resulting ground states can be written compactly as 
\begin{align} 
\ket{[(g,h)]} &= F^{[h]D}_y T^{[g]}_x \ket{\Psi_{\mathcal{D}(G)}} \nonumber\\
&= \prod_s A(s) \left( \sum_{(g_x,g_y)\in[(g,h)]} T_y^{g_y}T_x^{g_x} \right) \ket{\mathbf 1}. \label{eq:torus-gs-main} 
\end{align} 
The sum runs over the full simultaneous-conjugacy orbit of the commuting pair $[(g,h)]\in C_{\text{pairs}}$ with $C_{\text{pairs}}$ defined as
\begin{align}
    C_{\text{pairs}} = \{(g,h) \in G\times G| \ gh=hg,\notag \\
    \qquad (g,h)\sim (f g\bar{f},fh\bar{f}), f \in G \}. \label{Cpair}
\end{align}
We can interpret the form of the ground states of Eq.~\eqref{eq:torus-gs-main} as threading along the \(x\)-cycle a flux \(g_x\in G\), while along the \(y\)-cycle it inserts a flux \(g_y\in G\). However, for a non-Abelian group, these labels are not individually gauge-invariant. First, local gauge transformations generated by $A(s)$ conjugate the holonomy $g_x\rightarrow f g_x \bar{f}$, similarly for the other direction. This means that only the conjugacy class is invariant under application of $A(s)$.
Moreover, there is a non-trivial constraint (only non-trivially satisfied for non-Abelian theories) on the loop products in the case of a torus, since one can look at the loop resulting from the product along all the edges, which is contractible and thus has a value $1$. Nevertheless, it is also composed of the loop products in the $x$ and $y$ directions and their inverses meaning the loop products must commute
\begin{align}
    g_y^{-1}g_x^{-1}g_yg_x =1 \quad  \rightarrow \quad g_x g_y = g_y g_x.\label{toposec_comm}
\end{align}
Therefore only $(g_x,g_y) \sim (fg_xf^{-1},fg_yf^{-1})$ pairs are allowed, since only these conjugations preserve Eq.~\eqref{toposec_comm}. In this way one arrives at distinct ground states belonging to the conjugacy pair set of Eq.~\eqref{Cpair}. This flux description of the non-Abelian topological sectors coincides with the discussion of non-Abelian topological order by Henley \cite{henley2011,henley2013}. The topological ground state degeneracy of the quantum double $\mathcal{D}(G)$ for any finite group is then
\begin{align} 
\mathrm{GSD}_{T^2}\bigl(\mathcal D(G)\bigr) = |C_{\rm pairs}|, 
\end{align} 
We note that this number coincides with the number of anyon types, as we will explicitly illustrate for the $S_3$ example.

\begin{figure}[b]
    \centering
    \includegraphics[width=\linewidth]{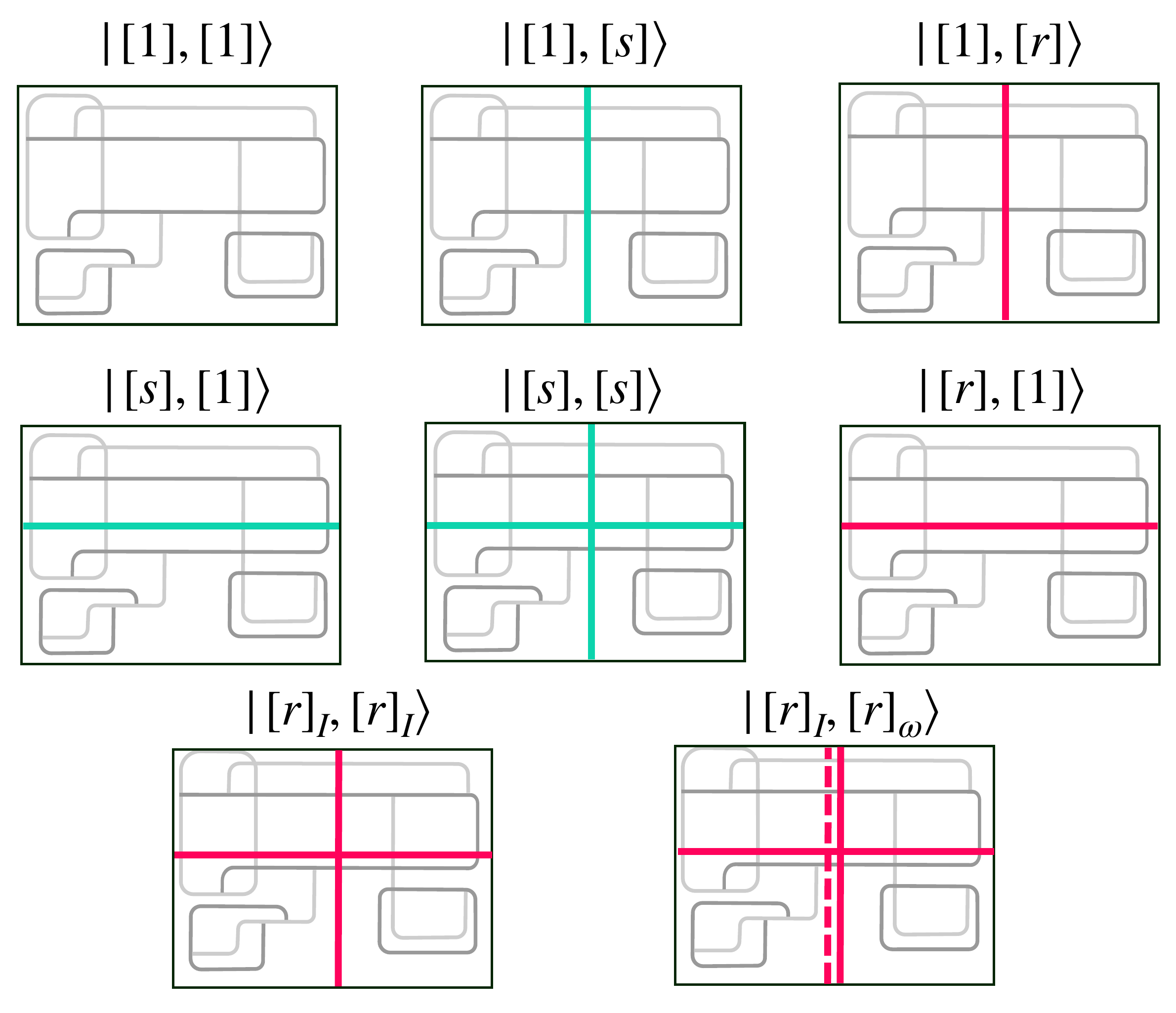}
    \caption{\textbf{Ground states of the $S_3$ quantum double on the torus} We denote the flux-free state as $\ket{[1],[1]}$ on top of which we apply $T^C_x$ or $T^C_y$ operators to construct orthogonal ground states threading a flux $C$ along the two handles of the torus. The flux line with $C=[s]$ is colored turquoise while $C=[r]$ is colored red. The last two states are distinguished by the irrep of the centralizer $\omega\in \text{Rep}(Z_{[r]})$ which labels an electric charge and therefore is generated by acting on the flux-free state with a dyonic 1-form operator $K^{[r]\omega}_y$ denoted with both a line and a dashed line.}
    \label{fig:TorusGS}
\end{figure}
\subsection{Example: Ground states of $\mathcal D(S_3)$ on a torus}

\label{subsubsec:S3-torus-gs-main}

We now illustrate the general torus construction for the quantum double of
$G=S_3$. The conjugacy classes are
$[1]=\{1\}$,
$[s]=\{s,sr,sr^2\}$,
$[r]=\{r,r^2\}$. The general result of the last section implies that for $S_3$, the degeneracy is given by $\abs{C_{\mathrm{pairs}}}$ this set has eight elements, in agreement with the eight anyon
types of $\mathcal D(S_3)$. Starting from the flux-free state
\begin{align}
\ket{[1],[1]}=\ket{\Psi_0}\equiv\ket{\Psi_{\mathcal{D}(S_3)}},
\end{align}
we first act with the magnetic 1-form symmetry operators
$T_x^C$ and $T_y^C$ along the two non-contractible cycles. A single magnetic
loop threads a conjugacy-class flux through one handle and generates
\begin{align}
\ket{[s],[1]}&=T_x^{[s]}\ket{\Psi_0},&
\ket{[r],[1]}&=T_x^{[r]}\ket{\Psi_0},
\\
\ket{[1],[s]}&=T_y^{[s]}\ket{\Psi_0},&
\ket{[1],[r]}&=T_y^{[r]}\ket{\Psi_0} .
\end{align}
Together with the flux-free state, these give the five sectors in which at least one of the two threaded fluxes is trivial. The remaining sectors require threading flux through both handles. The general commuting-pair condition implies that the two fluxes must commute. For $S_3$, the relevant centralizers are
$Z_{[1]}=S_3$,
$Z_{[s]}=\{1,s\}$,
$Z_{[r]}=\{1,r,r^2\}$. Thus two nontrivial fluxes can give a nonzero flat state only when they belong
to compatible classes. For different nontrivial classes, the intersection $C\cap Z_{\widetilde C}=\emptyset$, $C\neq \widetilde C $. Hence the only additional sectors come from the pairs
$([s],[s])$ and $([r],[r])$. For the transposition class, the commuting-pair orbit gives a single state,
\begin{align}
\ket{([s],[s])}
&=
T_y^{[s]}T_x^{[s]}\ket{\Psi_0}
\nonumber\\
&=
\prod_s A(s)
\left(
T_y^sT_x^s
+
T_y^{sr}T_x^{sr}
+
T_y^{sr^2}T_x^{sr^2}
\right)
\ket{\mathbf 1}.
\label{eq:S3-ss-main}
\end{align}
The centralizer charge does not produce an additional independent state in this
sector; it only changes the overall character factor. For this example, we choose to use the $K_\sigma^{CR}$ operators, but these are just generalized Fourier
transforms of $F_\sigma^{CD}$. The rotation class behaves differently. Since 
$Z_{[r]}=\{1,r,r^2\}\simeq \mathbb Z_3$, there are nontrivial centralizer charges
$R=I,\omega,\bar\omega$. The pure magnetic operator
$T_y^{[r]}T_x^{[r]}$ gives one state, but it does not fully resolve the
$([r],[r])$ sector. A second independent state is obtained by using a dyonic
closed ribbon carrying the nontrivial centralizer charge $\omega$:
\begin{align}
\ket{([r]_I,[r]_I)}
&=
T_y^{[r]}T_x^{[r]}\ket{\Psi_0}
\\
\ket{([r]_I,[r]_\omega)}
&=
K_y^{[r]\omega}T_x^{[r]}\ket{\Psi_0}
\nonumber\\
&=
K_y^{[r]\omega}K_x^{[r]I}\ket{\Psi_0} .
\label{eq:S3-rr-main}
\end{align}
Equivalently, the $([r],[r])$ sector is two-dimensional because the two commuting-pair orbits $
(r,r)\sim(r^2,r^2)$, and $(r,r^2)\sim(r^2,r)$
are distinct under simultaneous conjugation. Putting everything together, the torus ground-state subspace of
$\mathcal D(S_3)$ is spanned by
\begin{align}
&\ket{[1],[1]},\quad
\ket{[s],[1]},\quad
\ket{[r],[1]},\quad
\ket{[1],[s]},\quad
\ket{[1],[r]},
\nonumber\\
&\ket{([s],[s])},\quad
\ket{([r]_I,[r]_I)},\quad
\ket{([r]_I,[r]_\omega)} .
\label{eq:S3-eight-gs-main}
\end{align}
We exhausted the number of linearly independent states; it can be seen for example that $\ket{([r]_\omega,[r]_I)}$ is contained in the span (see App.~\ref{A:anomaly}). Therefore, we obtain eight states, exactly the same number as the number of anyons. We interpret this result as follows. All torus ground states of
$\mathcal D(S_3)$ can be generated from the non-contractible magnetic 1-form operators $T_\sigma^C$, as in the Abelian toric-code case, provided one adds a
single additional 1-form operator, $K_\sigma^{[r]\omega}$, which inserts a
combined flux $[r]$ and nontrivial charge $\omega$ along one non-contractible
direction, see Fig.~\ref{fig:TorusGS}. In short, the ground states are generated by the magnetic 1-form
symmetry together with a dyonic 1-form symmetry operator that resolves
the residual centralizer-charge data. For $\mathcal D(S_3)$, a single
additional dyonic operator, $K_\sigma^{[r]\omega}$, completes the construction
of the full eight-dimensional torus ground-state subspace.

\subsection{Detecting topological sectors with 1-form symmetry operators}

The previous form of the ground states both for the torus and the cylinder have the special property of being eigenstates of the electric 1-form operators $W^{\Gamma}(\mathcal{C})$ and the magnetic 1-form operators toggle between ground states. %In the toric code case the eigenbasis of magnetic 1-form ground states would be minimally entangled states. 
Here we fix the basis of the ground states to be given by Eq.~\eqref{eq:torus-gs-main} and ask the question of what operations detect the topological charge of each basis state, later on we deal with a general linear combination.

,We start with detecting the distinct ground states for $\mathcal{D}(S_3)$. For the cylinder, it is simple since measuring or applying $W^\Gamma_x$ along the non-periodic direction identifies all ground states as can be seen from Eq.~\eqref{CylGS}, which implies that by comparing the values of the character table to the measured eigenvalues by changing the $\Gamma$ irrep, all ground states are identified. For the torus, the situation is similar since we can use the anomaly Eq.~\eqref{Anomaly} to measure for each handle of the torus the $(W^\Gamma_{x},W^\Gamma_{y})$ operators and obtain the eigenvalues $(\chi_\Gamma(\mathcal{C}_{x}),\chi_\Gamma(\mathcal{C}_{y}))$. This leads to the following Table~\ref{tab:character-GS}, which indicates that almost all ground-state sectors are accounted for, but only seven sectors instead of eight are labeled. The two-fold degeneracy of the $(\chi_\Gamma([r]),\chi_\Gamma([r]))$ sector indicates that there is an additional operator that must distinguish them. Since we have used all the pure electric 1-form operators, it must contain some magnetic flux. 

\begin{table}[t]
\centering
% \begin{tabular}{|c|c|c|c|}
% \hline
% $W_y^\Gamma /\ W_x^\Gamma$ & $d_\Gamma$ & $\chi_\Gamma([s])$ & $\chi_\Gamma([r])$ \\
% \hline
% $d_\Gamma$ & $\ket{[1],[1]}$ & $\ket{[1],[s]}$ & $\ket{[1],[r]}$ \\
% \hline
% $\chi_\Gamma([s])$ & $\ket{[s],[1]}$ & $\ket{[s],[s]}$ & $-$  \\
% \hline
% $\chi_\Gamma([r])$ & $\ket{[r],[1]}$ & $-$ & $\ket{[r]_I,[r]_I},\ket{[r]_I,[r]_\omega}$ \\
% \hline
% \end{tabular}
\renewcommand{\arraystretch}{1.3}
\setlength{\tabcolsep}{8pt}
\begin{adjustbox}{max width=\columnwidth}
\begin{tabular}{cc|ccc}
\hline\hline
& & \multicolumn{3}{c}{\(W_x^\Gamma\)} \\
& & \(d_\Gamma\) & \(\chi_\Gamma([s])\) & \(\chi_\Gamma([r])\) \\
\hline
\multirow{3}{*}{\(W_y^\Gamma\)}
& \(d_\Gamma\)
& \(\ket{[1],[1]}\)
& \(\ket{[1],[s]}\)
& \(\ket{[1],[r]}\)
\\[0.25em]
& \(\chi_\Gamma([s])\)
& \(\ket{[s],[1]}\)
& \(\ket{[s],[s]}\)
& \(-\)
\\[0.25em]
& \(\chi_\Gamma([r])\)
& \(\ket{[r],[1]}\)
& \(-\)
& \(\begin{gathered}
    \ket{[r]_I,[r]_I},\\[-0.15em]
    \ket{[r]_I,[r]_\omega}
  \end{gathered}\)
\\
\hline\hline
\end{tabular}
\end{adjustbox}
\caption{Ground-state eigenvalues of the 1-form $\text{Rep}(S_3)$ Wilson loop operators for the irrep $\Gamma$ along the two handles of the torus for $\mathcal{D}(S_3)$.}
\label{tab:character-GS}
\end{table}
The strategy is to change the basis of closed ribbon operators to the orthogonal projectors introduced in Ref.~\cite{bombin_family_2008}, which we name $\Pi^{RC}_\sigma$. These operators give zero unless the state has a non-Abelian anyon $R,C$ enclosed by $\sigma$. It was shown that these projectors are a basis for the algebra of closed ribbons. Therefore, $\Pi^{RC}_\sigma$ can be expressed as a superposition of our dyon operators, see App.~\ref{A:closed-ribbon-projectors}. We use now the operator $\Pi^{R[r]}_x$ for $R=\omega$ and $R=I$ in the $(\chi_\Gamma([r]),\chi_\Gamma([r]))$ sector. 
Since the $\ket{([r]_I,[r]_\omega)}$ state contains a dyon line $(\omega,[r])$ threading the $y$ direction of the torus, see Fig.~\ref{fig:TorusGS}, we know that $\Pi^{R[r]}_x \ket{([r]_I,[r]_\omega)} = \ket{([r]_I,[r]_\omega)}$ but $\Pi^{R[r]}_x \ket{([r]_I,[r]_I)} = 0$ since the dyon line of the state is now $(I,[r])$. This procedure distinguishes the last two states. 

For a general quantum double the procedure is similar. We recall that the basis states are  $\ket{[(g,h)]}=F^{[h]D}_y T^{[g]}_x \ket{\Psi_{\mathcal{D}(G)}}$. They are labeled by conjugacy pairs $C_{\text{pair}}$ or alternatively a class $[g]$ and the conjugacy class of the centralizer $D\subset [g]\cap Z_{[h]}$ obtained for different classes $[h]$. Then, we know the Wilson loops will produce when acted upon these states the eigenvalues $(\chi_\Gamma([g]),\chi_\Gamma([h]))$ which will distinguish at most $\abs{\text{Class}(G)}^2$ states (if all anyon lines along $x$,$y$ are allowed) and at least $2\abs{\text{Class}(G)}-1$ (given by the states with one trivial flux line along $x$ or $y$). The rest of the states can be distinguished by changing the basis to use $\Pi^{RC}_\sigma$
and re-expressing the $F^{[h]D}_y$ operators in terms of the dyon 1-form operators $K^{R[h]}_y$ via the inverse transformation of Eq.~\eqref{Dyon-1-form}.

In fact, the last observation works for an arbitrary linear combination in the ground state space. Let us suppose the system is in the state 
$\ket{\Phi} = \sum_{[(g,h)]}\Phi_{[(g,h)]}\ket{[(g,h)]} $, with $\Phi_{[(g,h)]}$ arbitrary constants. We may rewrite it in  $\ket{\Phi} = \sum_{[g],[h],D } \Phi_{[(g,h)]}F^{[h]D}_yF^{[g][1]}_x\ket{\Psi_{\mathcal{D}(G)}}$. Now going to the anyon basis and changing to the projectors as shown in App.~\ref{A:closed-ribbon-projectors}, implies that  $\ket{\Phi} = \sum_{\frak a \frak b } \Phi_{\frak a,\frak b} K^{\frak a}_yK^{\frak b}_x\ket{\Psi_{\mathcal{D}(G)}}  $ is  $\ket{\Phi} = \sum_{\frak a, \frak b} \Phi'_{\frak a,\frak b} \Pi^{\frak a}_y K^{\frak b}_x\ket{\Psi_{\mathcal{D}(G)}} $. Therefore $\Pi^{\frak c}_y\ket{\Phi}= \sum_{ \frak b} \Phi'_{\frak c,\frak b} \Pi^{\frak c}_y K^{\frak b}_x\ket{\Psi_{\mathcal{D}(G)}}$, which explicitly ensures the state has a $\frak c$ anyon worldline threading along $x$ producing non-zero $\Pi^{\frak c}_y$ if $\Phi'_{\frak c,\frak b}\neq 0$, with $\Phi$ and $\Phi'$ related by the change of basis calculated in App.~\ref{A:closed-ribbon-projectors}. Similar reasoning works for the projector $\Pi^{\frak c}_x$.
We conclude then that the non-contractible non-invertible 1-form symmetry operators form a complete diagnostic algebra for the torus ground-state sectors.

We summarize from the previous results that if the system is in one of the $\ket{[(g,h)]}$ states then they will all be $\text{Rep}(G)^{(1)}$ symmetric states, as they are eigenstates of the $W_x^\Gamma,W_y^\Gamma$ electric 1-form operators. Nevertheless, since we constructed them from magnetic 1-form loops, applying the non-contractible 't Hooft loops $T^C_x,T^C_y$ will toggle to different $\ket{[(g,h)]}$ based on the magnetic fusion. Such a state has selected a fixed value of $W_x^\Gamma,W_y^\Gamma$ just like an Ising ferromagnetic ground state selects an up or down alignment. We have therefore shown that in the lattice the ground states for non-Abelian topological order have \textit{spontaneously broken the non-invertible 1-form $\text{Rep}(G)^{(1)}\times\text{Class}(G)^{(1)}$ magnetic symmetry}.

\section{Discussion and outlook} \label{sec:discussion}

In this work, we have studied and contributed to clarifying the physical meaning of non-invertible lattice 1-form symmetries in a large class of non-Abelian topological orders given by finite $G$ group quantum double models on a lattice. Starting from the exactly solvable fixed point, we constructed the electric and magnetic 1-form symmetries and showed how their algebraic properties differ from the 1-form operators in the Abelian topological order case. The electric 1-form symmetry operators realize the fusion algebra \(\mathrm{Rep}(G)\), while the magnetic ones realize the conjugacy-class algebra \(\mathrm{Class}(G)\) on the flux-free sector. For non-Abelian \(G\), these algebras are non-invertible as they correspond to non-Abelian anyon lines which can fuse into multiple channels.

From this generalized symmetry perspective the non-Abelian topological ordered flux-free ground state can be seen as a condensate of closed magnetic strings. Contractible magnetic loops act trivially on this condensate, whereas non-contractible magnetic loops move between distinct topological sectors labeled by conjugacy classes along the torus handles subject to a commutation relation. We showed that the emergent mixed 't Hooft anomaly allows us to resolve different topological sectors. Non-contractible Wilson loops measure the flux threaded through the complementary cycle, whereas magnetic loops generate the different symmetry-broken ground states. 
In a non-Abelian quantum double this information is not sufficient, because distinct sectors may carry the same conjugacy-class flux but differ by the centralizer electric charge. By including dyonic 1-form symmetry operators we constructed all ground states of the $G$-quantum double explicitly and showed how the non-contractible 1-form symmetry operators also allow us to resolve the full topological sector data.
The topological degeneracy on nontrivial manifolds is therefore naturally understood as spontaneous breaking of the non-invertible magnetic 1-form symmetry. 

Several directions follow. It will be important to understand how the
non-invertible 1-form symmetry structure emerges away from the exactly
solvable point, where loop operators become dressed while the topologically
ordered phase remains stable~\cite{HastingsWen2005,PaceWen2023}. 
An information theoretic perspective may shed light on these
properties~\cite{liu_1form_2025}. As a further step in complexity chiral lattice non-Abelian theories and twisted quantum doubles would give an additional level of understanding on 1-form symmetries.  Furthermore, Fredenhagen--Marcu order
parameters may be constructed to analyze the stability of the topological
phase~\cite{FredenhagenMarcu1983,GregorHuseMoessnerSondhi2011,xuFM}.
Knowledge of the higher-form symmetries can also be used to study
confinement transitions out of non-Abelian topological order, interpreted
as transitions in which the magnetic non-invertible 1-form symmetry is
restored~\cite{GaiottoKapustinSeibergWillett2015,
BhardwajBottiniSchaferNamekiTiwari2023,JiWen2020CategoricalSymmetry,
bombin_family_2008,XuSchuch2021NonAbelianTransitions,
XuGarreRubioSchuch2022}. More broadly, the developed framework may be
useful for characterizing non-Abelian topological states on quantum processors~\cite{Iqbal2024NonAbelianTO,XuSun,Evered,Will2025NonEquilibriumTopologicalOrder,
Lo2026UniversalS3}.

\begin{acknowledgments}
We are thankful to G. Delfino for useful comments on the draft. We also thank J. Boesl, J. Graf, Y.-J. Liu, and S. Moroz for discussions and collaborations on related projects. We acknowledge support from the Deutsche Forschungsgemeinschaft (DFG, German Research Foundation) under Germany’s Excellence Strategy--EXC--2111--390814868, TRR 360 – 492547816 and DFG grants No. KN1254/1-2, KN1254/2-1, the European Union (grant agreement No 101169765), as well as the Munich Quantum Valley, which is supported by the Bavarian state government with funds from the Hightech Agenda Bayern Plus.
\end{acknowledgments}

\section*{Data availability}
All data supporting the findings of this study are contained in this manuscript.

\addtocontents{toc}{\string\tocdepth@munge}

\appendix

\section{$G$-qudits} \label{A:qudits}Let $G$ be a finite group. The local Hilbert space at a lattice link $j$ is
$\mathcal{H}_j=\mathbb{C}[G]$, the complex group algebra of $G$. Its standard
orthonormal basis is $\{\ket{g}:g\in G\}$, so the local dimension is
$N_G=\abs{G}$. We define four families of linear operators
$X^g_+,X^g_-,Z^g_+,Z^g_-$, indexed by $g\in G$, by their action on basis
states
\begin{align}
    &X^g_+ \ket{h} = \ket{gh} , \qquad     X^g_- \ket{h} = \ket{hg^{-1}} \nonumber\\
   &Z^g_+ \ket{h} = \delta_{h,g}\ket{h} , \qquad   Z^g_- \ket{h} = \delta_{h^{-1},g}\ket{h}.
\end{align}
Let us define, as in the main text, $\bar{g}=g^{-1}$ for convenience. These operators satisfy the commutation relations:
\begin{align}
    &X^g_+ Z^h_+ = Z^{gh}_+X_+^g, \qquad  X^g_+ Z^h_- = Z^{h\bar{g}}_-X_+^g\nonumber\\
    &X^g_- Z^h_+ = Z^{h\bar{g}}_+X_-^g, \qquad  X^g_- Z^h_- = Z^{gh}_-X_-^g.
\end{align}
This follows from the definitions applied to an arbitrary basis state, for example
\begin{align}
    X^g_+ Z^h_+ \ket{f} = X^g_+ \delta_{h,f} \ket{f} = \delta_{gh,gf}X^g_+\ket{f} = Z^{gh}_+ X^g_+\ket{f}.
\end{align}
Similarly,
\begin{align}
    [Z^g_\pm,Z^h_\pm] = 0 , \quad [X_-^h,X_+^g]=0. \label{0comms}
\end{align}
The remaining commutators between the displayed families can be nonzero. In
contrast to the Abelian case, left (or right) multiplication operators with
different noncommuting labels also have a nonzero commutator
\begin{align}
    &[X_-^g,Z_-^h]= X_-^g(Z_-^h-Z_-^{\bar{g}h}),\label{Xcomms}\\
    &[X_-^g,Z_+^h]= (Z_+^{h\bar{g}}-Z^h_+)X_-^g, \nonumber\\
    &[X_+^g,Z_-^h]= X_+^g(Z_-^{h}-Z^{hg}_-),\nonumber\\
    &[X_+^h,X_+^g]= X_+^{hg}-X_+^{gh}, \quad [X_-^h,X_-^g]= X_-^{hg}-X_-^{gh}. \nonumber
\end{align}
Under Hermitian conjugation, the operators satisfy
\begin{align}
    (X^g_\pm)^\dagger = X^{\bar{g}}_\pm, \qquad     (Z^g_\pm)^\dagger = Z^{g}_\pm. \label{XZadjoint}
\end{align}
\section{Dual irrep basis} \label{A:irrepbasis}
We now introduce a new basis of states, obtained from a generalization of the Fourier transform. The basis can be constructed by noting that if we consider all irreducible representations of a finite group $G$ denoted by $\Gamma$ with matrix dimension $d_\Gamma$ then 
\begin{align}
\sum_\Gamma d_\Gamma^2 = \abs{G}.
\end{align}
For two irreducible representations $\Gamma$ and $\tilde{\Gamma}$, the great
orthogonality theorem states that
\begin{align}
\sum_{g}\Gamma(g)_{\alpha \beta}^*\tilde{\Gamma}(g)_{\gamma \delta} = \dfrac{\abs{G}}{d_\Gamma} \delta_{\Gamma,\tilde{\Gamma}} \delta_{\alpha \gamma} \delta_{\beta \delta},
\end{align}
where $\Gamma(g)_{\alpha\beta}$ is a matrix element of $\Gamma(g)$ and
$\alpha,\beta=1,\dots,d_\Gamma$. We define the dual irrep basis by
\begin{align}
\ket{\Gamma_{\alpha\beta}}= \sqrt{\tfrac{d_\Gamma}{\abs{G}}} \sum_g \Gamma(g)_{\alpha\beta} \ket{g},\\
\ket{g}= \sum_{\Gamma,\alpha,\beta} \sqrt{\tfrac{d_\Gamma}{\abs{G}}} \Gamma(g)_{\alpha\beta}^* \ket{\Gamma_{\alpha\beta}}.
\end{align}
The great orthogonality theorem shows that these states are orthonormal
\begin{align}
\braket{\Gamma_{\alpha\beta}}{\tilde{\Gamma}_{\gamma\delta}}= \delta_{\Gamma,\tilde{\Gamma}} \delta_{\alpha \gamma} \delta_{\beta \delta}.
\end{align}
These states block diagonalize the $X$ operators
\begin{align}
X^g_+\ket{\Gamma_{\alpha\beta}} = \sum_{\alpha'} \Gamma_{\alpha\alpha'}(\bar{g})\ket{\Gamma_{\alpha'\beta}},  \nonumber \\
X^g_-\ket{\Gamma_{\alpha\beta}} = \sum_{\beta'}
\ket{\Gamma_{\alpha\beta'}}\Gamma(g)_{\beta'\beta}.
\end{align}
Moreover, when the $Z$ operators act on the states, we get
\begin{align}
    Z^h_{+}\ket{\Gamma_{\alpha\beta}}& = \sqrt{d_\Gamma/\abs{G}} \sum_g \Gamma(g)_{\alpha\beta} \delta_{g,h}\ket{g} \nonumber\\
    &= \sqrt{\tfrac{d_\Gamma}{\abs{G}}}\Gamma(h)_{\alpha\beta} \ket{h} \nonumber\\
    &= \sum_{\tilde{\Gamma},\alpha',\beta} \tfrac{\sqrt{d_\Gamma d_{\tilde{\Gamma}}}}{\abs{G}} \tilde{\Gamma}(h)_{\alpha'\beta'}^*\Gamma(h)_{\alpha\beta} \ket{\tilde{\Gamma}_{\alpha'\beta'}}.
\end{align}
Let us define the dual-$Z$ operators as
\begin{align}
&Z^\Gamma_{\alpha\beta}\ket{h}=\Gamma(h)_{\alpha\beta}\ket{h}, \nonumber\\
&Z^\Gamma_{\alpha\beta} = \sum_h \Gamma(h)_{\alpha\beta} Z^h_+= \sum_h \Gamma^*(h)_{\beta\alpha} Z^h_-.
\end{align}
The adjoint of the operator is just the complex-conjugate representation 
\begin{align}
    &(Z^\Gamma_{\alpha\beta})^\dagger= (\sum_h \Gamma(h)_{\alpha\beta}\ketbra{h})^\dagger \nonumber  \\
    &= \sum_h\Gamma(h)^*_{\alpha\beta}\ketbra{h}=Z^{(\Gamma^*)}_{\alpha\beta}.
\end{align}
From the identity $\braket{g}{h}=\delta_{h,g}$ and the definition of the irrep basis in terms of the group basis we obtain
\begin{align}
    \sum_{\Gamma,\alpha,\beta}\tfrac{d_\Gamma}{\abs{G}}\Gamma(g)_{\alpha\beta}\Gamma(h)^*_{\alpha\beta}=\delta_{h,g}.
\end{align}
Now we can use this identity to invert the previous relation between $Z^\Gamma$ and $Z^g_+$ to obtain
\begin{align}
    Z^g_+ =  \sum_{\Gamma,\alpha,\beta}\tfrac{d_\Gamma}{\abs{G}}\Gamma(g)^*_{\alpha\beta}Z^{\Gamma}_{\alpha\beta} =\tfrac{1}{\abs{G}}\sum_\Gamma d_\Gamma \tr{\bm{\Gamma}^\dagger(g)\bm{Z}^{\Gamma}}, \label{Z+2gam}
\end{align}
analogously
\begin{align}
    Z^g_- =  \sum_{\Gamma,\alpha,\beta}\tfrac{d_\Gamma}{\abs{G}}Z^{\Gamma}_{\alpha\beta} \Gamma(g)_{\beta \alpha}=\tfrac{1}{\abs{G}}\sum_\Gamma d_\Gamma \tr{\bm{\Gamma} (g)\bm{Z}^{\Gamma}}.\label{Z-2gam}
\end{align}
The action of $Z^{\Gamma'}$ on a dual-basis state is
\begin{align}
Z^{\Gamma'}_{\alpha'\beta'}\ket{\Gamma_{\alpha \beta}}= \sqrt{\dfrac{d_\Gamma}{\abs{G}}} \sum_{g}(\Gamma'\otimes \Gamma)(g)_{\alpha'\alpha,\beta'\beta} \ket{g}.
\end{align}
Writing $\bm Z^\Gamma$ for the matrix with entries
$(\bm Z^\Gamma)_{\alpha\beta}=Z^\Gamma_{\alpha\beta}$, we also have the
well-defined matrix-operator identities
\begin{align}
    X^g_+\bm Z^\Gamma &= \Gamma(\bar g)\bm Z^\Gamma X_+^g, \nonumber\\
    X^g_-\bm Z^\Gamma &= \bm Z^\Gamma\Gamma(g)X_-^g.
\end{align}

\section{Algebraic properties of the $G$-Kitaev quantum double and anyonic excitations}\label{A:gKqdm}
Let $z=(s,p)$ be a site consisting of a vertex $s$ and an adjacent plaquette
$p$. For an edge $j$ incident on $s$, define $X^g(j,s)=X_-^g(j)$ when $j$ is
oriented away from $s$ and $X^g(j,s)=X_+^g(j)$ when it is oriented toward
$s$. Likewise, $Z^h(j,p)$ denotes $Z_+^h(j)$ or $Z_-^h(j)$ according to
whether the edge orientation agrees or disagrees with the chosen orientation
of $\partial p$. If the ordered boundary of $p$ is $(j_1,\ldots,j_k)$, define
\begin{align}
& A_g(s)=\prod_{j \in \operatorname{star}(s)} X^g(j,s), \\
& B_h(p)=\sum_{h_1\cdots h_k=h}
\prod_{m=1}^k Z^{h_m}(j_m,p),\\
& A(s)=\frac{1}{\abs{G}}\sum_{g\in G}A_g(s),
\qquad B(p)=B_1(p).
\end{align}
The averaged star operator $A(s)$ and the zero-flux plaquette operator $B(p)$
are projectors. Indeed, since $A_{g'}(s)A_g(s)=A_{g'g}(s)$,
\begin{align}
A(s)^2
&=\frac{1}{\abs{G}^2}\sum_{g,g'\in G}A_{g'}(s)A_g(s) \nonumber\\
&=\frac{1}{\abs{G}^2}\sum_{g,g'\in G}A_{g'g}(s)
=\frac{1}{\abs{G}}\sum_{h\in G}A_h(s)
=A(s).
\end{align}
The $B_h(z)$ are mutually orthogonal diagonal projectors, so
$B(p)^2=B(p)$. Moreover, $A_g(s)^\dagger=A_{\bar{g}}(s)$, and the sum over
$g$ is invariant under inversion. Thus
\begin{align}
&A(s)^2=A(s), \quad B(p)^2=B(p),\nonumber\\
&A(s)^\dagger=A(s), \quad 
B(p)^\dagger=B(p).
\end{align}
The projectors also commute
\begin{align}
&[A(s),B(p)] =0, \quad
[A(s),A(s')]=0, \nonumber  \\
& [B(p),B(p')]=0, \quad \text{for all} \; s,s',p,p'.
\end{align}
 Let $z=(s,p)$ and $z'=(s',p')$ denote the sites as in Fig.~\ref{fig:ribbon}. The subspace with
possible excitations only at $z$ and $z'$ is denoted $\mathcal{L}(z,z')$, and
its projector is
\begin{align}
    P_{z,z'}=\prod_{r\neq s,s'}A(r)\prod_{l\neq p,p'}B(l).
\end{align}
Operators that probe an excitation at $z$ must preserve
$\mathcal{L}(z,z')$. 

The operators $A_g(z)\equiv A_g(s) $ and $B_h(z)\equiv B_h(p)$ generate a local
algebra $\mathcal{D}(z)$ and commute with $P_{z,z'}$. This algebra contains
all local operators acting on $\mathcal{L}(z,z')$ near $z$~\cite{kitaev_fault-tolerant_2003}.
It is independent of the choice of $z$ up to isomorphism and is the quantum
double $\mathcal{D}(G)$. Its defining relations are
\begin{align}
    &A_f A_g=A_{fg},\quad  B_h B_i=\delta_{h,i}B_h, \nonumber \\
    &A_g B_h=B_{ghg^{-1}}A_g.
\end{align}
The operators $D_{(h,g)}=B_hA_g$ form a linear basis of $\mathcal{D}$. With
$\bm m=(h_1,g_1)$ and $\bm n=(h_2,g_2)$, their multiplication rule is
\cite{kitaev_fault-tolerant_2003}
\begin{align}
    D_{\boldsymbol{m}}D_{\boldsymbol{n}}
    &=\Omega_{\boldsymbol{m}\boldsymbol{n}}^{\boldsymbol{k}}D_{\boldsymbol{k}}, \nonumber \\
    \Omega_{(h_1,g_1)(h_2,g_2)}^{(h,g)}
    &=\delta_{h_1,g_1h_2\bar{g}_1}\delta_{h,h_1}\delta_{g,g_1g_2}.
\end{align}
The algebra is also closed under adjoints $A_g^{\dagger}=A_{g^{-1}} \quad B_h^{\dagger}=B_h \quad D_{(h, g)}^{\dagger}=D_{\left(\bar{g}h g, \bar{g}\right)}$ consequently the algebra is a finite-dimensional $C^*$-algebra and decomposes as $\mathcal{D}= \oplus_d L(\mathcal{K}_d)$ with $d$ running over irreducible representations of the quantum double algebra, representing the particle types and $L(\mathcal{K}_d)$ denoting the space of linear operators on the Hilbert space of "subtypes" of local degrees of freedom $\mathcal{K}_d$.

\section{Fusion of Wilson loops and electric 1-form properties}\label{A:Wilsonfusion}

Consider a loop $\mathcal C$ with $p$ edges and a basis state
$\ket{g_1,g_2,\dots,g_p}$, where $g_\ell$ labels the group element on edge
$\ell\in\mathcal C$. We absorb orientation reversals into
$g_\ell\mapsto g_\ell^{-1}$, so all factors below follow the loop
orientation. The product of two Wilson loops on the same contour acts as
\begin{align}
& W^{\Gamma_1}(\mathcal{C})W^{\Gamma_2}(\mathcal{C})
\ket{g_1,\dots,g_p}
\nonumber\\
& =
\tr \Big[\mathcal{P}\prod_{\ell\in \mathcal{C}}
\bm{Z}^{\Gamma_1}(\ell)\Big]
\tr \Big[\mathcal{P}\prod_{\ell\in \mathcal{C}}
\bm{Z}^{\Gamma_2}(\ell)\Big]
\ket{g_1,\dots,g_p}.
\end{align}
Equivalently,
\begin{align}
& W^{\Gamma_1}(\mathcal{C})W^{\Gamma_2}(\mathcal{C})
\ket{g_1,\dots,g_p}
\nonumber\\
& =
\tr [\Gamma_1(g_p)\cdots \Gamma_1(g_1)]
\tr[\Gamma_2(g_p)\cdots \Gamma_2(g_1)]
\ket{g_1,\dots,g_p}.
\end{align}
Writing the traces explicitly in components gives
\begin{align}
& W^{\Gamma_1}(\mathcal{C})W^{\Gamma_2}(\mathcal{C})
\ket{g_1,\dots,g_p}
\nonumber\\
& =
\Gamma_1(g_p)_{\alpha\beta_1}
\cdots
\Gamma_1(g_1)_{\beta_{p-1}\alpha}
\nonumber\\
&\quad \times
\Gamma_2(g_p)_{\alpha'\beta_1'}
\cdots
\Gamma_2(g_1)_{\beta_{p-1}'\alpha'}
\ket{g_1,\dots,g_p}.
\end{align}
Regrouping the matrix elements link by link
\begin{align}
& W^{\Gamma_1}(\mathcal{C})W^{\Gamma_2}(\mathcal{C})
\ket{g_1,\dots,g_p}
\nonumber\\
& =
\left(
\Gamma_1(g_p)_{\alpha\beta_1}
\Gamma_2(g_p)_{\alpha'\beta_1'}
\right)
\cdots
\nonumber\\
&\quad \times
\left(
\Gamma_1(g_1)_{\beta_{p-1}\alpha}
\Gamma_2(g_1)_{\beta_{p-1}'\alpha'}
\right)
\ket{g_1,\dots,g_p}.
\end{align}
This is the matrix element of the tensor-product representation:
\begin{align}
& W^{\Gamma_1}(\mathcal{C})W^{\Gamma_2}(\mathcal{C})
\ket{g_1,\dots,g_p}
\nonumber\\
& =
(\Gamma_1\otimes \Gamma_2)(g_p)_{
\alpha\alpha',\beta_1\beta_1'}
\cdots
\nonumber\\
&\quad \times
(\Gamma_1\otimes \Gamma_2)(g_1)_{
\beta_{p-1}\beta_{p-1}',\alpha\alpha'}
\ket{g_1,\dots,g_p}
\nonumber\\
& =
\tr[(\Gamma_1\otimes \Gamma_2)(g_p)
\cdots
(\Gamma_1\otimes \Gamma_2)(g_1)]
\ket{g_1,\dots,g_p}.
\end{align}
Since $\rho=\Gamma_1\otimes\Gamma_2$ is a representation,
$\rho(g_1)\rho(g_2)=\rho(g_1g_2)$. Moreover, there is a unitary $U$ such that
\begin{align}
U^\dagger\rho(g)U
=\bigoplus_\Gamma
\left(I_{N^\Gamma_{\Gamma_1\Gamma_2}}\otimes\Gamma(g)\right),
\end{align}
where $N^\Gamma_{\Gamma_1\Gamma_2}$ is the multiplicity of $\Gamma$ in the
tensor product. Therefore,
\begin{align}
& W^{\Gamma_1}(\mathcal{C})W^{\Gamma_2}(\mathcal{C})
\ket{g_1,\dots,g_p}
\nonumber\\
& =
\tr[(\Gamma_1\otimes \Gamma_2)(g_p\dots g_1)]
\ket{g_1,\dots,g_p}
\nonumber\\
& =
\sum_\Gamma N^{\Gamma}_{\Gamma_1\Gamma_2}
\tr{\Gamma(g_p\dots g_1)}
\ket{g_1,\dots,g_p}.
\end{align}
Thus $N^\Gamma_{\Gamma_1\Gamma_2}\in\mathbb Z_{\geq0}$ are the
Clebsch--Gordan multiplicities. This proves the Wilson-loop fusion rule
$W^{\Gamma_1}W^{\Gamma_2}=\sum_\Gamma
N^\Gamma_{\Gamma_1\Gamma_2}W^\Gamma$ on a fixed contour. We now calculate the commutator of the Wilson loop with the quantum
double Hamiltonian. First, we calculate the commutator of a closed
Wilson loop with the smallest $X$ loop operator, denoted
$T^g(\partial\tilde{p})$. If the Wilson line does not touch the edges
pierced by $\partial\tilde{p}$, then it commutes with
$T^g(\partial\tilde{p})$. Otherwise, it must pierce the dual loop twice.
Let us denote the two edges it touches by $1,2$. In the $G$ basis, and
suppressing all other edges of the lattice, we have
\begin{align}
& W^\Gamma(\mathcal{C})T^g(\partial \tilde{p})
\ket{g_1,g_2}
\nonumber\\
& =
\tr \Big[
\mathcal{P}
\prod_{\ell\in \mathcal{C}}
\bm{Z}^{\Gamma}(\ell)
\Big]
\prod_{\tilde{\ell}\in\tilde{\mathcal{C}}}
X^g_{O_{\tilde{\ell}}}(\tilde{\ell})
\ket{g_1,g_2}.
\end{align}
For the two crossed links this gives
\begin{align}
& W^\Gamma(\mathcal{C})T^g(\partial \tilde{p})
\ket{g_1,g_2}
\nonumber\\
& =
\tr[\cdots
\bm{Z}^\Gamma(2)
\bm{Z}^\Gamma(1)
\cdots]\,
X^g_{+}(1)X^g_{-}(2)
\nonumber\\
&\quad \times
\prod_{\tilde{\ell}\neq\{1,2\}}
X^g_{O_{\tilde{\ell}}}
\ket{g_1,g_2}.
\end{align}
Acting first with the $X$ operators gives
\begin{align}
& W^\Gamma(\mathcal{C})T^g(\partial \tilde{p})
\ket{g_1,g_2}
\nonumber\\
& =
\tr[\cdots
\Gamma(g_2\bar{g})\Gamma(gg_1)
\cdots]
\ket{gg_1,g_2\bar{g}}.
\end{align}
Using $\Gamma(g_2\bar{g})\Gamma(gg_1)
=\Gamma(g_2)\Gamma(g_1)$, this becomes
\begin{align}
& W^\Gamma(\mathcal{C})T^g(\partial \tilde{p})
\ket{g_1,g_2}
\nonumber\\
& =
\tr[\cdots
\Gamma(g_2)\Gamma(g_1)
\cdots]
\ket{gg_1,g_2\bar{g}}.
\end{align}
Equivalently,
\begin{align}
& W^\Gamma(\mathcal{C})T^g(\partial \tilde{p})
\ket{g_1,g_2}
\nonumber\\
& =
X^g_{+}(1)X^g_{-}(2)
\tr\Big[
\mathcal{P}
\prod_{\ell\in \mathcal{C}}
\bm{Z}^{\Gamma}(\ell)
\Big]
\ket{g_1,g_2}
\nonumber\\
& =
T^g(\partial \tilde{p})
W^\Gamma(\mathcal{C})
\ket{g_1,g_2}.
\end{align}
We assumed an orientation of the lattice with all edges pointing up or
right, and assumed that the Wilson loop traverses the dual square from
left to right. The same result holds for the other possible Wilson-loop
piercings. Therefore,
\begin{align}
[T^g(\partial \tilde{p}),W^\Gamma(\mathcal{C})]=0 .
\end{align}
This calculation concerns the elementary gauge-transformation loop
$T^g(\partial\tilde p)$. A generic naive product of $X$ operators along a
larger dual loop need not commute with a Wilson loop; a gauge-invariant
non-Abelian magnetic loop requires the ribbon construction below.
With these results in mind, we can now calculate the Hamiltonian
commutator. The quantum double Hamiltonian is
\begin{align}
H_\text{QD}
&=
\sum_{\tilde{p}}
\left(
1-\frac{1}{|G|}
\sum_g T^g(\partial \tilde{p})
\right)
\nonumber\\
&\quad +
\sum_p
\left(
1-\frac{1}{|G|}
\sum_{\tilde{\Gamma}}
d_{\tilde{\Gamma}}\,
W^{\tilde{\Gamma}}(\partial p)
\right).
\end{align}
Thus,
\begin{align}
& [H_\text{QD},W^\Gamma(\mathcal{C})]
\nonumber\\
& =
-\sum_{\tilde{p}}
\frac{1}{|G|}
\sum_g
[T^g(\partial \tilde{p}),W^\Gamma(\mathcal{C})]
\nonumber\\
&\quad
-\sum_p
\frac{1}{|G|}
\sum_{\tilde{\Gamma}}
d_{\tilde{\Gamma}}\,
[
W^{\tilde{\Gamma}}(\partial p),
W^\Gamma(\mathcal{C})
]
\nonumber\\
&=0 .
\end{align}

We next prove the projected gluing relation for Wilson loops. Geometric fusion
of adjacent loops is not an operator identity on the full Hilbert space; it
becomes an identity after projection to a locally defect-free subspace. On a group-element basis state $|\{g_\ell\}\rangle$,  $W^\Gamma(\mathcal C)$  acts diagonally as
\begin{align}
W^\Gamma(\mathcal C)|\{g_\ell\}\rangle
=
\chi_\Gamma\!\left(U_{\mathcal C}\right)
|\{g_\ell\}\rangle ,
\end{align}
where
\begin{align}
U_{\mathcal C}
=
\mathcal P\prod_{\ell\in\mathcal C}
g_\ell^{o_\ell},
\end{align}
where $o_\ell=\pm1$ according to whether the edge orientation agrees with the
orientation of the loop. Consider two elementary plaquette loops $\mathcal C_1$ and $\mathcal C_2$ sharing one edge with opposite orientations. Let
\begin{align}
\mathcal C=\mathcal C_1\oplus\mathcal C_2
\end{align}
denote the boundary of their union, with the common edge removed. Let
$\mathcal R$ contain the two plaquettes and their boundary vertices, and define
\begin{align}
P_{\mathcal R}^0
=
\prod_{s\subset\mathcal R}A(s)
\prod_{p\subset\mathcal R}B(p).
\end{align}
The plaquette projectors impose local flatness in $\mathcal R$. Therefore, for any basis configuration surviving the projection,
\begin{align}
U_{\mathcal C_1}=1,
\qquad
U_{\mathcal C_2}=1.
\end{align}
After transporting the two plaquette holonomies to a common base point, the
holonomy around the joined loop is their ordered product, up to an overall
conjugation determined by the base-point convention:
\begin{align}
U_{\mathcal C}
\sim
U_{\mathcal C_1}U_{\mathcal C_2}.
\end{align}
Here $\sim$ denotes equality up to conjugation, which does not change a
character. Thus, inside the image of $P_{\mathcal R}^0$,
\begin{align}
U_{\mathcal C}=1.
\end{align}
It follows that, after restriction to the defect-free subspace,
\begin{align}
W^\Gamma(\mathcal C_1)|_{P_{\mathcal R}^0}
&=d_\Gamma I,\nonumber\\
W^\Gamma(\mathcal C_2)|_{P_{\mathcal R}^0}
&=d_\Gamma I,\nonumber\\
W^\Gamma(\mathcal C)|_{P_{\mathcal R}^0}
&=d_\Gamma I.
\end{align}
Therefore
\begin{align}
W^\Gamma(\mathcal C_1)W^\Gamma(\mathcal C_2)
=
d_\Gamma\,W^\Gamma(\mathcal C)
\end{align}
as an identity on that subspace. Equivalently,
\begin{align}
P_{\mathcal R}^0
W^\Gamma(\mathcal C_1)W^\Gamma(\mathcal C_2)
P_{\mathcal R}^0
=
d_\Gamma\,
P_{\mathcal R}^0
W^\Gamma(\mathcal C)
P_{\mathcal R}^0 .
\label{eq:unnormalized-gluing}
\end{align}
For normalized Wilson loops
\begin{align}
\widetilde W^\Gamma(\mathcal C)
=
\frac{1}{d_\Gamma}W^\Gamma(\mathcal C),
\end{align}
this becomes the simpler gluing relation
\begin{align}
P_{\mathcal R}^0
\widetilde W^\Gamma(\mathcal C_1)
\widetilde W^\Gamma(\mathcal C_2)
P_{\mathcal R}^0
=
P_{\mathcal R}^0
\widetilde W^\Gamma(\mathcal C)
P_{\mathcal R}^0 .
\label{eq:normalized-gluing}
\end{align}
Eqs.~\eqref{eq:unnormalized-gluing} and~\eqref{eq:normalized-gluing} are projected identities: they express the emergent topological deformability of Wilson loops only after electric and magnetic defects in the region $\mathcal R$ have been removed.

\section{Open ribbon operators}\label{A:openribbon}

Magnetic loop operators must commute with every $A(s)$ and $B(p)$ term. A
naive product
\begin{align}
T^g(\widetilde{\mathcal C})
=\prod_{\tilde\ell\in\widetilde{\mathcal C}}
X^g_{O_{\tilde\ell}}(\tilde\ell)
\end{align}
does not have this property for a general non-Abelian group. In the horizontal--vertical
orientation convention, write the plaquette state as
$\ket{\psi_p}=\ket{x_0,y_2,x_1,y_1}$, ordered counterclockwise from the bottom edge. When
the path crosses the opposite edges $y_1$ and $y_2$, the two orderings give
\begin{align}
T^g(\widetilde{\mathcal C})B(p)\ket{\psi_p}
&=\delta_{1,\bar y_1\bar g\bar x_1gy_2x_0}
\ket{x_0,gy_2,x_1,gy_1},\nonumber\\
B(p)T^g(\widetilde{\mathcal C})\ket{\psi_p}
&=\delta_{1,\bar y_1\bar x_1y_2x_0}
\ket{x_0,gy_2,x_1,gy_1}.
\end{align}
Equality for every $x_1\in G$ would require
$\bar g\bar x_1g=\bar x_1$, or $g\in Z(G)$. Since $Z(S_3)=\{1\}$, the
naive operator is then trivial for $G=S_3$. The way around this obstruction is
to transport the group label along the ribbon. If the first crossed edge is
acted on by $h$, the second label $g_2$ must obey
\begin{align}
\bar h\bar x_1g_2=\bar x_1,
\qquad\text{hence}\qquad
g_2=x_1h\bar x_1.
\end{align}
Repeating this argument at the next plaquette yields
\begin{align}
g_3=x_2x_1h\bar x_1\bar x_2
=(x_2x_1)h(x_2x_1)^{-1}.
\end{align}
Thus the label at each crossed edge is obtained by parallel transport along
the direct-lattice part of the ribbon. This construction also commutes with
the interior star operators. For example, $A_f(s)$ sends
$y_2\mapsto fy_2$ and $x_1\mapsto fx_1$, and the transported action then gives
\begin{align}
y_2\mapsto fx_1h\bar x_1\bar f,\quad fy_2
=fx_1h\bar x_1y_2,
\end{align}
the same result as applying the ribbon action before $A_f(s)$. The diagrams
below show the direct and dual parts of the ribbon. The ribbon operator
$F_\rho^{(h,g)}$ combines these transported $X$ operations with a projector
onto a specified direct-lattice holonomy
\begin{align}
     &\scalebox{1.2}{$F_\rho^{(h,g)}$} \adjustbox{valign=c,raise=-0.05em}{\includegraphics[width=0.5\linewidth]{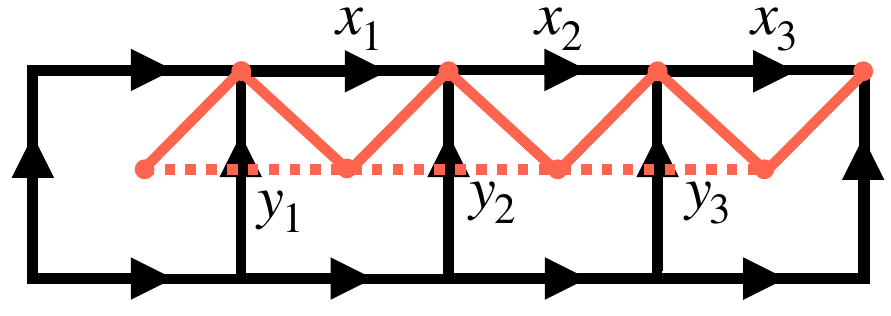}}{} \nonumber \\
     &=  \scalebox{1.3}{$\delta_{\bar{g},\prod\limits^{\leftarrow}_\ell x_\ell}$} \adjustbox{valign=c,raise=-0.05em}{\includegraphics[width=0.6\linewidth]{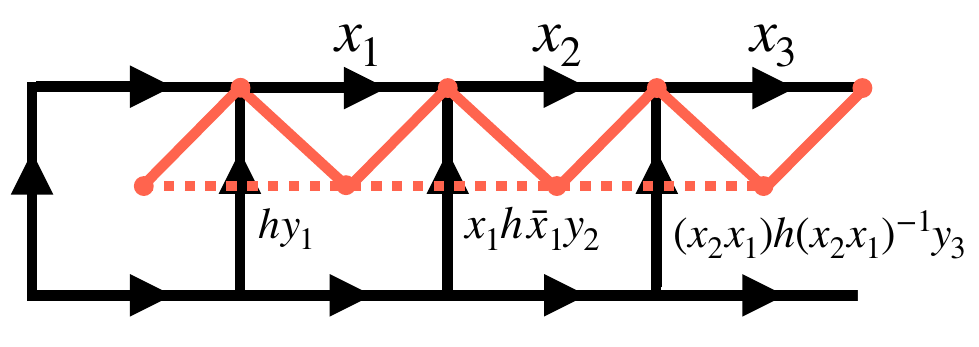}}.
\end{align}
The regular-character identity gives
\begin{align}
\delta_{\bar g,\prod\limits_\ell^{\leftarrow}x_\ell}
=\frac{1}{\abs{G}}\sum_\Gamma d_\Gamma
\chi_\Gamma\!\left(g\prod_\ell^{\leftarrow}x_\ell\right).
\end{align}
For $g=1$ the action becomes
\begin{align}
   &\scalebox{1.3}{$F_\rho^{(1,1)}$} \adjustbox{valign=c,raise=-0.05em}{\includegraphics[width=0.5\linewidth]{Figures/ribbons.pdf}}{} \notag\\
   &= \dfrac{1}{\abs{G}}\sum_\Gamma d_\Gamma W^\Gamma(\rho) \adjustbox{valign=c,raise=-0.05em}{\includegraphics[width=0.5\linewidth]{Figures/ribbons.pdf}}{}.
\end{align}
For a closed ribbon $\sigma$, Wilson loops are the character transforms of
ribbons with $h=1$:
\begin{align}
W^\Gamma(\mathcal C)
&=\sum_g\chi_\Gamma(\bar g)F_\sigma^{(1,g)}, \nonumber\\
F^{(1,g)}_\rho
&=\frac{1}{\abs{G}}\sum_\Gamma d_\Gamma
\tr\!\left[\Gamma(g)\mathcal P\prod_{\ell\in\rho}
\bm Z^\Gamma(\ell)\right].
\end{align}
This inversion follows by inserting the definition of $F$ and using
\begin{align}
    \dfrac{1}{\abs{G}}\sum_g \chi_{\Gamma}(ag)\chi_{\Gamma'}(\bar{g}b)
    = \frac{\delta_{\Gamma,\Gamma'}}{d_\Gamma}\chi_{\Gamma}(ab).
    \label{prod_chars}
\end{align}
An open ribbon $\rho(z,z')$ commutes with all star and plaquette terms except,
possibly, those at its endpoint sites $z$ and $z'$. A closed ribbon has no
endpoints and commutes with every Hamiltonian term.
\section{Translation invariance and closed ribbon operators}\label{A:closedribbons}
To obtain a magnetic 1-form operator from $F_\sigma^{(h,1)}$, its action on a
closed ribbon must be independent of the arbitrary base point. With base point
$x$, the first transported actions are
$y_1\mapsto hy_1$ and $y_2\mapsto x_1h\bar x_1y_2$. Moving the base point to
the next site instead starts with $y_2\mapsto hy_2$. A single fixed label would
therefore require $h=x_1h\bar x_1$ for every $x_1\in G$, i.e., $h\in Z(G)$.
For $S_3$, this leaves only $h=1$. We must instead sum closed ribbons over conjugate labels. Let $\mathcal T$
denote a one-site shift of the base point of a ribbon $\rho(x,x')$, where
$x=(s,p)$ and $x'=(s',p')$. If $h'$ denotes the label before the shift, then
$h'=x_1h\bar x_1$. Diagrammatically,
\begin{align}
&\mathcal{T}F^{(h',g)}_{\rho(x,x')}\mathcal{T}^{-1}\adjustbox{valign=c,raise=-0.05em}{\includegraphics[width=0.5\linewidth]{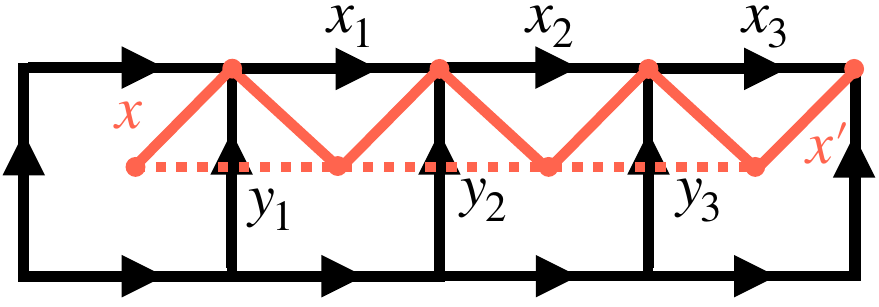}}{} \nonumber\\
&= \delta_{\bar{g},x_3x_2}\adjustbox{valign=c,raise=-0.05em}{\includegraphics[width=0.5\linewidth]{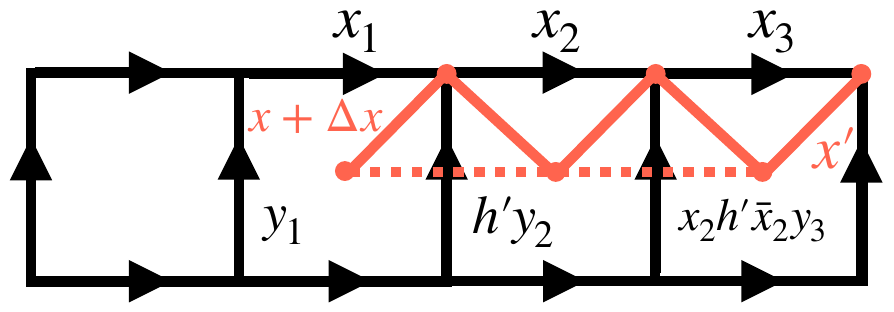}}{}.\nonumber
\end{align}
For the translated operator to represent the same closed ribbon $\sigma$, we
require
\begin{align}
&\mathcal{T}F^{(h',g')}_{\sigma(x)}\mathcal{T}^{-1}\adjustbox{valign=c,raise=-0.05em}{\includegraphics[width=0.6\linewidth]{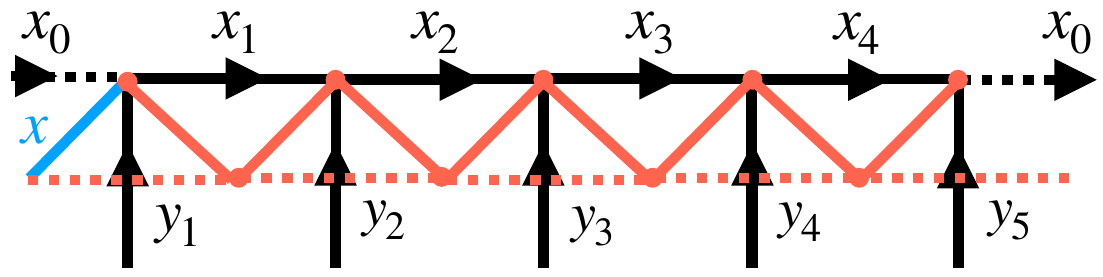}}{} \nonumber \\ &=\delta_{\bar{g}',x_1x_0x_4x_3x_2}\adjustbox{valign=c,raise=-0.05em}{\includegraphics[width=0.65\linewidth]{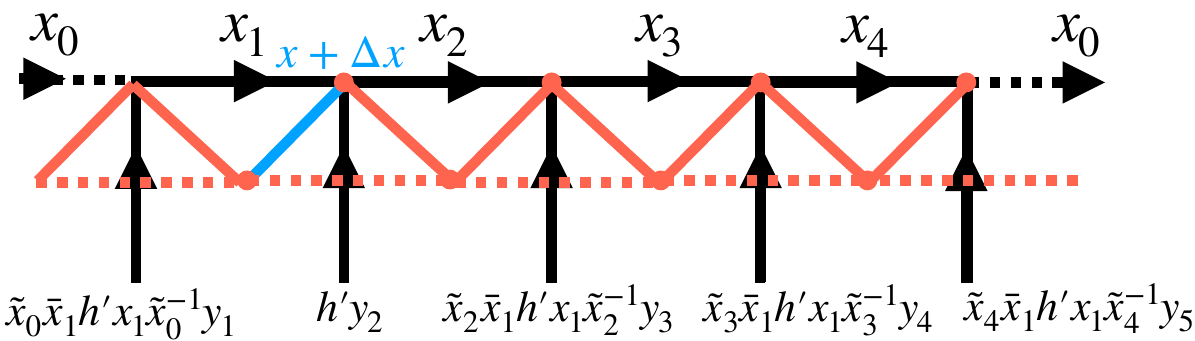}}{} \nonumber
\end{align}
\begin{align}
&=\delta_{\bar{g},x_0x_4x_3x_2x_1}\adjustbox{valign=c,raise=-0.05em}{\includegraphics[width=0.6\linewidth]{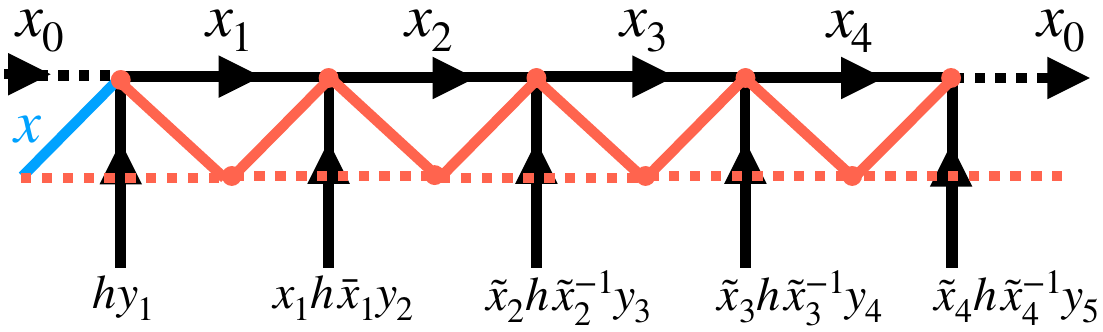}}{} \nonumber \\
&= F^{(h,g)}_{\sigma(x)}\adjustbox{valign=c,raise=-0.05em}{\includegraphics[width=0.6\linewidth]{Figures/ribbons_closed.pdf}}{}.
\end{align}
Define $\tilde x_i=x_ix_{i-1}\cdots x_1$. Matching the action on successive vertical edges gives
\begin{align}
    &h=\bar{x}_1 h'x_1,\nonumber\\
    &x_2h'\bar{x}_2 =\tilde{x}_2 \bar{x}_1 h'x_1 \tilde{x}_2^{-1}= \tilde{x}_2 h \tilde{x}_2^{-1}= x_2x_1 h (x_2x_1)^{-1},\nonumber\\
    &x_3x_2h'(x_3x_2)^{-1} = \tilde{x}_3 \bar{x}_1 h'x_1 \tilde{x}_3^{-1}= \tilde{x}_3 h \tilde{x}_3^{-1},
    \nonumber\\
    &\tilde{x}_4 \bar{x}_1 h'x_1 \tilde{x}_4^{-1}= \tilde{x}_4 h \tilde{x}_4^{-1},\nonumber\\
    & \tilde{x}_0 \bar{x}_1 h'x_1 \tilde{x}_0^{-1}=  h.
\end{align}
Here $\tilde x_0=x_0x_4x_3x_2x_1$ is the total direct-lattice holonomy. These
conditions are compatible when $\tilde x_0\in Z(h)$. Matching the Kronecker
deltas,
\begin{align}
\delta_{\bar g',x_1\tilde x_0\bar x_1}
=\delta_{\bar g,\tilde x_0},
\end{align}
 gives $g=\tilde x_0^{-1}=\bar x_1g'x_1$. In summary,
\begin{align}
    h=\bar{x}_1 h'x_1, \label{eq:rib_per} \nonumber\\
    \tilde{x}_0 h = h\tilde{x}_0, \nonumber\\
    g = \tilde{x}_0^{-1}=\bar{x}_1 g' x_1.
\end{align}
Thus $h$ and $h'$ belong to the same conjugacy class. Let $C=[r_C]$, choose
$k\in Z_C\equiv Z(r_C)$, and write an arbitrary state in the ribbon basis as
$\ket{\Phi}=\sum_{\{x_\ell,y_\ell\}}\phi(\{x_\ell,y_\ell\})
\ket{\{x_\ell,y_\ell\}}$. Then
\begin{align}
    &\mathcal{T}F^{(r_C,k)}_{\sigma(x)} \mathcal{T}^{-1}\ket{\Phi} \nonumber \\
    &=\sum_{\{x_\ell,y_\ell\}}\phi(\{x_\ell,y_\ell\}) \mathcal{T}F^{(r_C,k)}_{\sigma(x)} \mathcal{T}^{-1}\ket{\{x_\ell,y_\ell\}} \nonumber\\
    &=\sum_{\{x_\ell,y_\ell\}}\phi(\{x_\ell,y_\ell\}) F^{(\bar{x}_1r_Cx_1,\bar{x}_1 k x_1)}_{\sigma(x)} \ket{\{x_\ell,y_\ell\}},
\end{align}
because conjugation preserves the commutation relation $kr_C=r_Ck$. Choose representatives $p_i$ of the left cosets $G/Z_C$ such that
\begin{align}
c_i=\bar p_i r_Cp_i,
\qquad i=1,\ldots,\abs{C}.
\end{align}
Every $x\in G$ has a unique decomposition $x=fp_i$ with $f\in Z_C$ and an
appropriate $i$. The orbit--stabilizer theorem therefore gives
\begin{align}
    \abs{G} = \abs{C}\abs{Z_C}.
\end{align}
Let $D$ be a conjugacy class of $Z_C$. The sum
\begin{align}
  F^{CD}_\sigma= \dfrac{1}{\abs{D}}\sum_{i=1}^{\abs{C}} \sum_{k\in D} F^{(\bar{p}_ir_Cp_i,\bar{p}_i k p_i)}_{\sigma}
\end{align}
is base-point independent. To see this, fix $x_1$. For every $i$, write
$p_ix_1=f_{ij}p_j$ with $f_{ij}\in Z_C$. Conjugation by $x_1$ sends the
summand labeled by $(i,k)$ to the summand labeled by
$(j,f_{ij}^{-1}kf_{ij})$. The map $i\mapsto j$ is a permutation and
$f_{ij}^{-1}Df_{ij}=D$, so the double sum is unchanged. 

The centralizer of $c\in G$ is
$Z_c=\{g\in G:cg=gc\}$. Centralizers of elements in the same conjugacy class
are conjugate and hence isomorphic. An alternative basis is obtained by considering the irreps of $Z_C$, which we denote by $R$ of dimension $d_R$ in that case, we define the most general closed ribbon operator is
\begin{align}
K^{CR}_\sigma
&=\sum_{i=1}^{\abs{C}}\sum_{k\in Z_C}\chi_R(k)^*
F^{(\bar p_ir_Cp_i,\bar p_ikp_i)}_\sigma \nonumber\\
&=\sum_{i=1}^{\abs{C}}\sum_{D\in\operatorname{Class}(Z_C)}
\chi_{\bar R}(D)\sum_{k\in D}
F^{(\bar p_ir_Cp_i,\bar p_ikp_i)}_\sigma.
\end{align}
\section{Fusion of 't Hooft loops and magnetic 1-form properties } \label{A:magentic1form}
We begin with two properties of the modified $X$ operators:
\begin{align}
\tilde{X}_+^h(i)\tilde{X}_+^g(j)
&=\tilde{X}_+^g(j)\tilde{X}_+^h(i), \quad i\neq j,\nonumber\\
\tilde{X}_+^h(i)\tilde{X}_+^g(i)
&=\tilde{X}_+^{hg}(i).
\end{align}
To derive the pure magnetic ('t Hooft) operator, let
\begin{align}
U_\sigma=\prod_{\ell\in\sigma}^{\leftarrow}x_\ell,
\quad
c_i=\bar p_ir_Cp_i,
\quad
T_\sigma^{c_i}=\prod_j^{\leftarrow}\tilde X_+^{c_i}(j).
\end{align}
We now derive the form of the 't Hooft operators of Eq.~\eqref{thooft_loop} first let us analyze the action of the $K_\sigma^{CI}$ operators on a basis state
\begin{align}
&K^{CI}_\sigma \ket{\{x_\ell,y_\ell\}} \nonumber\\
&= \sum_{i, k\in Z_C} F^{(\bar{p}_ir_Cp_i,\bar{p}_i k p_i)}_{\sigma}\adjustbox{valign=c,raise=-0.05em}{\includegraphics[width=0.6\linewidth]{Figures/ribbons_closed.pdf}}{}\nonumber\\
 &= \sum_{i=1}^{\abs{C}} \sum_{k\in Z_C} \delta_{\bar{p}_i \bar{k}p_i,U_\sigma}T_\sigma^{c_i}\ket{\{x_\ell,y_\ell\}}\nonumber\\
&=\sum_{i=1}^{\abs{C}} \sum_{k\in Z_C} \delta_{ \bar{k},p_iU_\sigma\bar{p}_i}T_\sigma^{c_i}\ket{\{x_\ell,y_\ell\}}\nonumber \\
&=\sum_{i=1}^{\abs{C}} \sum_{k\in Z_C} \delta_{ k,p_i\bar{U}_\sigma\bar{p}_i}T_\sigma^{c_i}\ket{\{x_\ell,y_\ell\}} .
\end{align}
The argument $p_iU_\sigma^{-1}\bar p_i$ belongs to $Z_C$ precisely when
$U_\sigma\in Z_{c_i}$. This condition is enforced by
\begin{align}
P^{c_i}_{\mathrm{com}}(\sigma)\ket{\{x_\ell,y_\ell\}}
=\sum_{z\in Z_{c_i}}\delta_{U_\sigma,z}
\ket{\{x_\ell,y_\ell\}}.
\end{align}
On the support of this projector, the regular-character identity for $Z_C$
gives
\begin{align}
\delta_{k,a}
=\frac{1}{\abs{Z_C}}\sum_{R\in\operatorname{Irr}(Z_C)}
d_R\chi_R(\bar ka),
\qquad a\in Z_C.
\end{align}
Using $\chi_I(k)=1$ and the character identity in
Eq.~\eqref{prod_chars}, now applied to $Z_C$, yields
\begin{align}
&K^{CI}_\sigma \ket{\{x_\ell,y_\ell\}} \nonumber \\
&=\sum_{i=1}^{\abs{C}}\sum_{R\in\operatorname{Rep}(Z_C)}
\sum_{k\in Z_C}P^{c_i}_{\mathrm{com}}(\sigma)
\frac{d_R}{\abs{Z_C}}
\chi_R\!\left(\bar kp_iU_\sigma^{-1}\bar p_i\right)\nonumber\\
& \qquad \qquad \qquad \qquad \qquad \qquad \qquad \qquad \times T_\sigma^{c_i}\ket{\{x_\ell,y_\ell\}}\nonumber\\
&=\sum_{i=1}^{\abs{C}}P^{c_i}_{\mathrm{com}}(\sigma)
T_\sigma^{c_i}\ket{\{x_\ell,y_\ell\}}\nonumber\\
&=\sum_{g\in C}P^g_{\mathrm{com}}(\sigma)
\prod_j^{\leftarrow}\tilde X_+^g(j)
\ket{\{x_\ell,y_\ell\}}.
\label{eq:thooft-loop-appendix}
\end{align}
To obtain the fusion algebra, recall that the group algebra $\mathbb C[G]$
consists of formal sums $\sum_{g\in G}c_gg$. Its center, also called the class
algebra, has the conjugacy-class sums
\begin{align}
e_C=\sum_{g\in C}g
\end{align}
as a basis. They obey $he_C=e_Ch$ for every $h\in G$ and
\begin{align}
e_{C_1}e_{C_2}=\sum_C N^C_{C_1C_2}e_C,
\qquad N^C_{C_1C_2}\in\mathbb Z_{\geq0}.
\end{align}
Under the standard involution $g^*=g^{-1}$,
$e_C^*=e_{C^{-1}}$; the stronger statement $e_C^*=e_C$ holds only when
$C=C^{-1}$ (as it does for every conjugacy class of $S_3$).
These facts are standard properties of the class algebra
\cite{bombin_family_2008}. For a fixed closed ribbon $\sigma$, the map
\begin{align}
g\longmapsto \mathcal P\prod_{\tilde\ell\in\sigma}
\tilde X^g_{O_{\tilde\ell}}(\tilde\ell)
\end{align}
represents the group-algebra multiplication. Consequently, the class-summed
operators $T^C(\sigma)$ realize the same commutative fusion algebra when applied to the flux-free ground state. We note that other states with $U_\sigma\neq 1$ may project to zero when the operator $T^C_\sigma$ is applied. Ther torus ground state for example, already have magnetic loops around the handles and therefore may change the fusion of two  $T^C_\sigma$ operators.

\paragraph{Overlapping magnetic 1-form commutator.} We calculate the commutator of two general overlapping magnetic ribbons. Let
\(\sigma\) and \(\tau\) be two distinct closed ribbons which intersect at two
points. We denote the fixed-label component of the magnetic class operator by
\begin{equation}
    \mathcal T^h(\tau)
    =
    P^h_{\rm com}(\tau)T_h(\tau),
    \qquad
    T^C(\tau)=\sum_{h\in C}\mathcal T^h(\tau),
\end{equation}
and analogously for \(\sigma\). The projector \(P^h_{\rm com}(\tau)\) imposes
that the direct-lattice holonomy \(U_\tau\) along the ribbon commutes with the
magnetic label \(h\),
\begin{equation}
    P^h_{\rm com}(\tau)
    =
    \delta_{U_\tau h,hU_\tau}.
\end{equation}
Choose a base point on \(\tau\). If \(x_j\) are the direct-lattice variables
entering the parallel transport along \(\tau\), define
\begin{equation}
    u_0=1,
    \qquad
    u_j=x_jx_{j-1}\cdots x_1 .
\end{equation}
The fixed-label ribbon \(T^h(\tau)\) acts on the dual edge \(y_{j+1}\) with the
transported label
\begin{equation}
    L_j(h)=u_jhu_j^{-1},
    \qquad
    y_{j+1}\mapsto L_j(h)y_{j+1}.
\end{equation}
Thus the \(x_j\)'s enter only as parallel transports on the two sides of \(h\). Let \(T^g(\sigma)\) cross \(\tau\) at two positions \(p<q\). Let \(z_m\) be the
direct-lattice variables along \(\sigma\), with partial transports
\begin{equation}
    v_0=1,
    \qquad
    v_m=z_mz_{m-1}\cdots z_1 .
\end{equation}
If the two intersections occur after the partial transports \(v_{m_p}\) and
\(v_{m_q}\) along \(\sigma\), we define the corresponding transported
\(g\)-insertions by
\begin{equation}
    a_p=v_{m_p}g^{\epsilon_p}v_{m_p}^{-1},
    \qquad
    a_q=v_{m_q}g^{\epsilon_q}v_{m_q}^{-1},
\end{equation}
where \(\epsilon_p,\epsilon_q=\pm1\) are fixed by the local orientations of the
two crossings. In the orientation convention used below, the two crossings act
on the \(\tau\)-variables as
\begin{equation}
    x_p\mapsto a_px_p,
    \qquad
    x_q\mapsto x_qa_q .
\end{equation}
Other orientations only change the precise placement of inverses and left/right
multiplications; the structure of the commutator is unchanged. Define
\begin{equation}
    D_p=x_px_{p-1}\cdots x_1,
    \quad
    C_{pq}=x_{q-1}x_{q-2}\cdots x_{p+1}.
\end{equation}
Then the transformed partial transports along \(\tau\) can be written as
\begin{equation}
    u_j'=u_j\delta_j^{[g]},
\end{equation}
with
\begin{equation}
    \delta_j^{[g]}=
    \begin{cases}
        1, & j<p,\\[1mm]
        D_p^{-1}a_pD_p, & p\leq j<q,\\[1mm]
        D_p^{-1}C_{pq}^{-1}a_qC_{pq}a_pD_p, & j\geq q .
    \end{cases}
\end{equation}
Consequently, on the forward-transformed input configuration, the transported
\(h\)-label becomes
\begin{equation}
    L_j(h)=u_jhu_j^{-1}
  \; \longmapsto \;
    L'_j(h)
    =
    u_j\delta_j^{[g]}h
    \left(\delta_j^{[g]}\right)^{-1}u_j^{-1}.
\end{equation}
The corresponding forward ribbon action is therefore
\begin{equation}
    y_{j+1}\mapsto
    u_j\delta_j^{[g]}h
    \left(\delta_j^{[g]}\right)^{-1}u_j^{-1}y_{j+1}.
\end{equation}
The same crossing changes the holonomy entering the projector
\(P^h_{\rm com}(\tau)\). Writing
\begin{equation}
    U_\tau
    =
    B_q x_q C_{pq}x_pD_{p-1},
    \quad
    D_{p-1}=x_{p-1}\cdots x_1,
\end{equation}
where \(B_q\) denotes the remaining ordered product after \(x_q\), the action of
\(T^g(\sigma)\) gives
\begin{equation}
    U_\tau
    \mapsto
    U'_\tau
    =
    B_q x_q a_q C_{pq}a_px_pD_{p-1}.
\end{equation}
Equivalently,
\begin{equation}
    U'_\tau
    =
    U_\tau\Delta_\tau^{[g]},
\end{equation}
with
\begin{equation}
    \Delta_\tau^{[g]}
    =
    D_p^{-1}C_{pq}^{-1}a_qC_{pq}a_pD_p .
\end{equation}
In general,
\begin{equation}
    U'_\tau\neq U_\tau ,
\end{equation}
because the two insertions occur at different positions in the ordered product
and carry different \(\sigma\)-ribbon parallel transports. Care is required with the direction of this transformation: the primed
expressions above describe the active, forward change of a basis
configuration. The operator that appears when commuting $T^g(\sigma)$ to the
right is instead the conjugated (pullback) operator. Define it, and the
corresponding conjugated ribbon, by
\begin{align}
    P_{\rm com}^{h,[g]}(\tau)
    \equiv T^g(\sigma)P_{\rm com}^{h}(\tau)T^g(\sigma)^{-1},
    \nonumber \\
    T^{h,{[g]}}(\tau)
    \equiv T^g(\sigma)T^h(\tau)T^g(\sigma)^{-1}.
\end{align}
Equivalently, on an input configuration $\ket{\mathbf \{x_\ell,y_\ell\}}$,
\begin{align}
P_{\rm com}^{h,[g]}(\tau)T^g(\sigma)\ket{\mathbf \{x_\ell,y_\ell\}}
=\delta_{U_\tau h,hU_\tau}
T^g(\sigma)\ket{\mathbf \{x_\ell,y_\ell\}}.
\end{align}
These definitions give the exact operator identity
\begin{equation}
    T^g(\sigma)P_{\rm com}^{h}(\tau)
    =
    P_{\rm com}^{h,[g]}(\tau)T^g(\sigma).
\end{equation}
The local crossing relation for the fixed-label projected ribbon is then
\begin{equation}
    T^g(\sigma)\mathcal T^h(\tau)
    =
    P_{\rm com}^{h,[g]}(\tau)
    T^{h,{[g]}}(\tau)T^g(\sigma).
\end{equation}
Hence
\begin{equation}
\begin{split}
    [T^g(\sigma),\mathcal T^h(\tau)]
    &=
    \left[
        P_{\rm com}^{h,[g]}(\tau)T^{h,{[g]}}(\tau)
        -
        P_{\rm com}^{h}(\tau)T^h(\tau)
    \right]  \\
    &\hspace{1.0cm}\times T^g(\sigma).
\end{split}
\end{equation}
Including also the projector on the \(\sigma\)-ribbon,
\(\mathcal T^g(\sigma)=P^g_{\rm com}(\sigma)T^g(\sigma)\), gives the symmetric
fixed-label expression
\begin{equation}
\begin{split}
    [\mathcal T^g(\sigma),\mathcal T^h(\tau)]
    &=
    P^g_{\rm com}(\sigma)
    P_{\rm com}^{h,[g]}(\tau)
    T^{h,{[g]}}(\tau)T^g(\sigma)
    \\
    &
    -
    P^h_{\rm com}(\tau)
    P_{\rm com}^{g,[h]}(\sigma)
    T^{g,{[h]}}(\sigma)T^h(\tau).
\end{split}
\label{eq:fixed_label_overlap_commutator}
\end{equation}
Here \(T_g^{[h]}(\sigma)\) and \(P_{\rm com}^{g,[h]}(\sigma)\) are defined in
the same way, with the roles of \(\sigma\) and \(\tau\) exchanged. For the class-summed magnetic operators, the microscopic commutator is therefore
\begin{align}
    &[T^{C_1}(\sigma),T^{C_2}(\tau)]\nonumber\\
    &=\sum_{g\in C_1}
    \sum_{h\in C_2}
    \Big[
    P^g_{\rm com}(\sigma)
    P_{\rm com}^{h,[g]}(\tau)
    T_h^{[g]}(\tau)T_g(\sigma)
   \nonumber \\
    &\qquad -P^h_{\rm com}(\tau)
    P_{\rm com}^{g,[h]}(\sigma)
    T_g^{[h]}(\sigma)T_h(\tau)
    \Big]. \label{eq:overlap_commutator}
\end{align}

This formula makes explicit that two fixed microscopic representatives of
overlapping magnetic ribbons need not commute on the full Hilbert space. The
commutator contains two effects: the segment-dependent twist of the transported
magnetic labels and the corresponding change of the holonomy projectors.

For contractible ribbons, the topological commutativity is recovered after
projection to a region with no electric defects. Let \(R\) be a region through
which one ribbon can be deformed away from the other, and define
\begin{equation}
    \Pi_R=\prod_{s\subset R}A(s).
\end{equation}
In the defect-free subspace, ribbon deformations through \(R\) are exact:
\begin{align}
    \Pi_R T^{h,{[g]}}(\tau)\Pi_R
    =
    \Pi_R T^h(\tau)\Pi_R,
    \\
    \Pi_R P_{\rm com}^{h,[g]}(\tau)\Pi_R
    =
    \Pi_R P_{\rm com}^{h}(\tau)\Pi_R .
\end{align}
Consequently,
\begin{equation}
    \Pi_R
    [T^{C_1}(\sigma),T^{C_2}(\tau)]
    \Pi_R
    =
    0 .
\end{equation}
Thus overlapping contractible magnetic loops commute as topological operators
in the no-electric-defect sector, although their fixed microscopic
representatives can have a nonzero commutator on the full Hilbert space.

\paragraph{Projected gluing of magnetic loop operators.} We now prove the
projected gluing relation for magnetic loop operators. The non-Abelian case
differs from the Abelian one: the actions on the common edge do not cancel on
the full Hilbert space. Their residual action is a local gauge transformation
inside the region swept out by the deformation and disappears only after
projection to the defect-free subspace.

Let $\sigma_1$ and $\sigma_2$ be two adjacent elementary dual-lattice loops sharing one direct edge with opposite induced orientations. Let
\begin{align}
\sigma=\sigma_1\oplus\sigma_2
\end{align}
denote the joined loop obtained by removing the common segment. Let $\mathcal R$
contain the two elementary loops and their shared edge, and define
\begin{align}
P_{\mathcal R}^0
=
\prod_{s\subset\mathcal R}A(s)
\prod_{p\subset\mathcal R}B(p).
\end{align}
This projector imposes both gauge invariance and flatness in $\mathcal R$.

For fixed labels $g,h\in G$, the two elementary magnetic loops act on the
shared edge as
\begin{align}
X_+^g X_-^h |a\rangle
=
|g a h^{-1}\rangle ,
\label{eq:common-edge-action}
\end{align}
where $a\in G$ is the group element on the common edge. Thus the common-edge action does not cancel even when $g=h$; in that case it conjugates the internal edge,
\begin{align}
|a\rangle\mapsto |gag^{-1}\rangle .
\end{align}
This is the obstruction to an exact gluing identity on the full Hilbert space.

However, inside the image of $P_{\mathcal R}^0$, such an internal conjugation is a gauge redundancy. The reason is that the star projectors $A(s)$ average over local gauge transformations at the vertices in $\mathcal R$. Hence two configurations that differ only by a gauge transformation supported in the interior of $\mathcal R$ have the same projection under $P_{\mathcal R}^0$. The plaquette projectors $B(p)$ additionally impose local flatness, so the parallel transport relating the two base points of the adjacent loops is path-independent in $\mathcal R$.

Consequently, after projection, the product depends only on the conjugacy
classes of the labels. If $h$ is conjugate to $g$, the internal action in
Eq.~\eqref{eq:common-edge-action} can be absorbed by a gauge transformation
supported in $\mathcal R$. A formula for a single fixed representative is not
canonical, because the gauge transformation also changes that representative.
The invariant statement is therefore the conjugacy-class-summed relation
below. Labels in different conjugacy classes leave an interior mismatch and
are removed by the defect-free projection.

We now pass to the physical magnetic symmetry operators. For a conjugacy class $C\subset G$, define
\begin{align}
T^C(\sigma)
=
\sum_{g\in C}
P^g_{\mathrm{com}}(\sigma)T^g(\sigma),
\end{align}
where $P^g_{\mathrm{com}}(\sigma)$ imposes the base-point consistency condition for a closed non-Abelian ribbon. In the locally flat region $\mathcal R$, these consistency projectors are transported consistently across the elementary gluing move, so they do not obstruct the argument above.

Writing $\mathcal T_g(\sigma)=P^g_{\mathrm{com}}(\sigma)T^g(\sigma)$, the
class-summed gluing relation is
\begin{align}
&P_{\mathcal R}^0\,
T^C(\sigma_1)T^C(\sigma_2)\,
P_{\mathcal R}^0
\nonumber\\
&=
\sum_{g,h\in C}
P_{\mathcal R}^0\,
\mathcal T_g(\sigma_1)\mathcal T_h(\sigma_2)\,
P_{\mathcal R}^0
\nonumber\\
&=
|C|\sum_{g\in C}
P_{\mathcal R}^0\,
\mathcal T_g(\sigma_1\oplus\sigma_2)\,
P_{\mathcal R}^0
\nonumber\\
&=
|C|\,
P_{\mathcal R}^0\,
T^C(\sigma_1\oplus\sigma_2)\,
P_{\mathcal R}^0 .
\label{eq:unnormalized-magnetic-gluing}
\end{align}
The factor $|C|$ appears because, for each representative on the joined loop,
the sum over the second elementary loop contains $|C|$ conjugate labels that
are gauge-equivalent after projection. Equivalently, the normalized operators
$\widetilde T^C=T^C/|C|$ obey
\begin{align}
P_{\mathcal R}^0\widetilde T^C(\sigma_1)
\widetilde T^C(\sigma_2)P_{\mathcal R}^0
=P_{\mathcal R}^0\widetilde T^C(\sigma_1\oplus\sigma_2)
P_{\mathcal R}^0.
\end{align}
Thus magnetic loops can be glued and deformed only projectively inside the defect-free subspace. The obstruction on the full Hilbert space is the residual common-edge action
\begin{align}
X_+^gX_-^h|a\rangle=|gah^{-1}\rangle ,
\end{align}
which is not the identity for non-Abelian $G$. The defect-free projector turns
this residual internal action into a gauge redundancy, while the
conjugacy-class sum removes the choice of representative.

\section{Details on the generalized  mixed 't Hooft anomaly and ground state construction}\label{A:anomaly}
We compute the product of $W_x^\Gamma$ and $T_y^C$ away from any defects ($A(s),B(p)= 1$).  Define first the product of group elements along the direct lattice loop
\begin{align}
    g_x=\mathcal{P}\prod_{\ell\in \mathcal{C}_x}g_\ell.
\end{align}
 Since for the fixed-point model all excitations can be built on top of the ground-state subspace, we first consider the action of the operators on the flux-free ground state
\begin{align}
    \ket{\Psi_{\mathcal{D}(G)}}=P_\text{GS}\ket{\mathbf{1}}
    =
    \prod_s A(s)\ket{\mathbf{1}} .
\end{align}
For this state the non-contractible holonomy along the \(x\)-cycle is trivial, up to conjugation, so that
\begin{align}
    g_x=1 .
\end{align}
Therefore the Wilson loop acts as
\begin{align}
    W_x^\Gamma \ket{\Psi_{\mathcal{D}(G)}}
    =
    \tr \Gamma(1)\ket{\Psi_{\mathcal{D}(G)}}
    =
    d_\Gamma \ket{\Psi_{\mathcal{D}(G)}},
\end{align}
where
\begin{align}
    d_\Gamma=\dim \Gamma=\chi_\Gamma(1).
\end{align}
Now consider the action of a magnetic class operator \(T_y^C\), with
\begin{align}
    T_y^C
    =
    \sum_{g\in C}
    P^{g}_{\mathrm{com}}(\tilde{\mathcal{C}}_y)\,T_y^g .
\end{align}
Here $\chi_\Gamma(C)$ denotes the common value of $\chi_\Gamma(g)$ for
$g\in C$. Acting first with $T_y^C$ inserts a representative $g\in C$ through
the $x$-cycle. Hence
\begin{align}
    &W_x^\Gamma T_y^C \ket{\Psi_{\mathcal{D}(G)}}
    =
    \sum_{g\in C}
    P^{g}_{\mathrm{com}}(\tilde{\mathcal{C}}_y)
    W_x^\Gamma T_y^g
    \ket{\Psi_{\mathcal{D}(G)}} \nonumber\\
    &=
    \sum_{g\in C}
    P^{g}_{\mathrm{com}}(\tilde{\mathcal{C}}_y)
    \tr\!\left[\Gamma(g)\Gamma(g_x)\right]
    T_y^g
    \ket{\Psi_{\mathcal{D}(G)}} \nonumber
\end{align}
\begin{align}
    &=
    \sum_{g\in C}
    P^{g}_{\mathrm{com}}(\tilde{\mathcal{C}}_y)
    \tr\Gamma(g)
    T_y^g
    \ket{\Psi_{\mathcal{D}(G)}} \nonumber\\
    &=
    \chi_\Gamma(C)
    \sum_{g\in C}
    P^{g}_{\mathrm{com}}(\tilde{\mathcal{C}}_y)
    T_y^g
    \ket{\Psi_{\mathcal{D}(G)}} \nonumber\\
    &=
    \chi_\Gamma(C)
    T_y^C
    \ket{\Psi_{\mathcal{D}(G)}} .
\end{align}
On the other hand,
\begin{align}
    T_y^C W_x^\Gamma \ket{\Psi_{\mathcal{D}(G)}}
    &=
    d_\Gamma
    T_y^C
    \ket{\Psi_{\mathcal{D}(G)}} .
\end{align}
Combining both equations gives
\begin{align}
    W_x^\Gamma T_y^C \ket{\Psi_{\mathcal{D}(G)}}
    =
    \frac{\chi_\Gamma(C)}{d_\Gamma}
    T_y^C W_x^\Gamma
    \ket{\Psi_{\mathcal{D}(G)}} .
\end{align}
Thus, the commutation eigenvalue is not the bare character, but the normalized central character
\begin{align}
    \omega_\Gamma(C)
    =
    \frac{\chi_\Gamma(C)}{d_\Gamma}.
\end{align}
In particular, for the identity conjugacy class \(C=[1]\),
\begin{align}
    \omega_\Gamma([1])
    =
    \frac{\chi_\Gamma(e)}{d_\Gamma}
    =
    1,
\end{align}
so that \(T_y^{[1]}\) commutes trivially with \(W_x^\Gamma\), as it should. For a ground state carrying magnetic flux $C'$ along the $y$-cycle, the fusion rule gives
\begin{align}
&W_x^\Gamma T_y^C\ket{[1],C'}\nonumber\\
&=W_x^\Gamma T_y^CT_y^{C'}\ket{\Psi_{\mathcal D(G)}}\nonumber\\
&=W_x^\Gamma\sum_{\widetilde C}N^{\widetilde C}_{CC'}
T_y^{\widetilde C}\ket{\Psi_{\mathcal D(G)}}\nonumber\\
&=\sum_{\widetilde C}N^{\widetilde C}_{CC'}
\frac{\chi_\Gamma(\widetilde C)}{d_\Gamma}
T_y^{\widetilde C}W_x^\Gamma\ket{\Psi_{\mathcal D(G)}}.
\end{align} 
This gives the same type of mixed 't Hooft anomaly as before except that it applies to each fusion channel. We next express the dyonic ribbon as a mixed Wilson--'t Hooft operator. Let
$U_\sigma=\mathcal P\prod_{\ell\in\sigma}x_\ell$. Then
\begin{align}
&K^{CR}_\sigma \ket{ \{ x_\ell,g_\ell \} } \nonumber \\
&=\sum_{i=1}^{\abs{C}}\sum_{k\in Z_C}\chi_R(\bar k)
\delta_{\bar k,p_iU_\sigma\bar p_i}
T_\sigma^{c_i}\ket{\{x_\ell,g_\ell\}}\nonumber\\
&=\sum_{i=1}^{\abs{C}}\sum_{k\in Z_C}\chi_R(\bar k)
\frac{1}{\abs{G}}\sum_\Gamma d_\Gamma
\chi_\Gamma(\bar p_ikp_iU_\sigma)
T_\sigma^{c_i}\ket{\{x_\ell,g_\ell\}}\nonumber\\
&=\sum_{i=1}^{\abs{C}}\sum_{k\in Z_C}\chi_R(\bar k)
\frac{1}{\abs{G}}\sum_\Gamma d_\Gamma
\tr\!\left[\Gamma(\bar p_ikp_i)
\mathcal P\prod_{\ell\in\sigma}\bm Z^\Gamma(\ell)\right]\nonumber\\
& \hspace{5cm} \times T_\sigma^{c_i}\ket{\{x_\ell,g_\ell\}}.
\end{align}
Define
\begin{align}
\bm W^\Gamma(\sigma)
&=\mathcal P\prod_{\ell\in\sigma}\bm Z^\Gamma(\ell),\\
\xi^{CR}_{\Gamma\alpha\beta,i}
&=\frac{d_\Gamma}{\abs{G}}\sum_{k\in Z_C}
\chi_R(\bar k)\Gamma(\bar p_ikp_i)_{\beta\alpha}.
\end{align}
Since
$\tr[\Gamma(\bar p_ikp_i)\bm W^\Gamma]
=\sum_{\alpha,\beta}\Gamma(\bar p_ikp_i)_{\beta\alpha}
W^\Gamma_{\alpha\beta}$, we obtain
\begin{align}
K^{CR}_\sigma
=\sum_{i=1}^{\abs{C}}\sum_{\Gamma,\alpha,\beta}
\xi^{CR}_{\Gamma\alpha\beta,i}
W^\Gamma_{\alpha\beta}(\sigma)T^{c_i}_\sigma.
\label{eq:dyon-mixWT-appendix}
\end{align}

We now construct the torus ground-state subspace for a general $\mathcal D(G)$
quantum double, starting from the flux-free state
\begin{align}
\ket{\Psi_{\mathcal D(G)}}=\prod_s A(s)\ket{\mathbf 1},
\end{align}
where $\ket{\mathbf 1}$ is the product state with every edge in the identity
element. Acting with closed ribbons along the two non-contractible cycles gives
the candidate state
\begin{align}
F_y^{\widetilde C\widetilde D}
F_x^{CD}
\ket{\Psi_{\mathcal D(G)}} .
\label{eq:general-torus-ribbon-state}
\end{align}
Because closed ribbon operators commute with the Hamiltonian, any nonzero
state of the form \eqref{eq:general-torus-ribbon-state} lies in the ground-state
subspace. Not every choice of $C,D,\widetilde C,\widetilde D$ gives a nonzero
state, however. First, the action of a single $x$-cycle ribbon gives
\begin{align}
F_x^{CD}\ket{\Psi_{\mathcal D(G)}}
&=
F_x^{CD}\prod_s A(s)\ket{\mathbf 1}
\nonumber\\
&=
\prod_s A(s)F_x^{CD}\ket{\mathbf 1}
\nonumber\\
&=
\delta_{D,[1]}\prod_s A(s)T_x^C\ket{\mathbf 1}
\nonumber\\
&=
\delta_{D,[1]}T_x^C\ket{\Psi_{\mathcal D(G)}} .
\label{eq:first-ribbon-action}
\end{align}
Thus the first ribbon is nonzero only when $D=[1]$, the identity class in $Z_C$. It therefore creates a pure magnetic flux sector labeled by the conjugacy class $C$. We next apply a second ribbon along the $y$-direction:
\begin{align}
F_y^{\widetilde C\widetilde D}T_x^C
\ket{\Psi_{\mathcal D(G)}}
=
\prod_s A(s)
F_y^{\widetilde C\widetilde D}T_x^C
\ket{\mathbf 1}.
\label{eq:second-ribbon-start}
\end{align}
Let $r_{\widetilde C}$ be a representative of $\widetilde C$, and choose elements $p_i\in G$ such that
\begin{align}
h_i=\bar p_i r_{\widetilde C}p_i,
\qquad
i=1,\ldots,|\widetilde C|,
\end{align}
run over the conjugacy class $\widetilde C$. Using the definition of the magnetic ribbon operator $T_y^{h_i}$, we obtain
\begin{align}
&F_y^{\widetilde C\widetilde D}T_x^C\ket{\mathbf 1}
\nonumber\\
&=
\frac{1}{|\widetilde D|}
\sum_{g\in C}
\sum_{i=1}^{|\widetilde C|}
\sum_{k\in\widetilde D}
\delta_{\bar p_i\bar k p_i,\bar g}\,
T_y^{h_i}T_x^g
\ket{\mathbf 1}.
\label{eq:second-ribbon-action-compact}
\end{align}
Here the product of direct-lattice links along the $y$-cycle appears with inverses because of the orientation convention. The Kronecker delta imposes
\begin{align}
g=\bar p_i k p_i .
\label{eq:g-k-condition}
\end{align}
Let $Z_{\widetilde C}=Z(r_{\widetilde C})$. Since
$k\in\widetilde D\subset Z_{\widetilde C}$ commutes with
$r_{\widetilde C}$, Eq.~\eqref{eq:g-k-condition} implies
\begin{align}
g h_i = h_i g .
\label{eq:commuting-condition}
\end{align}
Therefore, the second ribbon gives a nonzero flat state only when the two
threaded fluxes commute. More precisely, for the chosen representatives the
allowed centralizer class satisfies
\begin{align}
\bar p_i\widetilde Dp_i\subseteq C\cap Z_{h_i},
\label{eq:D-condition}
\end{align}
for an admissible (equivalently, any consistently relabeled) index $i$. Conversely, imposing the commutation condition reconstructs the centralizer sum. For any function $f(g,h_i)$, one has
\begin{align}
&\sum_{g,i}
\sum_{k\in\widetilde D}
\delta_{k,p_i g\bar p_i}
f(g,h_i)
\nonumber\\
&\hspace{0.5cm}
=
\sum_i
\sum_{g\in \bar p_i\widetilde D p_i}
\delta_{g h_i,h_i g}
f(g,h_i).
\label{eq:centralizer-sum-compact}
\end{align}
Thus the ribbon operator selects precisely the part of the first flux class that commutes with the second flux. We can now label the nonzero states by a commuting pair $(g,h)$. Taking
$\widetilde C=[h]$ and $C=[g]$, with $h$ chosen as the representative used to
define $Z_h$, the nonzero ribbon states obey
\begin{align}
F_y^{[h]D}T_x^{[g]}
\ket{\Psi_{\mathcal D(G)}}
=
\prod_s A(s)
F_y^{[h]D}T_x^{[g]}
\ket{\mathbf 1},
\label{eq:nonzero-ribbon-state}
\end{align}
with
\begin{align}
D\subseteq [g]\cap Z_h .
\end{align}
The resulting states are naturally written as simultaneous-conjugacy orbits:
\begin{align}
\ket{[(g,h)]}
&=
F_y^{[h]D}T_x^{[g]}
\ket{\Psi_{\mathcal D(G)}}
\nonumber\\
&=
\prod_s A(s)
\left(
\sum_{(g_x,g_y)\in[(g,h)]}
T_y^{g_y}T_x^{g_x}
\right)
\ket{\mathbf 1}.
\label{eq:commuting-pair-state}
\end{align}
Here
\begin{align}
[(g,h)]
=
\{(xg\bar x,xh\bar x):x\in G\},
\qquad
gh=hg .
\end{align}
Equivalently, the torus sectors are labeled by
\begin{align}
C_{\mathrm{pairs}}
=
\frac{
\{(g,h)\in G\times G:\ gh=hg\}
}{
G\text{ conjugation}
}.
\label{eq:Cpairs}
\end{align}
Thus the closed-ribbon construction gives one independent ground state for each commuting pair modulo simultaneous conjugation:
\begin{align}
\mathrm{GSD}_{T^2}\bigl(\mathcal D(G)\bigr)
=
|C_{\mathrm{pairs}}|.
\end{align}
\paragraph{Torus ground states for $S_3$.} We can further apply the 't Hooft operators $T_x^C$ or $T_y^C$, with
$C=\{[1],[s],[r]\}$. These operators generate five distinct ground states,
which we denote by
\begin{align}
\ket{[1],[1]}&=\ket{\Psi_{\mathcal D(G)}}, \nonumber \\
\ket{[s],[1]}&=T_x^{[s]}\ket{\Psi_{\mathcal D(G)}},\nonumber  \\
\ket{[r],[1]}&=T_x^{[r]}\ket{\Psi_{\mathcal D(G)}}, \nonumber \\
\ket{[1],[s]}&=T_y^{[s]}\ket{\Psi_{\mathcal D(G)}}, \nonumber \\
\ket{[1],[r]}&=T_y^{[r]}\ket{\Psi_{\mathcal D(G)}} .
\end{align}
To continue building states on top of these, we need the $K^{CR}$ operators.
On the state $\ket{\mathbf 1}$, we always have
\begin{align}
K_\sigma^{CI}\ket{\mathbf 1}
=
T_\sigma^C\ket{\mathbf 1}.
\end{align}
Let us compute the action of $K_\sigma^{CR}$ on
$\ket{[1],[1]}=\ket{\Psi_{\mathcal D(G)}}$, where the ribbon $\sigma$ is
assumed to be non-contractible along either the $x$ or $y$ direction. Since
\begin{align}
K_\sigma^{CR}\ket{\Psi_{\mathcal D(G)}}
\nonumber=
\sum_{i=1}^{|C|}
\sum_{k\in Z_C}
\chi_R(\bar k)\,
\delta_{\bar k,\,
p_i\left(\prod_\ell x_\ell\right)\bar p_i}\nonumber\\
\left[
\prod_j^{\leftarrow}
\widetilde X_+^{\bar p_i r_C p_i}(j)
\right]
\prod_s A(s)\ket{\mathbf 1},
\nonumber
\end{align}
we obtain then
\begin{align}
&K_\sigma^{CR}\ket{\Psi_{\mathcal D(G)}}\nonumber\\
&=
\prod_s A(s)
\sum_{i=1}^{|C|}
\sum_{k\in Z_C}
\chi_R(\bar k)\,
\delta_{\bar k,1}
\left[
\prod_j^{\leftarrow}
\widetilde X_+^{\bar p_i r_C p_i}(j)
\right]
\ket{\mathbf 1}
\nonumber\\
&=
d_R\prod_s A(s)
\sum_{i=1}^{|C|}
\left[
\prod_j^{\leftarrow}
\widetilde X_+^{\bar p_i r_C p_i}(j)
\right]
\ket{\mathbf 1}
\nonumber\\
&=
d_R\prod_s A(s)T_\sigma^C\ket{\mathbf 1}.
\label{eq:K-on-flux-free}
\end{align}
We used the fact that $1\in Z_C$ for any class $C$, since the identity
commutes with all group elements. We also used that the ribbon operators
commute with all $A(s)$. Thus including the centralizer irrep
$R\in \mathrm{Rep}(Z_C)$ does not generate additional states when the operator
acts once on $\ket{\Psi_{\mathcal D(G)}}$: the action of the $K^{CR}$ operators
is the same as that of the magnetic 1-form operators, up to the factor $d_R$. 

The situation changes when ribbon operators are applied along the two distinct
handles of the torus. As we now show, not all states $\ket{(C_1,C_2)}$ are
allowed. Let the first operator carry the trivial irrep $R=I$, with $d_I=1$.
Then the second ribbon operator acts as
\begin{align}
&K_y^{\widetilde C R}K_x^{CI}\ket{\Psi_{\mathcal D(G)}}=
K_y^{\widetilde C R}
\prod_s A(s)T_x^C\ket{\mathbf 1}
\nonumber\\
&=
\prod_s A(s)K_y^{\widetilde C R}
T_x^C\ket{\mathbf 1}
\nonumber\\
&=\prod_s A(s)
\sum_{j=1}^{|\widetilde C|}
\sum_{i=1}^{|C|}
\sum_{k\in Z_{\widetilde C}}
\chi_R(\bar k)\,
\delta_{\bar k,\,
o_j\left(\prod_\ell x_\ell\right)\bar o_j}
\nonumber\\
&\hspace{1.0cm}\times
\left[
\prod_{j_y}^{\leftarrow}
\widetilde X_+^{\bar o_j r_{\widetilde C}o_j}(j_y)
\right]
\left[
\prod_{j_x}^{\leftarrow}
\widetilde X_+^{c_i}(j_x)
\right]
\ket{\mathbf 1}
\nonumber\\
&=
\prod_s A(s)
\sum_{j=1}^{|\widetilde C|}
\sum_{i=1}^{|C|}
\sum_{k\in Z_{\widetilde C}}
\chi_R(\bar k)\,
\delta_{\bar k,o_jc_i\bar o_j}
\nonumber\\
&\hspace{1.0cm}\times
\left[
\prod_{j_y}^{\leftarrow}
\widetilde X_+^{\bar o_j r_{\widetilde C}o_j}(j_y)
\right]
\left[
\prod_{j_x}^{\leftarrow}
\widetilde X_+^{c_i}(j_x)
\right]
\ket{\mathbf 1}.
\label{eq:two-K-action}
\end{align}
Here we defined
\begin{align}
\widetilde c_j=\bar o_j r_{\widetilde C}o_j\in \widetilde C,
\qquad
c_i\in C .
\end{align}
We used the fact that the oriented holonomy along the complementary ribbon is
$c_i$ after $T_x^C$ threads that flux through the $x$-cycle. The factor $\delta_{\bar k,o_jc_i\bar o_j}$ implies that
$k\in C^{-1}=\{g^{-1}:g\in C\}$. A nonzero term also requires
$k\in Z_{\widetilde C}=Z(r_{\widetilde C})$. Therefore,
\begin{align}
C^{-1}\cap Z_{\widetilde C}=\emptyset
\quad\Rightarrow\quad
K_y^{\widetilde C R}K_x^{CI}
\ket{\Psi_{\mathcal D(G)}}=0 .
\end{align}
For the two distinct nontrivial classes, $[s]\cap Z_r=\emptyset$ and
$[r]\cap Z_s=\emptyset$. Thus a sector with nontrivial flux through both cycles is possible only when the two fluxes have the same class. For $S_3$, every conjugacy class is closed under inversion. Its conjugacy
classes and the centralizers of our chosen representatives are
\begin{align}
[1]&=\{1\}, &
[s]&=\{s,sr,sr^2\}, &
[r]&=\{r,r^2\},
\\
Z_1&=S_3, &
Z_s&=\{1,s\}, &
Z_r&=\{1,r,r^2\}.
\end{align}
For the transposition class, $[s]\cap Z_s=\{s\}$. Taking
$r_{[s]}=s$, we obtain
\begin{align}
&K_y^{[s]R}K_x^{[s]I}\ket{\Psi_{\mathcal D(G)}}
\nonumber\\
&=
\prod_s A(s)
\sum_{j=1}^{|[s]|}
\sum_{i=1}^{|[s]|}
\sum_{k\in Z_s}
\chi_R(\bar k)\,
\delta_{\bar k,o_jc_i\bar o_j}
\nonumber\\
&\hspace{1.0cm}\times
\left[
\prod_{j_y}^{\leftarrow}
\widetilde X_+^{\bar o_j r_{[s]}o_j}(j_y)
\right]
\left[
\prod_{j_x}^{\leftarrow}
\widetilde X_+^{c_i}(j_x)
\right]
\ket{\mathbf 1}
\nonumber
\end{align}
\begin{align}
&=
\prod_s A(s)
\sum_{j=1}^{|[s]|}
\sum_{i=1}^{|[s]|}
\chi_R(\bar s)\,
\delta_{\bar o_j\bar s o_j,c_i}
\nonumber\\
&\hspace{1.0cm}\times
\left[
\prod_{j_y}^{\leftarrow}
\widetilde X_+^{\bar o_j r_{[s]}o_j}(j_y)
\right]
\left[
\prod_{j_x}^{\leftarrow}
\widetilde X_+^{c_i}(j_x)
\right]
\ket{\mathbf 1}
\nonumber\\
&=
\chi_R(\bar s)\prod_s A(s)
\left(
T_y^sT_x^s
+
T_y^{sr}T_x^{sr}
+
T_y^{sr^2}T_x^{sr^2}
\right)
\ket{\mathbf 1}.
\label{eq:ss-sector}
\end{align}
Thus the irrep of the centralizer contributes only the normalization factor
$\chi_R(\bar s)$. We define the sixth torus ground state as
\begin{align}
\ket{([s],[s])}
&=
T_y^{[s]}T_x^{[s]}\ket{\Psi_{\mathcal D(G)}}
=
K_y^{[s]I}K_x^{[s]I}\ket{\Psi_{\mathcal D(G)}}
\nonumber\\
&=
\prod_s A(s)
\left(
T_y^sT_x^s
+
T_y^{sr}T_x^{sr}
+
T_y^{sr^2}T_x^{sr^2}
\right)
\ket{\mathbf 1}.
\label{eq:sixth-gs}
\end{align}
The last nontrivial class is $[r]$. We take $r_{[r]}=r$, for which
\begin{align}
&K_y^{[r]R}K_x^{[r]I}\ket{\Psi_{\mathcal D(G)}}
\nonumber\\
&=
\prod_s A(s)
\sum_{j=1}^{|[r]|}
\sum_{i=1}^{|[r]|}
\sum_{k\in Z_{[r]}}
\chi_R(\bar k)\,
\delta_{\bar k,o_jc_i\bar o_j}
\nonumber\\
&\hspace{1.0cm}\times
\left[
\prod_{j_y}^{\leftarrow}
\widetilde X_+^{\bar o_jro_j}(j_y)
\right]
\left[
\prod_{j_x}^{\leftarrow}
\widetilde X_+^{c_i}(j_x)
\right]
\ket{\mathbf 1}.
\label{eq:rr-start}
\end{align}
In contrast to the previous applications of 1-form operators, there are now
two distinct allowed group elements, $k=r$ and $k=r^2$. These constrain the
flux threaded in both handles of the torus through
\begin{align}
\bar k=o_jc_i\bar o_j .
\end{align}
If $c_i=r$, then we must have either
$o_jr\bar o_j=\bar r=r^2$, or
$o_jr\bar o_j=(r^2)^{-1}=r$. In the second case $o_j=1$, while in the first
case $o_j=s$, using $srs=r^{-1}=r^2$ and $\bar s=s$. Performing the same
calculation for $c_i=r^2$, we obtain the analogous two possibilities.

Therefore,
\begin{align}
&K_y^{[r]R}K_x^{[r]I}\ket{\Psi_{\mathcal D(G)}}
\nonumber\\
&=
\prod_s A(s)
\Big[
\chi_R(r^2)
\left(
T_y^{r^2}T_x^r
+
T_y^rT_x^{r^2}
\right)
\nonumber\\
&\hspace{1.0cm}
+
\chi_R(r)
\left(
T_y^rT_x^r
+
T_y^{r^2}T_x^{r^2}
\right)
\Big]
\ket{\mathbf 1}.
\label{eq:rr-sector}
\end{align}
Thus threading a flux along the complementary direction produces a
two-dimensional subspace generated by
\begin{align}
\ket{\Phi_1}
&=
\prod_s A(s)
\left(
T_y^{r^2}T_x^r
+
T_y^rT_x^{r^2}
\right)
\ket{\mathbf 1},
\nonumber\\
\ket{\Phi_2}
&=
\prod_s A(s)
\left(
T_y^rT_x^r
+
T_y^{r^2}T_x^{r^2}
\right)
\ket{\mathbf 1}.
\end{align}
These states make the symmetry between applying the 1-form operator along the
$x$ or $y$ direction first manifest. They can also be written as linear combinations of
\begin{align}
K_y^{[r]R}K_x^{[r]I}\ket{\Psi_{\mathcal D(G)}},
\end{align}
with $R=I,\omega,\bar\omega$, the irreps of
$Z_{[r]}=\mathbb Z_3$.
\begin{align}
\zeta=e^{2\pi i/3},
\qquad
\chi_\omega(r^k)=\zeta^k,
\qquad
\chi_{\bar\omega}(r^k)=\zeta^{-k},
\end{align}
for $k=0,1,2$. Not all states
$K_y^{[r]R}K_x^{[r]I}\ket{\Psi_{\mathcal D(G)}}$ are linearly independent. We
choose $R=I$ and $R=\omega$ to define the last two torus ground states.

In summary, the ground-state subspace of the $G=S_3$ quantum double model on the torus is spanned by
\begin{align}
\ket{[1],[1]}&=\ket{\Psi_{\mathcal D(G)}}, \nonumber\\
\ket{[s],[1]}&=T_x^{[s]}\ket{\Psi_{\mathcal D(G)}}, \nonumber\\
\ket{[r],[1]}&=T_x^{[r]}\ket{\Psi_{\mathcal D(G)}}, \nonumber\\
\ket{[1],[s]}&=T_y^{[s]}\ket{\Psi_{\mathcal D(G)}}, \nonumber\\
\ket{[1],[r]}&=T_y^{[r]}\ket{\Psi_{\mathcal D(G)}}, \nonumber\\
\ket{([s],[s])}
&=
T_y^{[s]}T_x^{[s]}\ket{\Psi_{\mathcal D(G)}}
\nonumber\\
&=
K_y^{[s]I}K_x^{[s]I}\ket{\Psi_{\mathcal D(G)}}, \nonumber\\
\ket{([r]_I,[r]_I)}
&=
T_y^{[r]}T_x^{[r]}\ket{\Psi_{\mathcal D(G)}}
\nonumber\\
&=
K_y^{[r]I}K_x^{[r]I}\ket{\Psi_{\mathcal D(G)}}, \nonumber\\
\ket{([r]_I,[r]_\omega)}
&=
K_y^{[r]\omega}T_x^{[r]}\ket{\Psi_{\mathcal D(G)}}
\nonumber\\
&=
K_y^{[r]\omega}K_x^{[r]I}\ket{\Psi_{\mathcal D(G)}} .
\label{eq:S3-torus-basis}
\end{align}

\section{Dyonic 1-form operator fusion and topological charge projectors}
\label{A:closed-ribbon-projectors}

In this appendix we prove that the closed-ribbon line operators
$K_\sigma^{CR}$ form a basis of the topological closed-ribbon algebra.
We also fix their normalization and distinguish them from the orthogonal
charge projectors of Ref.~\cite{bombin_family_2008}.  All relations below
are exact identities for operators supported on the same proper closed
ribbon.  We label the simple objects of $\mathcal D(G)$ by
$\mathfrak a=(C,R)$, where $C$ is a conjugacy class, $c\in C$ is a fixed
representative, and $R\in\operatorname{Rep}(Z_c)$.  We write
\begin{align}
d_{\mathfrak a}=|C|d_R,
\qquad d_R=\dim R,
\qquad \mathfrak 0=([1],I).
\label{eq:anyonic-data}
\end{align}

\paragraph{Projector and line bases.}
Choose representatives $p_i$ of the left cosets $Z_c\backslash G$ and
representatives $q_j$ of the right cosets $G/Z_{c'}$.  Thus
\begin{align}
c_i&=p_i^{-1}cp_i,
&
k_i&=p_i^{-1}kp_i,
\nonumber\\
c'_j&=q_jc'q_j^{-1},
&
m_j&=q_jmq_j^{-1}.
\label{eq:transported-labels}
\end{align}
In the $F_\sigma^{h,g}$ convention of the main text, Bomb\'{\i}n's
projector with its original representation label is
\begin{align}
\Pi_{\sigma,\mathrm B}^{(C',R')}
=
\frac{d_{R'}}{|Z_{c'}|}
\sum_{j=1}^{|C'|}\sum_{m\in Z_{c'}}
\chi_{R'}(m)^*
F_\sigma^{m_j,c'_j}.
\label{eq:Bombin-projector-explicit}
\end{align}
For a proper closed ribbon these operators are Hermitian, mutually
orthogonal, and complete. For comparison with the modular-matrix convention used below, define
$\widetilde R'$ to be the conjugate centralizer representation,
\begin{align}
\chi_{\widetilde R'}(m)=\chi_{R'}(m)^*,
\qquad m\in Z_{c'}.
\label{eq:conjugate-centralizer-representation}
\end{align}
We use the oriented relabeling
\begin{align}
\Pi_\sigma^{\mathfrak b}
&:=
\Pi_{\sigma,\mathrm B}^{(C',\widetilde R')}
\nonumber\\
&=
\frac{d_{R'}}{|Z_{c'}|}
\sum_{j=1}^{|C'|}\sum_{m\in Z_{c'}}
\chi_{R'}(m)
F_\sigma^{m_j,c'_j},
\qquad \mathfrak b=(C',R').
\label{eq:Bombin-oriented-projector}
\end{align}
Here $d_{\widetilde R'}=d_{R'}$.  Since
$R'\mapsto\widetilde R'$ permutes $\operatorname{Rep}(Z_{c'})$, the
relabelled operators satisfy
\begin{align}
(\Pi_\sigma^{\mathfrak b})^\dagger
&=\Pi_\sigma^{\mathfrak b},
\nonumber\\
\Pi_\sigma^{\mathfrak b}\Pi_\sigma^{\mathfrak d}
&=\delta_{\mathfrak b,\mathfrak d}\Pi_\sigma^{\mathfrak b},
&
\sum_{\mathfrak b}\Pi_\sigma^{\mathfrak b}
&=\mathbf 1.
\label{eq:oriented-projector-algebra}
\end{align}

By contrast, the line operator defined in the main text is
\begin{align}
K_\sigma^{\mathfrak a}
=
\sum_{i=1}^{|C|}\sum_{k\in Z_c}
\chi_R(k)^*F_\sigma^{c_i,k_i}.
\label{eq:K-line-F-basis}
\end{align}
It is a quantum-character, or Verlinde-line, element rather than a
primitive idempotent.  The reversal of the two $F$ labels relative to
Eq.~\eqref{eq:Bombin-projector-explicit} is essential.  Its normalization
is canonical: for a contractible ribbon surrounding the vacuum,
\begin{align}
K_\sigma^{(C,R)}|\mathfrak 0\rangle
&=d_R T_\sigma^C|\mathfrak 0\rangle
=|C|d_R\widetilde T_\sigma^C|\mathfrak 0\rangle
\nonumber\\
&=d_{(C,R)}|\mathfrak 0\rangle.
\label{eq:line-vacuum-normalization}
\end{align}

\paragraph{Exact same-ribbon calculation.}
The change of basis follows directly from the microscopic ribbon
product
\begin{align}
F_\sigma^{h,g}F_\sigma^{h',g'}
=\delta_{g,g'}F_\sigma^{hh',g}.
\label{eq:F-same-ribbon-product}
\end{align}
Throughout this appendix, brackets denote the group commutator,
\begin{align}
[u,v]:=uvu^{-1}v^{-1},
\label{eq:group-commutator}
\end{align}
so $[u,v]=1$ is equivalent to $uv=vu$.

Multiplying Eqs.~\eqref{eq:K-line-F-basis} and
\eqref{eq:Bombin-oriented-projector}, and using
Eq.~\eqref{eq:F-same-ribbon-product}, gives
\begin{align}
K_\sigma^{\mathfrak a}\Pi_\sigma^{\mathfrak b}
&=
\frac{d_{R'}}{|Z_{c'}|}
\sum_{\substack{i,k,j,m\\k_i=c'_j}}
\chi_R(k)^*\chi_{R'}(m)
F_\sigma^{c_i m_j,c'_j}.
\label{eq:KPi-F-product}
\end{align}
The condition $k_i=c'_j$, together with $k\in Z_c$, implies
\begin{align}
[c_i,c'_j]=1,
\label{eq:commuting-transported-fluxes}
\end{align}
and hence $c_i\in Z_{c'_j}$. For fixed $j$, let
\begin{align}
H_j:=Z_{c'_j}=q_jZ_{c'}q_j^{-1}.
\end{align}
The operators $F_\sigma^{h,c'_j}$ with $h\in H_j$ form a copy of
$\mathbb C[H_j]$, because
\begin{align}
F_\sigma^{h,c'_j}F_\sigma^{h',c'_j}
=F_\sigma^{hh',c'_j}.
\end{align}
The terms of $K_\sigma^{\mathfrak a}$ that survive multiplication by
the $j$th block of $\Pi_\sigma^{\mathfrak b}$ therefore define
\begin{align}
z_{\mathfrak a,j}
:=
\sum_{\substack{i,\;k\in Z_c\\k_i=c'_j}}
\chi_R(k)^*c_i
\in\mathbb C[H_j].
\label{eq:central-element-z}
\end{align}
We now show that $z_{\mathfrak a,j}$ is central.  Let $y\in H_j$.
Because the $p_i$ represent $Z_c\backslash G$, there are a unique index
$i'$ and an element $a\in Z_c$ such that
\begin{align}
p_i y^{-1}=a p_{i'}.
\end{align}
Set $k':=a^{-1}ka\in Z_c$.  Then
\begin{align}
c_{i'}&=y c_i y^{-1},
&
p_{i'}^{-1}k'p_{i'}&=c'_j.
\end{align}
Thus conjugation by $y$ maps an allowed pair $(i,k)$ to another allowed
pair $(i',k')$.  Since $\chi_R$ is a class function on $Z_c$,
\begin{align}
\chi_R(k')=\chi_R(a^{-1}ka)=\chi_R(k).
\end{align}
Conjugation by $y$ consequently only permutes the terms in
$z_{\mathfrak a,j}$ without changing their coefficients.  Hence
\begin{align}
y z_{\mathfrak a,j}y^{-1}=z_{\mathfrak a,j}
\qquad (y\in H_j),
\end{align}
which proves
\begin{align}
z_{\mathfrak a,j}\in Z\!\left(\mathbb C[H_j]\right).
\end{align}
Let $\widetilde R'_j$ be the representation of $H_j$ transported from
$\widetilde R'$:
\begin{align}
\widetilde R'_j(y)
:=\widetilde R'(q_j^{-1}yq_j),
\qquad y\in H_j.
\end{align}
Its primitive central idempotent is
\begin{align}
e_{\widetilde R'_j}
&=
\frac{d_{R'}}{|H_j|}
\sum_{y\in H_j}
\chi_{\widetilde R'_j}(y^{-1})y
\nonumber\\
&=
\frac{d_{R'}}{|Z_{c'}|}
\sum_{m\in Z_{c'}}\chi_{R'}(m)m_j.
\label{eq:transported-central-idempotent}
\end{align}
Under the algebra identification
$y\mapsto F_\sigma^{y,c'_j}$, this is precisely the $j$th block of
$\Pi_\sigma^{\mathfrak b}$.

Because $z_{\mathfrak a,j}$ is central,
$\rho_{\widetilde R'_j}(z_{\mathfrak a,j})$ commutes with every matrix
$\rho_{\widetilde R'_j}(y)$.  Schur's lemma therefore implies
\begin{align}
\rho_{\widetilde R'_j}(z_{\mathfrak a,j})
=\lambda_{\mathfrak b}(\mathfrak a)\mathbf 1_{d_{R'}}.
\end{align}
Taking the trace determines the scalar:
\begin{align}
\lambda_{\mathfrak b}(\mathfrak a)
=\frac{\chi_{\widetilde R'_j}(z_{\mathfrak a,j})}{d_{R'}}.
\end{align}
Equivalently,
\begin{align}
z_{\mathfrak a,j}e_{\widetilde R'_j}
=
\frac{\chi_{\widetilde R'_j}(z_{\mathfrak a,j})}{d_{R'}}
e_{\widetilde R'_j}.
\label{eq:central-character-action}
\end{align}
Using
\begin{align}
\chi_{\widetilde R'_j}(c_i)
&=\chi_{\widetilde R'}(q_j^{-1}c_iq_j)
\nonumber\\
&=\chi_{R'}(q_j^{-1}c_iq_j)^*,
\end{align}
we obtain
\begin{align}
K_\sigma^{\mathfrak a}\Pi_\sigma^{\mathfrak b}
&=\lambda_{\mathfrak b}(\mathfrak a)
\Pi_\sigma^{\mathfrak b},
\label{eq:KPi-eigenvalue}\\
\lambda_{\mathfrak b}(\mathfrak a)
&=
\frac{1}{d_{R'}}
\sum_{\substack{i,\;k\in Z_c\\k_i=c'_j}}
\chi_R(k)^*
\chi_{R'}(q_j^{-1}c_iq_j)^*.
\label{eq:lambda-coset-sum}
\end{align}
To rewrite this as a group sum, set $x=p_iq_j$.  Then
$k=xc'x^{-1}$ and $q_j^{-1}c_iq_j=x^{-1}cx$.  Replacing the chosen
representatives $p_i$ by all elements of their left cosets introduces
the multiplicity $|Z_c|$.  Therefore
\begin{align}
\lambda_{\mathfrak b}(\mathfrak a)
&=
\frac{1}{|Z_c|d_{R'}}
\sum_{\substack{x\in G\\{}[c,xc'x^{-1}]=1}}
\chi_R(xc'x^{-1})^*
\chi_{R'}(x^{-1}cx)^*.
\label{eq:lambda-group-sum}
\end{align}
This form also makes the independence of $j$ manifest.In the convention of Ref.~\cite{coste_finite_2000}, the modular matrix
of $\mathcal D(G)$ and its vacuum row are
\begin{align}
S_{\mathfrak a\mathfrak b}
&=
\frac{1}{|Z_c||Z_{c'}|}
\sum_{\substack{x\in G\\{}[c,xc'x^{-1}]=1}}
\chi_R(xc'x^{-1})^*
\chi_{R'}(x^{-1}cx)^*,
\nonumber\\
S_{\mathfrak0\mathfrak b}
&=\frac{d_{R'}}{|Z_{c'}|}
=\frac{|C'|d_{R'}}{|G|}.
\label{eq:S-and-vacuum-row}
\end{align}
Comparison with Eq.~\eqref{eq:lambda-group-sum} yields the exact
same-ribbon identity
\begin{align}
K_\sigma^{\mathfrak a}\Pi_\sigma^{\mathfrak b}
=
\frac{S_{\mathfrak a\mathfrak b}}
     {S_{\mathfrak0\mathfrak b}}
\Pi_\sigma^{\mathfrak b}.
\label{eq:exact-KPi-S}
\end{align}
No ground-state or defect-free projector is required in this identity. If Bomb\'{\i}n's original label is retained instead, the exact relation
is
\begin{align}
K_\sigma^{\mathfrak a}
\Pi_{\sigma,\mathrm B}^{(C',R')}
=
\frac{S_{\mathfrak a,(C',\widetilde R')}}
     {S_{\mathfrak0,(C',R')}}
\Pi_{\sigma,\mathrm B}^{(C',R')}.
\label{eq:Bombin-unrelabelled-transform}
\end{align}
Thus the conjugation of the centralizer-representation label is an
orientation convention, not a change in the set of charge sectors.

Three checks of Eq.~\eqref{eq:exact-KPi-S} are
\begin{align}
\lambda_{\mathfrak b}(\mathfrak0)&=1,
&
\lambda_{\mathfrak0}(\mathfrak a)&=d_{\mathfrak a},
\nonumber\\
\lambda_{(C',R')}([1],\Gamma)
&=\chi_\Gamma(C')^*.
\label{eq:eigenvalue-checks}
\end{align}
The complex conjugate in the last equality follows from the orientation
of $K_\sigma^{[1]\Gamma}$ in Eq.~\eqref{eq:K-line-F-basis}.  Reversing
the electric loop replaces $\chi_\Gamma(C')^*$ by
$\chi_\Gamma(C')$.  Thus a mixed Wilson--'t Hooft relation written with
$\chi_\Gamma(C')$ uses the opposite loop orientation.

\paragraph{Explicit invertible transformation.}
Since the projectors resolve the identity,
Eq.~\eqref{eq:exact-KPi-S} gives
\begin{align}
K_\sigma^{\mathfrak a}
&=\sum_{\mathfrak b}
\mathsf L_{\mathfrak a\mathfrak b}
\Pi_\sigma^{\mathfrak b},
&
\mathsf L_{\mathfrak a\mathfrak b}
&:=\frac{S_{\mathfrak a\mathfrak b}}
          {S_{\mathfrak0\mathfrak b}}.
\label{eq:K-Pi-forward}
\end{align}
For the quantum double,
\begin{align}
\mathsf L_{(C,R),(C',R')}
=\frac{|G|}{|C'|d_{R'}}
S_{(C,R),(C',R')}.
\label{eq:L-explicit}
\end{align}
Unitarity of the full modular matrix gives
\begin{align}
\Pi_\sigma^{\mathfrak b}
&=\sum_{\mathfrak a}
(\mathsf L^{-1})_{\mathfrak b\mathfrak a}
K_\sigma^{\mathfrak a},
\nonumber\\
(\mathsf L^{-1})_{\mathfrak b\mathfrak a}
&=S_{\mathfrak0\mathfrak b}S_{\mathfrak a\mathfrak b}^*
\nonumber\\
&=\frac{|C'|d_{R'}}{|G|}
S_{\mathfrak a,(C',R')}^*,
\qquad \mathfrak b=(C',R').
\label{eq:K-Pi-inverse}
\end{align}
Indeed,
\begin{align}
\sum_{\mathfrak b}
\mathsf L_{\mathfrak a\mathfrak b}
(\mathsf L^{-1})_{\mathfrak b\mathfrak c}
=\sum_{\mathfrak b}
S_{\mathfrak a\mathfrak b}S_{\mathfrak c\mathfrak b}^*
=\delta_{\mathfrak a,\mathfrak c}.
\end{align}
This invertible transformation proves that the $K_\sigma^{CR}$ form a
basis of the same topological closed-ribbon algebra.  Its dimension is
\begin{align}
\dim\mathcal K_\sigma
=\sum_{C\in\operatorname{Conj}(G)}
\bigl|\operatorname{Rep}(Z_c)\bigr|.
\end{align}

\paragraph{Fusion of the line basis.}
The Verlinde formula~\cite{Verlinde} for the full modular matrix is
\begin{align}
N_{\mathfrak a\mathfrak c}^{\mathfrak d}
=
\sum_{\mathfrak b}
\frac{S_{\mathfrak a\mathfrak b}
S_{\mathfrak c\mathfrak b}
S_{\mathfrak d\mathfrak b}^*}
{S_{\mathfrak0\mathfrak b}}.
\label{eq:DG-Verlinde}
\end{align}
Equivalently, the eigenvalues of the line operators obey
\begin{align}
\frac{S_{\mathfrak a\mathfrak b}}
     {S_{\mathfrak0\mathfrak b}}
\frac{S_{\mathfrak c\mathfrak b}}
     {S_{\mathfrak0\mathfrak b}}
=
\sum_{\mathfrak d}
N_{\mathfrak a\mathfrak c}^{\mathfrak d}
\frac{S_{\mathfrak d\mathfrak b}}
     {S_{\mathfrak0\mathfrak b}}.
\end{align}
Together with the orthogonality of the projectors, this gives
\begin{align}
K_\sigma^{\mathfrak a}K_\sigma^{\mathfrak b}
=\sum_{\mathfrak c}
N_{\mathfrak a\mathfrak b}^{\mathfrak c}
K_\sigma^{\mathfrak c}.
\label{eq:full-line-fusion}
\end{align}
Here $N_{\mathfrak a\mathfrak b}^{\mathfrak c}\in\mathbb Z_{\geq0}$
is the multiplicity with which the simple anyon $\mathfrak c$ occurs in
$\mathfrak a\otimes\mathfrak b$:
\begin{align}
\mathfrak a\otimes\mathfrak b
\simeq
\bigoplus_{\mathfrak c}
N_{\mathfrak a\mathfrak b}^{\mathfrak c}\,\mathfrak c.
\end{align}
Thus these coefficients retain both the magnetic conjugacy-class labels
and the electric centralizer-representation labels.
\paragraph{Consistency with electric fusion.}
For $\mathfrak a=([1],\Gamma)$,
Eq.~\eqref{eq:eigenvalue-checks} gives
\begin{align}
W_\sigma^\Gamma=K_\sigma^{[1]\Gamma}
=\sum_{C'}\chi_\Gamma(C')^*
\sum_{R'\in\operatorname{Rep}(Z_{c'})}
\Pi_\sigma^{(C',R')}.
\label{eq:Wilson-projector-transform}
\end{align}
Orthogonal multiplication of the $\Pi^{CR}$ therefore does not imply
single-channel fusion for the line basis.  Instead, the ordinary
character-product identity, followed by complex conjugation, gives
\begin{align}
W_\sigma^{\Gamma_1}W_\sigma^{\Gamma_2}
=\sum_{\Gamma_3}
N_{\Gamma_1\Gamma_2}^{\Gamma_3}
W_\sigma^{\Gamma_3}.
\end{align}
Hence the $K^{CR}$ set the line, or fusion, basis, whereas the
$\Pi^{CR}$ the primitive-idempotent basis.

\bibliography{main}

\end{document}